%% file: arxiv-v7.tex
\documentclass[conference,compsoc]{IEEEtran}
\usepackage{booktabs}
\usepackage{tikz}
\usepackage{url}
\usepackage{pdfpages}
\usepackage{afterpage}
\usepackage{color, colortbl}
\usepackage{graphicx}
\usepackage{amsmath}
\usepackage{import}
\usepackage{pifont}
\usepackage{subfigure}
\usepackage{footmisc}
\usepackage{changepage}
\usepackage{hyperref}
\definecolor{Gray}{gray}{0.9}
\definecolor{LightCyan}{rgb}{0.88,1,1}
\pdfoutput=1

\makeatletter
\DeclareRobustCommand*\textsubscript[1]{%
  \@textsubscript{\selectfont#1}}
\def\@textsubscript#1{%
  {\m@th\ensuremath{_{\mbox{\fontsize\sf@size\z@#1}}}}}
\makeatother

\usepackage[flushleft]{threeparttable}

\begin{document}

\title{Benchmarking State-of-the-Art Deep Learning Software Tools}

\author{\IEEEauthorblockN{Shaohuai Shi, Qiang Wang, Pengfei Xu, Xiaowen Chu}
\IEEEauthorblockA{Department of Computer Science, Hong Kong Baptist University
\\\{csshshi, qiangwang, pengfeixu, chxw\}@comp.hkbu.edu.hk}
}

\maketitle

\begin{abstract}
Deep learning has been shown as a successful machine learning method for a variety of tasks, and its popularity results in numerous open-source deep learning software tools. Training a deep network is usually a very time-consuming process. To address the computational challenge in deep learning, many tools exploit hardware features such as multi-core CPUs and many-core GPUs to shorten the training time. However, different tools exhibit different features and running performance when training different types of deep networks on different hardware platforms, which makes it difficult for end users to select an appropriate pair of software and hardware. In this paper, we aim to make a comparative study of the state-of-the-art GPU-accelerated deep learning software tools, including Caffe, CNTK, MXNet, TensorFlow, and Torch. We first benchmark the running performance of these tools with three popular types of neural networks on two CPU platforms and three GPU platforms. We then benchmark some distributed versions on multiple GPUs. Our contribution is two-fold. First, for end users of deep learning tools, our benchmarking results can serve as a guide to selecting appropriate hardware platforms and software tools. Second, for software developers of deep learning tools, our in-depth analysis points out possible future directions to further optimize the running performance.
\end{abstract}


\IEEEpeerreviewmaketitle

\section{Introduction} \label{introduction}
In the past decade, deep learning has been successfully applied in diverse areas including computer vision, speech recognition, natural language processing, etc. The success of deep learning is attributed to its high representational ability of input data, by using various layers of artificial neurons \cite{lecun2015deep}. GPUs have played a key role in the success of deep learning by significantly reducing the training time \cite{deng2012three}. In order to increase the efficiency in developing deep learning methods, there are a number of open-source deep learning toolkits including Caffe from UC Berkeley \cite{jia2014caffe}, CNTK from Microsoft \cite{yu2014introduction}, TensorFlow from Google \cite{abadi2015tensorflow}, Torch \cite{collobert2011torch7}, MXNet \cite{chen2015mxnet}, and many other tools like Theano \cite{team2016theano}, Baidu's PaddlePaddle \cite{baidupaddle}, etc. All these tools support multi-core CPUs and many-core GPUs. One of the main tasks of deep learning is to learn a number of weights in each layer of network, which can be implemented by vector or matrix operations. TensorFlow uses Eigen \cite{eigen2016} as accelerated matrix operation library, while Caffe, CNTK, MXNet and Torch employ OpenBLAS \cite{openblas2013}, Intel MKL \cite{intel2007intel} or cuBLAS \cite{toolkit20114} to speed up matrix related calculations. All the mentioned tools import cuDNN \cite{chetlur2014cudnn}, which is a GPU-accelerated deep learning library, for neural network computing. However, because of the difference of optimization methods by vendors, these tools exhibit different running performance even when training the same neural network on the same hardware platform. Furthermore, the performance of a tool also changes a lot when training different types of networks, or using different types of hardware.

Given the diversity of deep learning software tools and the below hardware platforms, it could be difficult for end users to select an appropriate platform to carry out their deep learning tasks. In this paper, we benchmark a set of state-of-the-art GPU-accelerated deep learning tools (i.e., Caffe, CNTK, MXNet, TensorFlow and Torch) using three major types of deep neural networks (i.e., fully connected neural networks (FCNs) \cite{lecun1989backpropagation}, convolutional neural networks (CNNs) \cite{krizhevsky2012imagenet}\cite{tang2015document}\cite{graves2013speech}, and recurrent neural networks (RNNs) \cite{zaremba2014recurrent}\cite{graves2014towards}\cite{tang2015document}), and discuss their advantages and disadvantages on both contemporary CPUs and GPUs in terms of running time performance. Synthetic data and real-world data are both included in our performance evaluation. Our hardware platforms include two types of CPU (i.e., desktop CPU Intel i7-3820 and server-grade CPU Intel Xeon E5-2630) and three types of Nvidia GPU (i.e., GTX 980, GTX 1080 and Telsa K80 with Maxwell, Pascal and Kepler architecture respectively). We also use two Tesla K80 cards (i.e., a total of 4 GK210 GPUs) to evaluate the multi-GPU performance of each tool. For each network type, we choose a small-size network and a large-size network for evaluation.\footnote{Our source code and experimental data can be downloaded from \url{http://www.comp.hkbu.edu.hk/~chxw/dlbench.html}.} Our major findings are summarized as follows\footnote{The software tools are being upgraded frequently. The findings are based on our own experimental platforms and only apply to the software versions specified in the paper. We will regularly update our benchmarking results for new software versions on our website.}:
\begin{enumerate}
\item In general, the performance does not scale very well on many-core CPUs. In many cases, the performance of using 16 CPU cores is only slightly better than that of using 4 or 8 CPU cores. Using CPU computing platform, TensorFlow has a relatively better scalability compared with other tools.
\item With a single GPU platform, Caffe, CNTK and Torch perform better than MXNet and TensorFlow on FCNs; MXNet is outstanding in CNNs, especially the larger size of networks, while Caffe and CNTK also achieve good performance on smaller CNN; For RNN of LSTM, CNTK obtains excellent time efficiency, which is up to 5-10 times better than other tools.
\item With the parallelization of data during training, all the multi-GPU versions have a considerable higher throughput and the convergent speed is also accelerated. CNTK performs better scaling on FCN and AlexNet, while MXNet and Torch are outstanding in scaling CNNs.
\item GPU platform has a much better efficiency than many-core CPUs. All tools can achieve significant speedup by using contemporary GPUs.
\item Among the three GPU platforms, GTX1080 performs the best in most cases, due to its highest computational power.
\item To some extent, the performance is also affected by the design of configuration files. For example, CNTK allows the end users to fine-tune the system and trade off GPU memory for better computing efficiency, and MXNet gives the users to configure the auto-tune setting in using cuDNN library.
\end{enumerate}

The rest of the paper is organized as follows. Section \ref{backgroundandrelatedwork} presents the background and related work. Section \ref{experimentalmethods} introduces our benchmarking platform and methodology. Experimental results are presented in Section \ref{results}, followed by our discussion in Section \ref{discussion}. We conclude the paper and introduce our future work in Section \ref{conclusionandfuturework}.

\section{Background and Related Work} \label{backgroundandrelatedwork}
With the fast development of deep learning techniques, numerous deep neural networks including fully connected neural networks (FCNs), convolutional neural networks (CNNs), recurrent neural networks (RNNs), restricted boltzmann machine (RBM) have been developed for different applications \cite{schmidhuber2015deep}. In this paper, we focus on analyzing the running performance (or time speed) and also the convergent speed of data parallelization algorithms with three types of neural networks, namely FCNs, CNNs and RNNs. FCN has a long history dated back to 1980s when the backpropagation (BP) algorithm \cite{hecht1989theory} was first developed. And for CNN and RNN, they have been revealed strong power on the applications of image recognition and natural language processing respectively \cite{tang2015document}\cite{graves2013speech}.

\begin{figure*}[htbp]
  \centering     
  \subfigure[Fully Connected Network]
  {
    \includegraphics[width=0.24\linewidth]{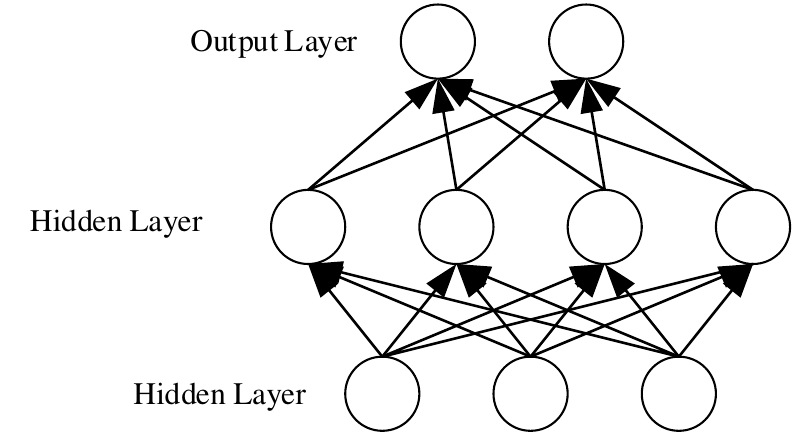}
    \label{fig:dnna}
  }
  \subfigure[Convolutional Neural Network (AlexNet \cite{krizhevsky2012imagenet})]
  {
    \includegraphics[width=0.45\linewidth]{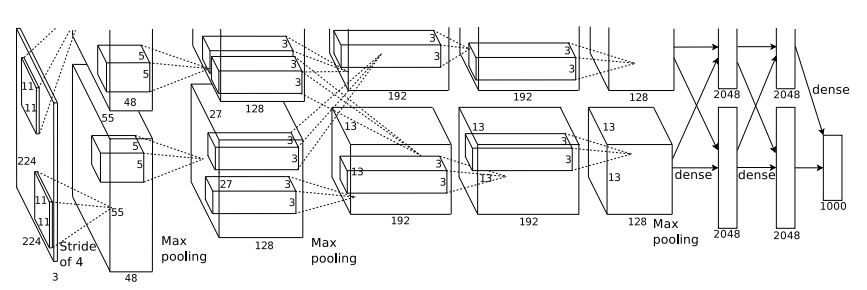}
    \label{fig:dnnb}
  }
  \subfigure[Recurrent Neural Network.]
  {
    \includegraphics[width=0.24\linewidth]{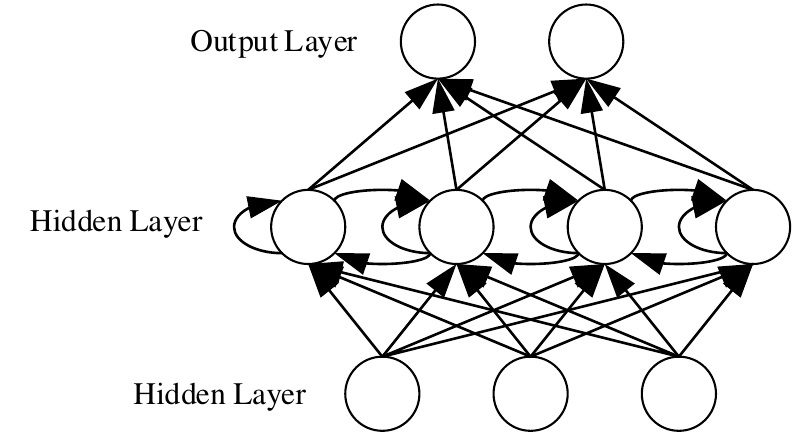}
    \label{fig:dnnc}
  }
  \caption{Examples of deep learning models }\label{fig:dnn_model}
\end{figure*}

FCN is a feed-forward neural network, the first successful use of which is the ZIP codes recognition by Yann LeCun et al. in 1989 \cite{lecun1989backpropagation}. Assume that an FCN \textit{F} has \textit{I} input neurons and \textit{O} output neurons, and there are \textit{H} hidden layers, where the number of neurons is \textit{N\textsubscript{i}} (\textit{i} = 1, 2, ..., \textit{H}). The total number of weights in \textit{F} is:
\begin{equation}
\label{equation:1}
{I\times{N_1} + \displaystyle\sum_{i=1}^{H-2} {N_i\times N_{i+1}} + N_{H-1}\times O}.
\end{equation}

To reduce the total number of parameters in each layer, CNNs build a convolutional layer by using a set of kernels, and the parameters of each kernel are shared across entire field (e.g., a channel of color image). Eq. (\ref{equation:1}) shows that fully-connected layers easily lead to huge amount of parameters to learn. CNNs are developed to reduce the total number of parameters in each layer. Starting from the LeNet architecture, CNNs have accomplished a lot of successful tasks including ImageNet classification \cite{krizhevsky2012imagenet}, face recognition \cite{lawrence1997face}, and object detection \cite{zbontar2016stereo}. An example of CNN is shown in Fig. \ref{fig:dnnb} whose number of parameters is up to 61 millions \cite{krizhevsky2012imagenet}.

RNN allows cyclical connections of the units in the network \cite{graves2014towards}\cite{tang2015document}\cite{graves2013speech}, as illustrated in Fig. \ref{fig:dnnc}. RNN can link the entire historical input sequence to each output and find the relationship between the contextual features of inputs and the output. With this characteristic, RNN can maintain the information given by the former inputs similar with memory during the training period of one single sample. Furthermore, to address the training difficulty of vanished and exploding gradients, Long-short Term Memory (LSTM) \cite{hochreiter1997long}\cite{zaremba2014recurrent} has been proposed to record and discard the information properly. RNN with LSTM units has been proved one of the most successful in handling tasks of speech recognition and natural language processing \cite{tang2015document}\cite{graves2013speech}.

With the growing success of deep learning, there come out many popular open source GPU-accelerated software tools, among which Caffe, CNTK, MXNet, TensorFlow and Torch are examples of the most active and popular ones.

Caffe is developed by Berkeley Vision and Learning Center (BVLC) and has become open source since 2014. The authors \cite{jia2014caffe} claim that Caffe can process 40 million images per day with GPU-accelerated version on a single NVIDIA K40 or Titan GPU. After integrated with cuDNN, it achieves speedup about 1.3x on NVIDIA K40 card \cite{chetlur2014cudnn}.

CNTK is a unified computational network toolkit developed by Microsoft Research, which supports most popular networks. At December 2015, the official reported a performance result benchmarking on a fully connected 4-layer neural network compared to Caffe, TensorFlow, Theano and Torch, and the result shows that CNTK with multiple GPUs on single node or cross multiple machines achieve a much better speed (about 1.5x speedup) than the other compared toolkits.

MXNet is a multi-language supported deep learning framework, which aims to provide much more efficient and flexible programming interfaces to maximize efficiency and productivity.

TensorFlow is developed by Google which has integrated most common units in deep learning framework using data flow graphs. It supports many up-to-date networks such as CNNs, RNNs with different settings. TensorFlow is designed for remarkable flexibility, portability, and high efficiency of equipped hardware.

Torch is a scientific computing framework which provides data structures for the most useful components in machine learning algorithms such as multi-dimensional tensors and mathematical operations over them.

To accelerate the training speed of deep neural networks, both CPUs SSE techniques and float points SIMD models are used to implement deep learning algorithms \cite{vanhoucke2011improving}, which achieve 3x speedup over optimized floating-point baseline. Andre Viebke et al. also exploit thread- and SIMD-parallelism Intel Xeon Phi processors to speedup training of CNNs \cite{viebke2015potential}. Considering parallel algorithms for GPU, Jeffrey Dean et al. \cite{dean2012large} proposed a large scaled distribute deep networks and developed two algorithms (i.e., Downpour SGD and Sandblaster L-BFGS) that can be easily running on computing clusters with thousands of machines including GPU machines. Another way to accelerate training speed is to reduce the number of learning parameters, Song Han et al. \cite{han2015learning} use the method of pruning redundant connections to reduce parameters without losing network representational ability, which could reduce the number of parameters of AlexNet from 61 millions to 6.7 millions. Bahrampour et al. \cite{bahrampour2015comparative} did the similar work with us, but they only used a single architecture of GPU (i.e., NVIDIA Maxwell Titan X) and old version softwares (e.g., cuDNN v2, v3). Our previous work also discussed the benchmark results with older version softwares on a single GPU \cite{shi2016benchmarking}. In this paper, we use three major architectures of GPU and benchmark on some newer networks (e.g., ResNet-50) and softwares (e.g., cuDNN v5), and we also go insight into the source codes to analyze performance. Furthermore, the performance across multiple GPUs in a single machine is also compared in the paper.

The scalability of training is also critical for a deep learning framework since a single GPU has relatively small memory which limits the network size. Support of multiple GPUs becomes a standard in recent deep learning tools. To utilize multiple GPUs, distributed synchronous stochastic gradient descent (SGD) method has been widely used and achieved very good scaling performance \cite{das2016distributed}\cite{chen2016revisiting}. In terms of scalability, we focus on evaluating the time and convergence rate of data synchronization method. In data parallelization model, a mini-batch of samples with size $M$ is split into $M/N$ fragments for $N$ workers, and each worker performs forward and backward on its assigned samples independently with a duplicated model. When all workers finish, the gradients are aggregated and the model is updated. In practice, different tools implement synchronous SGD algorithm differently.

\begin{itemize}
\item Caffe. Tree reduction strategy is used in Caffe for data communications among the GPUs \cite{caffemultigpu}. For example, suppose there are 4 GPUs labelled as 0, 1, 2 and 3. First, GPU 0 exchanges gradients with GPU 1 while GPU 2 exchanges gradients with GPU 3. Then GPU 0 and GPU 2 will exchange gradients with each other. Afterwards, GPU 0 will calculate the updated model, and then transfers the updated model to GPU 2; and then GPU 0 transfers the model to GPU 1 and GPU 2 transfers the model to GPU 3 at the same time.
\item CNTK. It uses MPI for data communications among GPUs. CNTK supports four types of parallel SGD algorithms (i.e., DataParallelSGD, BlockMomentumSGD, ModelAveragingSGD and DataParallelASGD) \cite{cntkmultiplenodes}. In data parallel SGD, which is the focus of this paper, the tool distributes each mini-batch over $N$ workers. The resulting partial gradients in each worker with specified number of bits each gradient (1-bit SGD algorithm \cite{seide20141}) are then exchanged and aggregated after each mini-batch.
\item MXNet. It also partitions a mini-batch of samples to all GPUs. Each GPU performs forward and backward on a mini-batch size of $M/N$. The gradients are then summed over all GPUs before updating the model, which can be performed on both CPU and GPU.
\item TensorFlow. It places an individual model replica on each GPU. And it splits the mini-batch across the GPUs. There is a duplicated model in the CPU side, which is updated synchronously after all GPUs finish processing a mini-batch of data.
\item Torch. The data parallelization mechanism is similar to MXNet, and it puts the operation of gradients aggregation on the GPU side, which reduces the overhead of PCI-e data transfer.
\end{itemize}

\section{Experimental Methods} \label{experimentalmethods}
Processing time and convergence rate are two main factors that users concern when training a deep learning model. So we mainly measure these two metrics to evaluate each tool. On one hand, one popular and effective way to evaluate the running performance is to measure the time duration of an iteration that processes a mini-batch of input data. In practice, after a certain round of iterations or the convergence of learning, the training progress will be terminated. Therefore, we benchmark these tools by using a range of mini-batch sizes for different types of network. For each mini-batch size, we run numerous iterations and evaluate their average running speed. On the other hand, since data parallelization may affect the convergence rate, we also compare the convergence rates for the case of multiple GPUs.

We use both synthetic data sets and real-world data sets in the experiments. Synthetic data sets are mainly used for evaluating the running time performance, and the real-world data sets are used to measure the convergence rates. The methods of time measurement for each tool are as follows:
\begin{itemize}
\item Caffe: use ``caffe train'' command to train a specified network, and then calculate the average time difference between two consecutive iterations.
\item CNTK: similar to Caffe, but we exclude the first epoch which includes the time of disk I/O.
\item MXNet: use internal timing function which outputs the time details of each epoch and iteration.
\item TensorFlow: use timing function to calculate the average iteration time in source scripts.
\item Torch: the same as TensorFlow.
\end{itemize}

All the tools provide very flexible programming APIs or configuration options for performance optimization. For example, in CNTK, we may specify ``maxTempMemSizeInSamplesForCNN'' option in configuration file to control the size of temporary memory used by CNNs which may result in slightly worse efficiency but less memory requirement. MXNet, TensorFlow and Torch also have rich set of APIs for users to choose for the computation tasks. In other words, there may exist different APIs performing the same operations. As a result, we need to point it out that the performance results reported in our experiments are based on our understanding of usage of these tools and are not necessarily the best that can be achieved.

The software versions and related libraries in our experiments are shown in Table \ref{table:software_new}. Results on older versions can be found in \cite{shi2016benchmarking}.
\begin{table}[!ht]
\centering
\caption{The softwares used for experiments.}
\label{table:software_new}
\begin{tabular}{llllll}
\toprule
Software  &  Marjor Version & GitHub Commit ID   & cuDNN \\\midrule
Caffe     &  1.0.0		& 4ba654f    & v5.1 \\
CNTK      &  1.72	  	& f686879    & v5.1 \\
MXNet 	  &  0.7.0 		& 34b2798    & v5.1 \\
TensorFlow &  0.11		& 47dd089    & v5.1 \\
Torch     &  7			& 5c7c762    & v5.1 \\
\bottomrule
\end{tabular}
\end{table}

\begin{table}[htbp]
\begin{threeparttable}
\centering
\caption{The experimental setup of neural networks for synthetic data.}
\label{table:networksetup}
\begin{tabular}{c|c|c|c|c|c}
\toprule
\multicolumn{2}{c|}{Networks} & Input & Output & Layers & Parameters   \\\cline{1-6}
FCN   & FCN-S     & 26752 & 26752 & 5 & \textasciitilde 55 millions    \\\cline{1-6}
CNN	  & AlexNet-S   & 150528& 1000 & 4 &  \textasciitilde 61 millions    \\\cline{2-6}
      & ResNet-50 & 150528& 1000 & 50 & \textasciitilde 3.8 billions   \\\cline{1-6}
\end{tabular}
	\begin{tablenotes}
      \item[]Note: FCN-S has 4 hidden layers with 2048 nodes each layer; and both batch normalization operation and dropout operations are excluded in AlexNet-S; For testing CNN, the input data has the dimension of color image from ImageNet (i.e., 224x224x3) and the output dimension is the number of class of ImageNet.
    \end{tablenotes}
  \end{threeparttable}
\end{table}

\begin{table}[htbp]
\begin{threeparttable}
\centering
\caption{The experimental setup of neural networks for real data.}
\label{table:networksetup_new}
\begin{tabular}{c|c|c|c|c|c}
\toprule
\multicolumn{2}{c|}{Networks} & Input & Output & Layers & Parameters   \\\cline{1-6}
FCN   & FCN-R     & 784 & 10 & 5 & \textasciitilde 13 millions     \\\cline{1-6}
CNN   & AlexNet-R   & 3072& 10 & 4 & \textasciitilde 81 thousands \\\cline{2-6}
      & ResNet-56 & 3072& 10 & 56 & \textasciitilde 0.85 millions \\\cline{1-6}
RNN   & LSTM   & 10000 & 10000 & 2 & \textasciitilde 13 millions  \\\cline{1-6}
\end{tabular}
	\begin{tablenotes}
      \item[]Note: FCN-R has 3 hidden layers with the number of nodes: 2048, 4096 and 1024 respectively. The architecture of AlexNet-R is the same with the AlexNet for Cifar10 in the original paper \cite{krizhevsky2012imagenet} except that the local response normalization (LRN) operation is excluded which is not supported by CNTK. For ResNet-56, we exploit the architecture from the original paper \cite{he2015deep}.
    \end{tablenotes}
  \end{threeparttable}
\end{table}

\begin{table}[!ht]
\begin{threeparttable}
\centering
\caption{The experimental setup of hardware.}
\label{table:hardwaresetup}
\begin{tabular}{lllll}
\toprule
Computational Unit  &  Cores  &   Memory  & OS & CUDA \\\midrule
Intel CPU i7-3820 & 4 & 64 GB & Ubuntu 14.04 & -\\
Intel CPU E5-2630x2 & 16 & 128 GB & CentOS 7.2 & -\\
GTX 980 & 2048 & 4 GB & Ubuntu 14.04 & 8.0 \\
GTX 1080 & 2560 & 8 GB & Ubuntu 14.04 & 8.0\\
Telsa K80 GK210 & 2496 & 12 GB & CentOS 7.2 & 8.0\\
\bottomrule
\end{tabular}
	\begin{tablenotes}
      \item[]Note: There are 2 GK210 GPUs on a K80 card, but only one GPU is used for testing on single GPU performance comparison.
    \end{tablenotes}
  \end{threeparttable}
\end{table}

\begin{table}[!ht]
\begin{threeparttable}
\centering
\caption{The experimental hardware setting for data parallelization.}
\label{table:multigpuetup}
\begin{tabular}{llllll}
\toprule
GPUs  &   CPU & Memory &  PCIe  & OS & CUDA \\\midrule
K80x2 & E5-2630v4 & 128GB & PCIe 3.0  & CentOS 7.2 & 8.0\\
\bottomrule
\end{tabular}
	\begin{tablenotes}
      \item[]Note: There are two GK210 GPUs on a K80 card, so one K80 card can be used to conduct two-GPU parallelization experiments and two K80 cards are used to conduct four-GPU parallelization experiments.
    \end{tablenotes}
  \end{threeparttable}
\end{table}

\begin{table*}[htbp]
\centering
\caption{The conducted experiments with the combination of networks, tools and hardware.}
\label{table:conductedexperiments}
\begin{tabular}{ccccccccc}
\cline{1-9}
Network      &  Tool    & \multicolumn{7}{c}{Hardware} \\\cline{1-9}
& & \multicolumn{2}{c}{CPU} & \multicolumn{3}{c}{Single GPU} & \multicolumn{2}{c}{Multiple GPUs} \\\cline{3-9}
& & i7-3820 & E5-2630v3 & GTX 980 & GTX 1080 & K80 & 2 GPUs & 4 GPUs \\\cline{1-9}
	  & Caffe & \ding{51} & \ding{51} & \ding{51} & \ding{51} & \ding{51} & \ding{55} & \ding{55} \\\cline{2-9}
	  & CNTK & \ding{51} & \ding{51} & \ding{51} & \ding{51} & \ding{51} & \ding{55} & \ding{55} \\\cline{2-9}
FCN-S & MXNet & \ding{51} & \ding{51} & \ding{51} & \ding{51} & \ding{51} & \ding{55} & \ding{55} \\\cline{2-9}
	  & TensorFlow & \ding{51} & \ding{51} & \ding{51} & \ding{51} & \ding{51} & \ding{55} & \ding{55} \\\cline{2-9}
      & Torch & \ding{51} & \ding{51} & \ding{51} & \ding{51} & \ding{51} & \ding{55} & \ding{55} \\
\hline \hline
      & Caffe & \ding{51} & \ding{51} & \ding{51} & \ding{51} & \ding{51} & \ding{55} & \ding{55} \\\cline{2-9}
	  & CNTK & \ding{51} & \ding{51} & \ding{51} & \ding{51} & \ding{51} & \ding{55} & \ding{55} \\\cline{2-9}
AlexNet-S & MXNet & \ding{51} & \ding{51} & \ding{51} & \ding{51} & \ding{51} & \ding{55} & \ding{55} \\\cline{2-9}
	  & TensorFlow & \ding{51} & \ding{51} & \ding{51} & \ding{51} & \ding{51} & \ding{55} & \ding{55} \\\cline{2-9}
      & Torch & \ding{51} & \ding{51} & \ding{51} & \ding{51} & \ding{51} & \ding{55} & \ding{55} \\
\hline \hline
      & Caffe & \ding{51} & \ding{51} & \ding{51} & \ding{51} & \ding{51} & \ding{55} & \ding{55} \\\cline{2-9}
	  & CNTK & \ding{55} & \ding{55} & \ding{51} & \ding{51} & \ding{51} & \ding{55} & \ding{55} \\\cline{2-9}
ResNet-50 & MXNet & \ding{51} & \ding{51} & \ding{51} & \ding{51} & \ding{51} & \ding{55} & \ding{55} \\\cline{2-9}
	  & TensorFlow & \ding{51} & \ding{51} & \ding{51} & \ding{51} & \ding{51} & \ding{55} & \ding{55} \\\cline{2-9}
      & Torch & \ding{51} & \ding{51} & \ding{51} & \ding{51} & \ding{51} & \ding{55} & \ding{55} \\
\hline \hline
      & Caffe & \ding{51} & \ding{51} & \ding{51} & \ding{51} & \ding{51} & \ding{51} & \ding{51} \\\cline{2-9}
	  & CNTK & \ding{51} & \ding{51} & \ding{51} & \ding{51} & \ding{51} & \ding{51} & \ding{51} \\\cline{2-9}
FCN-R & MXNet & \ding{51} & \ding{51} & \ding{51} & \ding{51} & \ding{51} & \ding{51} & \ding{51} \\\cline{2-9}
	  & TensorFlow & \ding{51} & \ding{51} & \ding{51} & \ding{51} & \ding{51} & \ding{51} & \ding{51} \\\cline{2-9}
      & Torch & \ding{51} & \ding{51} & \ding{51} & \ding{51} & \ding{51} & \ding{51} & \ding{51} \\
\hline \hline
      & Caffe & \ding{51} & \ding{51} & \ding{51} & \ding{51} & \ding{51} & \ding{51} & \ding{51} \\\cline{2-9}
	  & CNTK & \ding{51} & \ding{51} & \ding{51} & \ding{51} & \ding{51} & \ding{51} & \ding{51} \\\cline{2-9}
AlexNet-R & MXNet & \ding{51} & \ding{51} & \ding{51} & \ding{51} & \ding{51} & \ding{51} & \ding{51} \\\cline{2-9}
	  & TensorFlow & \ding{51} & \ding{51} & \ding{51} & \ding{51} & \ding{51} & \ding{51} & \ding{51} \\\cline{2-9}
      & Torch & \ding{51} & \ding{51} & \ding{51} & \ding{51} & \ding{51} & \ding{51} & \ding{51} \\
\hline \hline
      & Caffe & \ding{51} & \ding{51} & \ding{51} & \ding{51} & \ding{51} & \ding{51} & \ding{51} \\\cline{2-9}
	  & CNTK & \ding{55} & \ding{55} & \ding{51} & \ding{51} & \ding{51} & \ding{51} & \ding{51} \\\cline{2-9}
ResNet-56 & MXNet & \ding{51} & \ding{51} & \ding{51} & \ding{51} & \ding{51} & \ding{51} & \ding{51} \\\cline{2-9}
	  & TensorFlow & \ding{55} & \ding{55} & \ding{51} & \ding{51} & \ding{51} & \ding{51} & \ding{51} \\\cline{2-9}
      & Torch & \ding{51} & \ding{51} & \ding{51} & \ding{51} & \ding{51} & \ding{51} & \ding{51} \\
\hline \hline
      & Caffe & \ding{55} & \ding{55} & \ding{55} & \ding{55} & \ding{55} & \ding{55} & \ding{55} \\\cline{2-9}
	  & CNTK & \ding{51} & \ding{51} & \ding{51} & \ding{51} & \ding{51} & \ding{55} & \ding{55} \\\cline{2-9}
LSTM  & MXNet & \ding{55} & \ding{55} & \ding{51} & \ding{51} & \ding{51} & \ding{55} & \ding{55} \\\cline{2-9}
	  & TensorFlow & \ding{51} & \ding{51} & \ding{51} & \ding{51} & \ding{51} & \ding{55} & \ding{55} \\\cline{2-9}
      & Torch & \ding{51} & \ding{51} & \ding{51} & \ding{51} & \ding{51} & \ding{55} & \ding{55} \\\cline{1-9}
\end{tabular}
\end{table*}

\textit{Neural networks and data sets}. For synthetic data testing, a large neural network (FCN-S) with around 55 million parameters is used to evaluate the performance of FCN; and we choose the classical AlexNet \cite{krizhevsky2012imagenet} and ResNet-50 \cite{he2015deep} that were performed for ImageNet \cite{deng2009imagenet} as representatives of CNNs. For real-world data experiments, a smaller FCN (FCN-R) is constructed for MNIST \cite{lecun1998mnist} data set; an AlexNet architecture named AlexNet-R and ResNet-56 are used for Cifar10 \cite{krizhevsky2014cifar} data set. For RNNs, considering that the main computation complexity is related to the length of input sequence, we select 2 LSTM \cite{zaremba2014recurrent} layers for testing, with input length of 32. The details of the network configuration are shown in Table \ref{table:networksetup} and Table \ref{table:networksetup_new}.

\textit{Hardware platforms}. We use two types of multi-core CPUs, one quad-core desktop CPU (i.e., Intel i7-3820 CPU @ 3.60GHz) and two 8-core server-grade CPUs (i.e., Intel Xeon CPU E5-2630 v3 @ 2.40GHz), to test the performance of tools with different number of threads; and three generations of GPU cards, NVIDIA GTX 980 @ 1127MHz with Maxwell architecture, GTX 1080 @ 1607MHz with Pascal architecture, and Telsa K80 @ 562MHz with Kepler architecture, are used to compare the performance on accelerators. Notice that we only use one of the two GK210 chips of K80 GPU on single GPU comparison, and the GPU autoboost feature is disabled to make our results repeatable. In order to avoid host memory dependency of neural network size, the two test machines are equipped with 64GB memory and 128GB memory respectively. The details of hardware configurations are shown in Table \ref{table:hardwaresetup}. The data parallelization experiments are conducted on two Tesla K80 cards which have a total of 4 GK210 GPUs. For multiple GPU experiments, the system configuration is shown in Table \ref{table:multigpuetup}.

The conducted experiments with the combination of networks, tools and hardware are shown in Table \ref{table:conductedexperiments}.

\section{Experimental Results} \label{results}
We present the experimental results in three sub-sections: CPU results, single GPU results and multi-GPU results. For CPU results and single GPU results, we mainly focus on the running time performance; while for multi-GPU case, we also present the convergence rates. The major results on different platforms are also shown in Table \ref{table:totalresultsSingleGPU} and Table \ref{table:totalresultsMultiGPU}.

\input{totalresultsSingleGPU}
\input{totalresultsMultiGPU}

\subsection{CPU Results}

The experiments are tested on CPU platforms with specific mini-batch size that can achieve the best running time performance according to our previous study in \cite{shi2016benchmarking}. The mini-batch sizes used for different networks are shown in Table \ref{table:minibatchsettingforcpu}.

\begin{table}[!ht]
\centering
\caption{The size of mini-batch used for different networks on CPU platforms.}
\label{table:minibatchsettingforcpu}
\begin{tabular}{cc}
\toprule
Network & Mini-batch size \\\midrule
FCN-S	& 64 \\
AlexNet-S & 16 \\
ResNet-50 & 16 \\
FCN-R & 1024 \\
AlexNet-R & 128 \\
ResNet-56 & 128\\
LSTM & 256 \\
\bottomrule
\end{tabular}
\end{table}

\subsubsection{Synthetic Data}~\\

\begin{figure}[htbp]
  \centering
  \centering     
  \subfigure[Results on i7-3820.]
  {
    \includegraphics[width=0.46\linewidth]{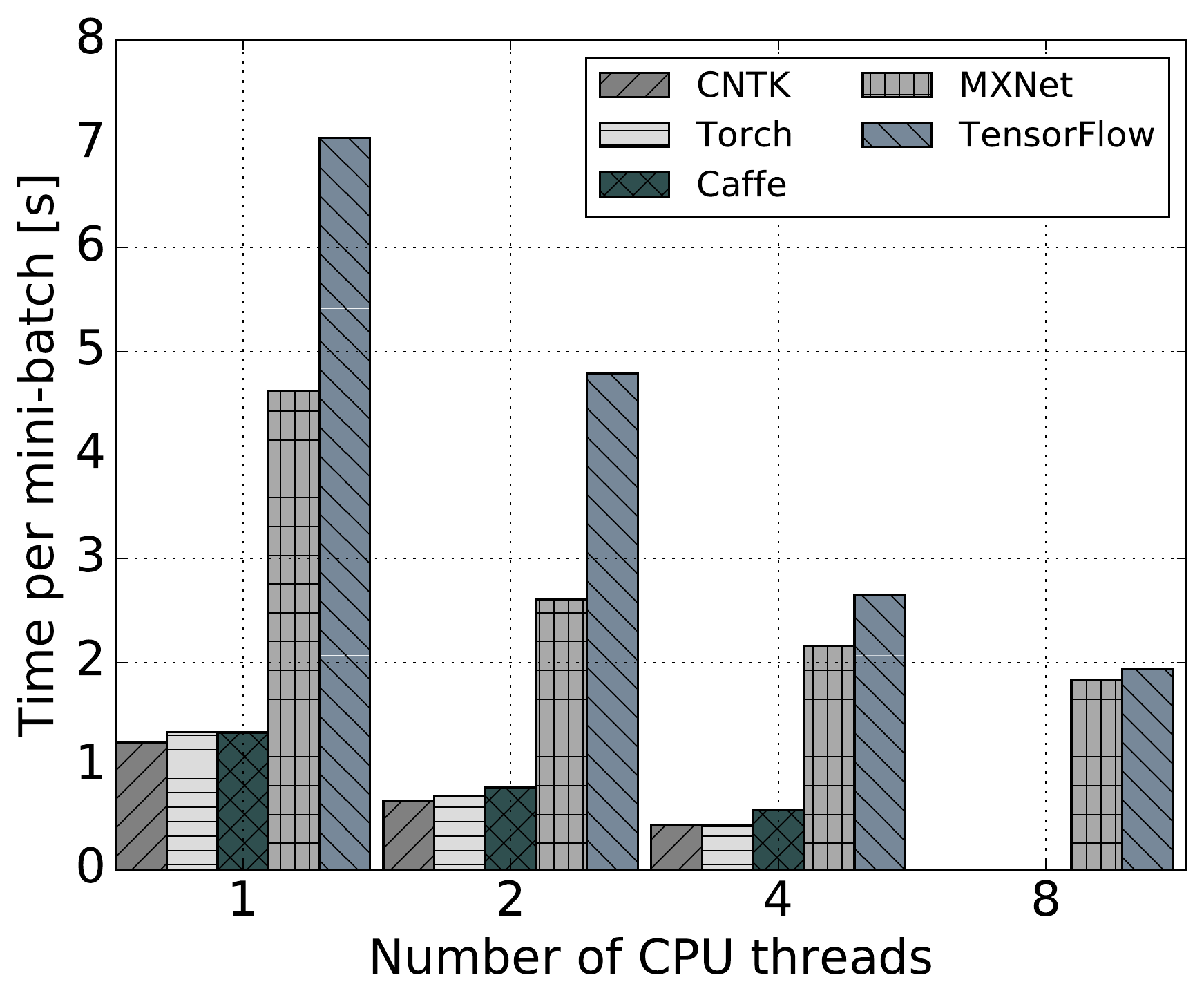}
    \label{fig:syncpulinesfcn5a}
  }
  \subfigure[Results on E5-2630.]
  {
    \includegraphics[width=0.46\linewidth]{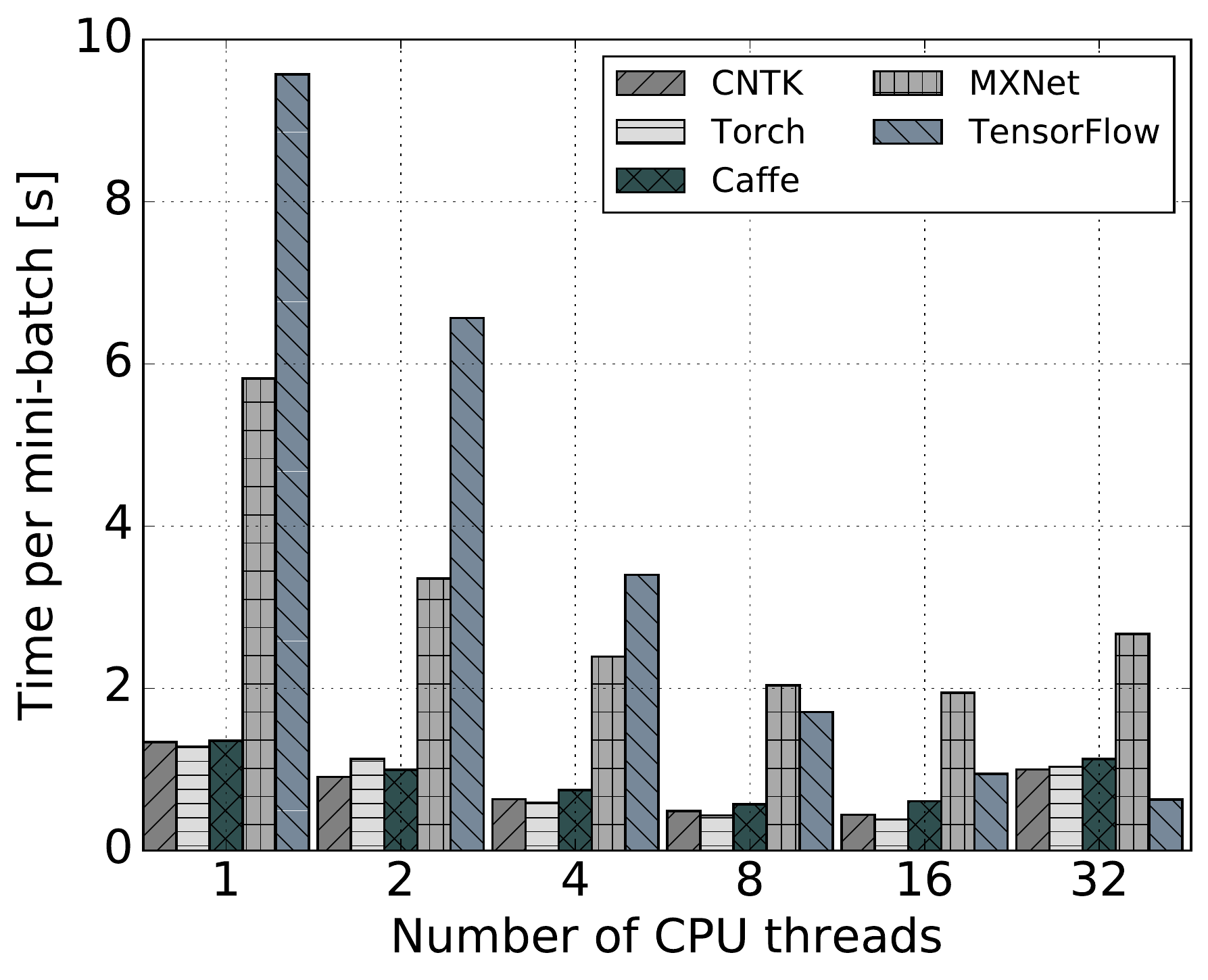}
    \label{fig:syncpulinesfcn5b}
  }
  \caption{FCN-S performance comparison on CPU platform with a mini-batch size of 64. (The lower the better.)}\label{fig:syncpulinesfcn5}
\end{figure}

\textit{FCN-S}. The running time results of mini-batch size 64 with different number of threads are shown in Fig. \ref{fig:syncpulinesfcn5}. On i7-3820 with 4 physical cores, Torch, CNTK, and Caffe have better performance than the other two tools under 1, 2, and 4 threads; CNTK and Torch cannot scale to 8 threads due to memory limitation, and Caffe also performs very poor with 8 threads due to memory issue. On the contrary, MXNet and TensorFlow can still improve the performance by using 8 threads. On the server with dual E5-2630 (a total of 16 physical cores), Torch, CNTK, and Caffe also perform much better than the other two tools under 1, 2, 4, 8, and 16 threads; however, TensorFlow performs the best with 32 threads.

\begin{figure}[htbp]
  \centering
  \subfigure[Results on i7-3820.]
  {
    \includegraphics[width=0.46\linewidth]{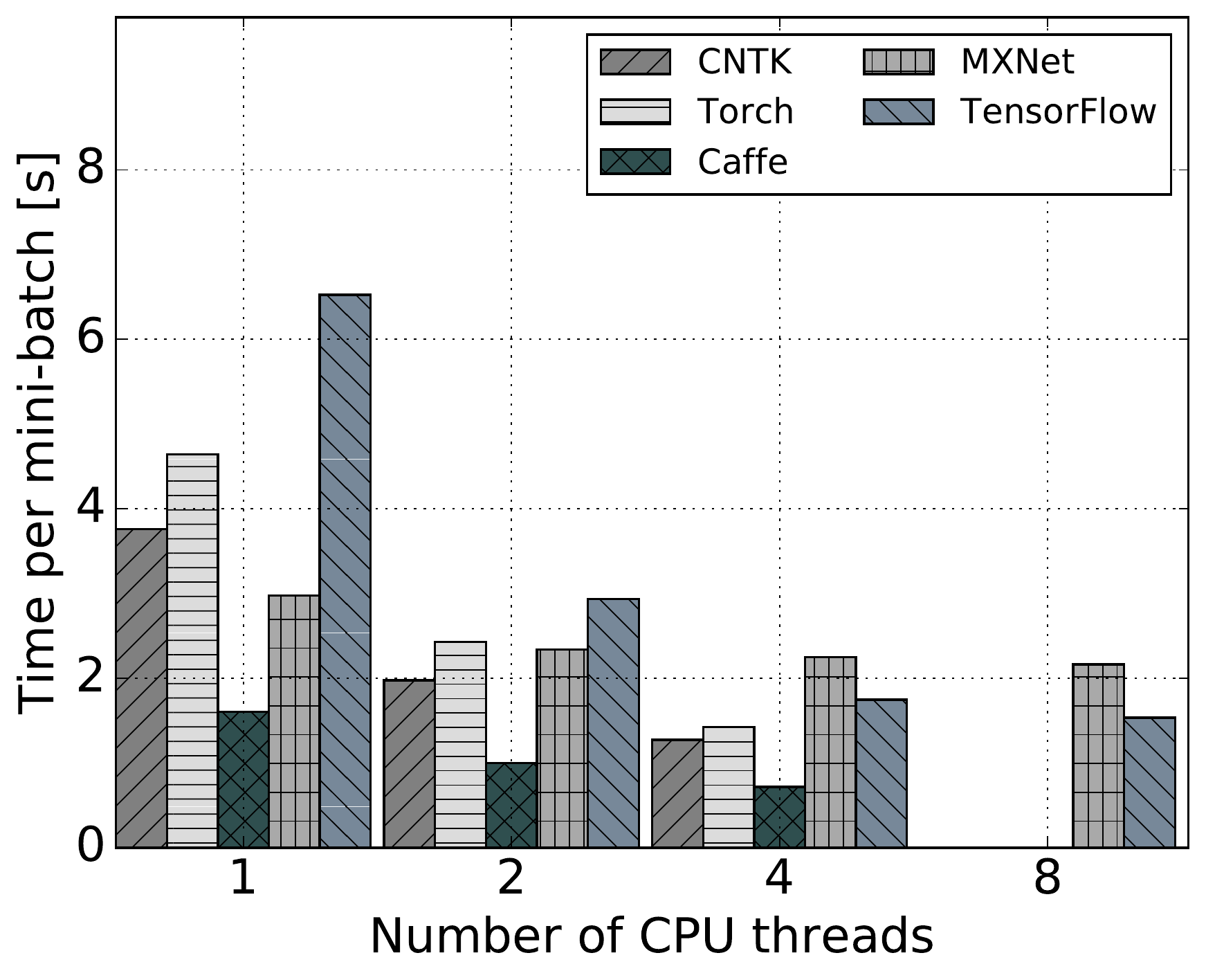}
    \label{fig:syncpulinesalexneta}
  }
  \subfigure[Results on E5-2630.]
  {
    \includegraphics[width=0.46\linewidth]{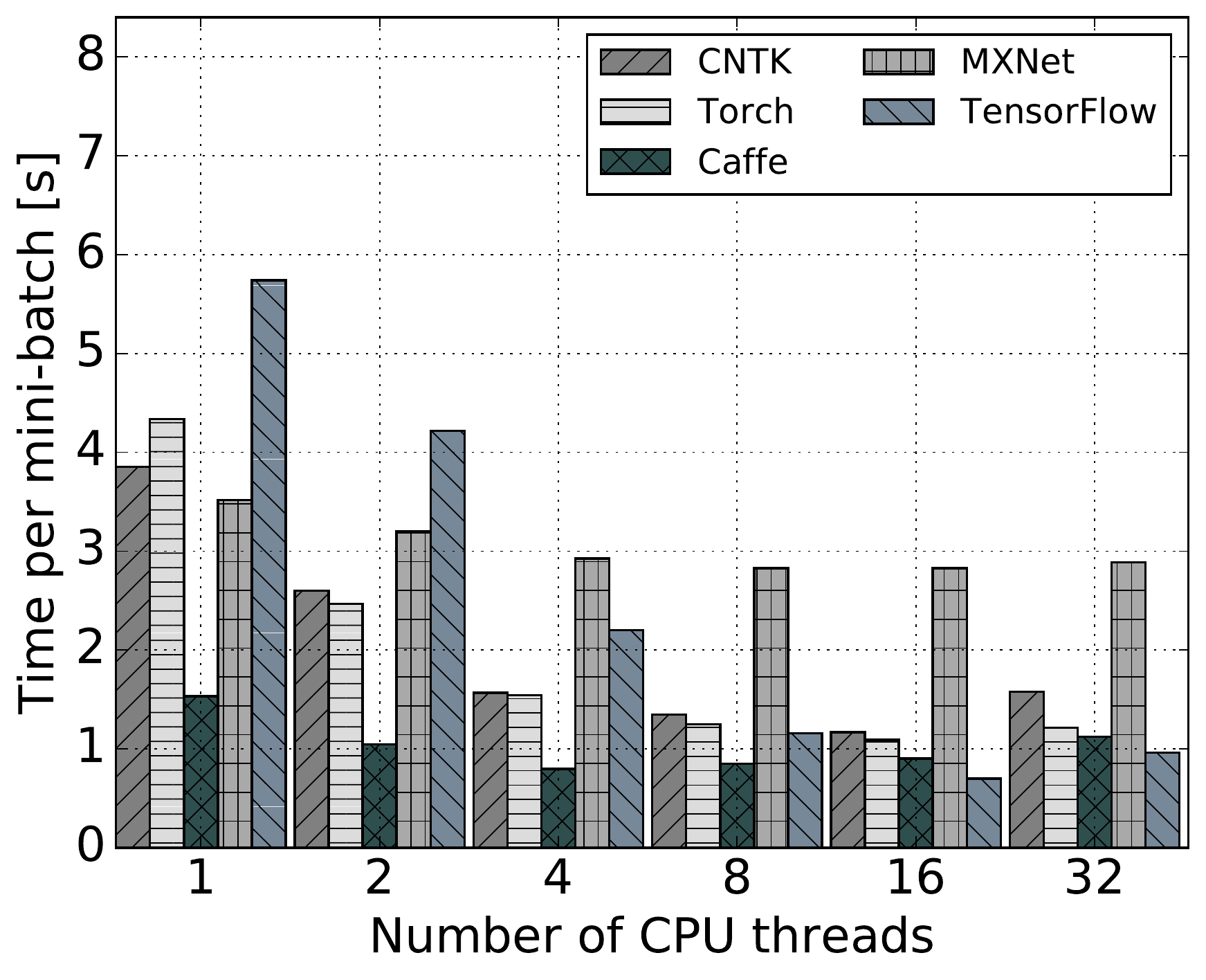}
    \label{fig:syncpulinesalexnetb}
  }
  \caption{AlexNet-S performance comparison on CPU platform with a mini-batch size of 16. (The lower the better.)}\label{fig:syncpulinesalexnet}
\end{figure}

\textit{AlexNet-S}.
The running time results of mini-batch size 16 with different number of threads are shown in Fig. \ref{fig:syncpulinesalexnet}. On i7-3820, Caffe surpasses the other four tools under 1, 2, 4 threads, but fails to run with 8 threads. CNTK and Torch have close performance with 1-4 threads and they also fail to execute with 8 threads. On E5-2630v3, Caffe achieves the best performance with 1, 2, 4, and 8 threads. Among all configurations, TensorFlow achieves the best performance by using 16 threads.

\begin{figure}[htbp]
  \centering
  \subfigure[Results on i7-3820.]
  {
    \includegraphics[width=0.46\linewidth]{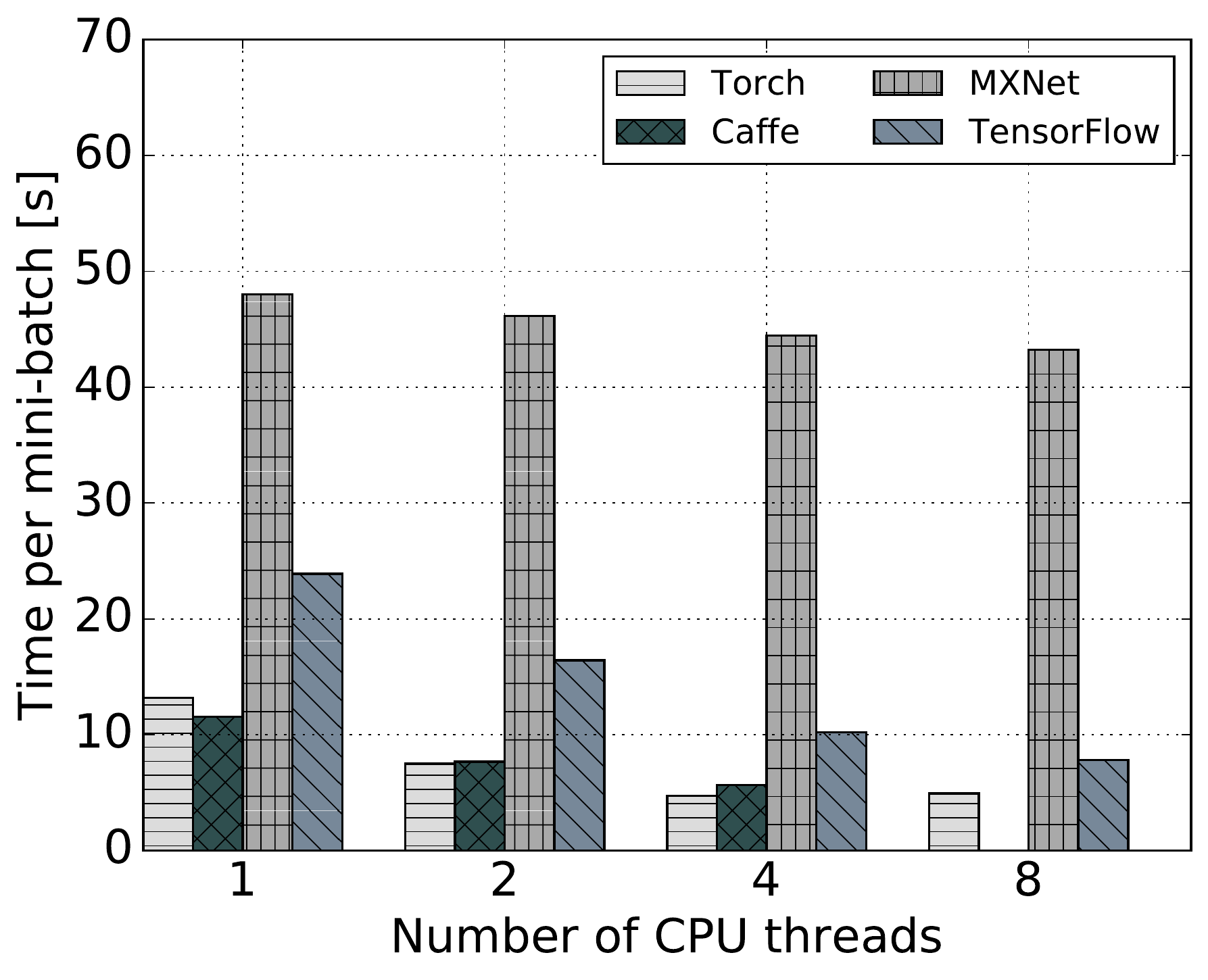}
    \label{fig:syncpulinesresneta}
  }
  \subfigure[Results on E5-2630.]
  {
    \includegraphics[width=0.46\linewidth]{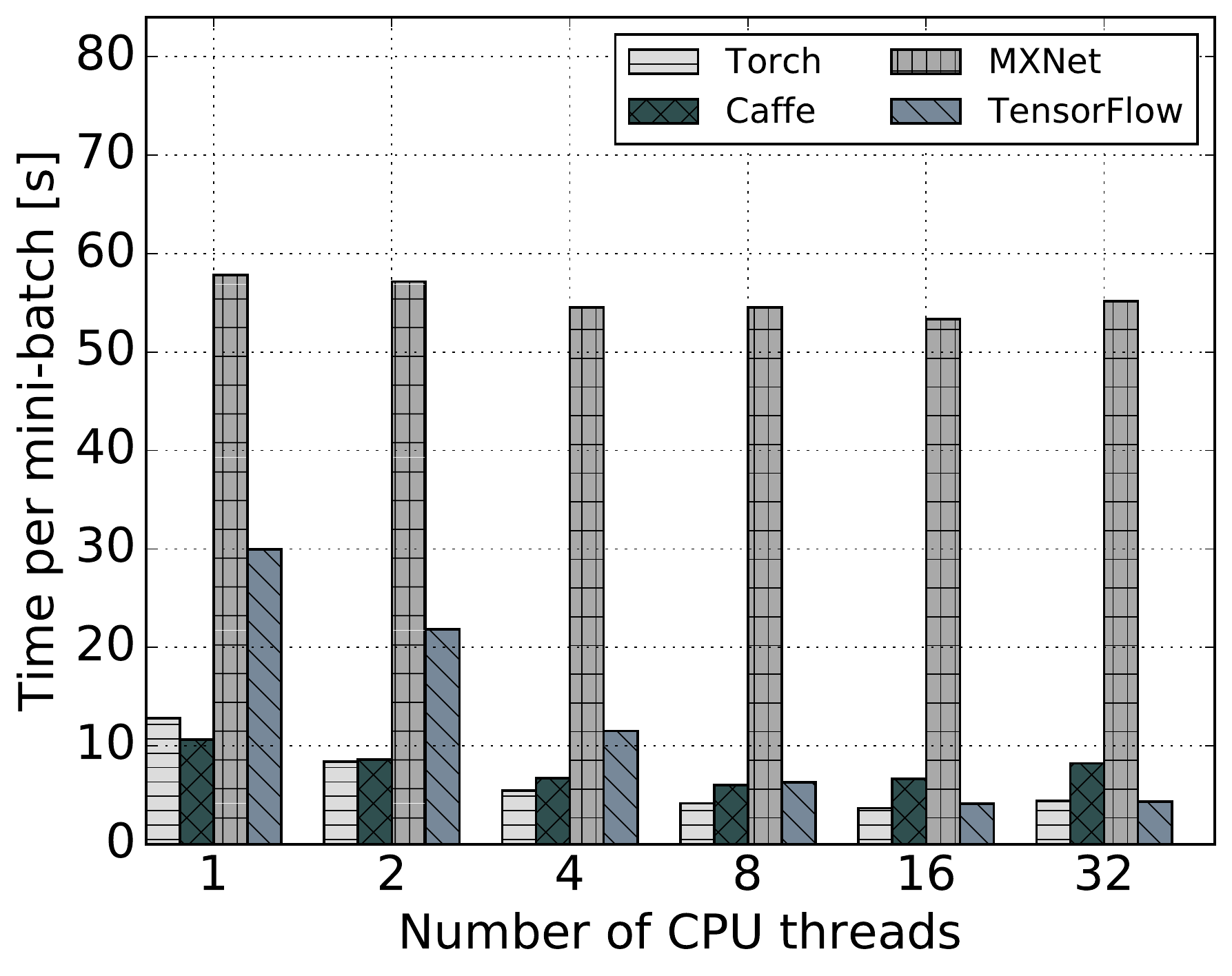}
    \label{fig:syncpulinesresnetb}
  }
  \caption{ResNet-50 performance comparison on CPU platform with a mini-batch size of 16. (The lower the better.)}\label{fig:syncpulinesresnet}
\end{figure}

\textit{ResNet-50}.
Since CNTK does not support batch normalization operation in CPU version, the result of ResNet with CNTK is not included. The performance comparison is shown in Fig. \ref{fig:syncpulinesresnet}. On i7-3820, Caffe and Torch performs better than others with 1, 2, and 4 threads. On E5-2630, Caffe and Torch again outperform others with 1, 2, and 4 threads, while TensorFlow's performance catches up with 8, 16, and 32 threads.

\subsubsection{Real Data}~\\

\begin{figure}[htbp]
  \centering
  \subfigure[Results on i7-3820.]
  {
    \includegraphics[width=0.46\linewidth]{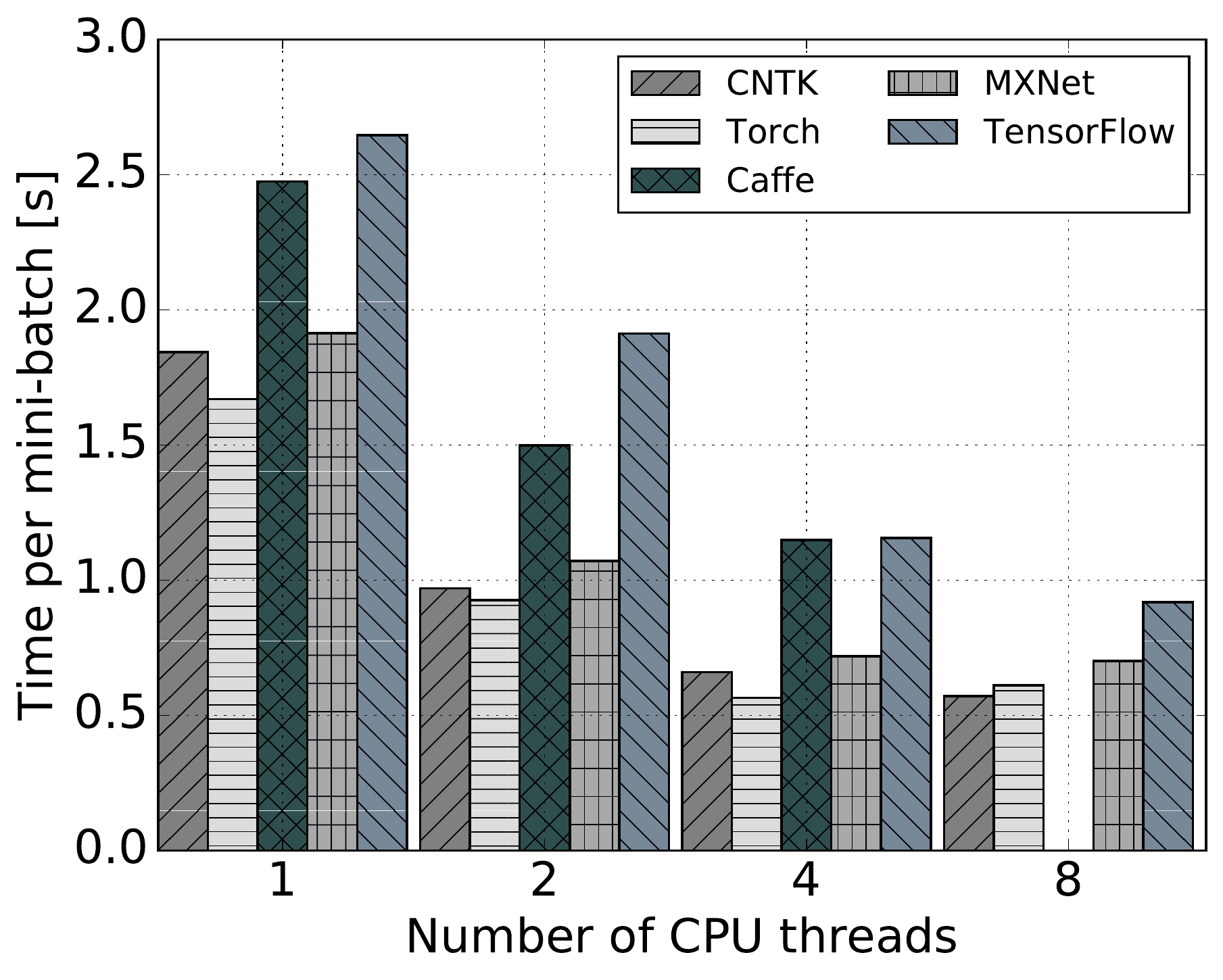}
    \label{fig:realcpulinesfcn5a}
  }
  \subfigure[Results on E5-2630.]
  {
    \includegraphics[width=0.46\linewidth]{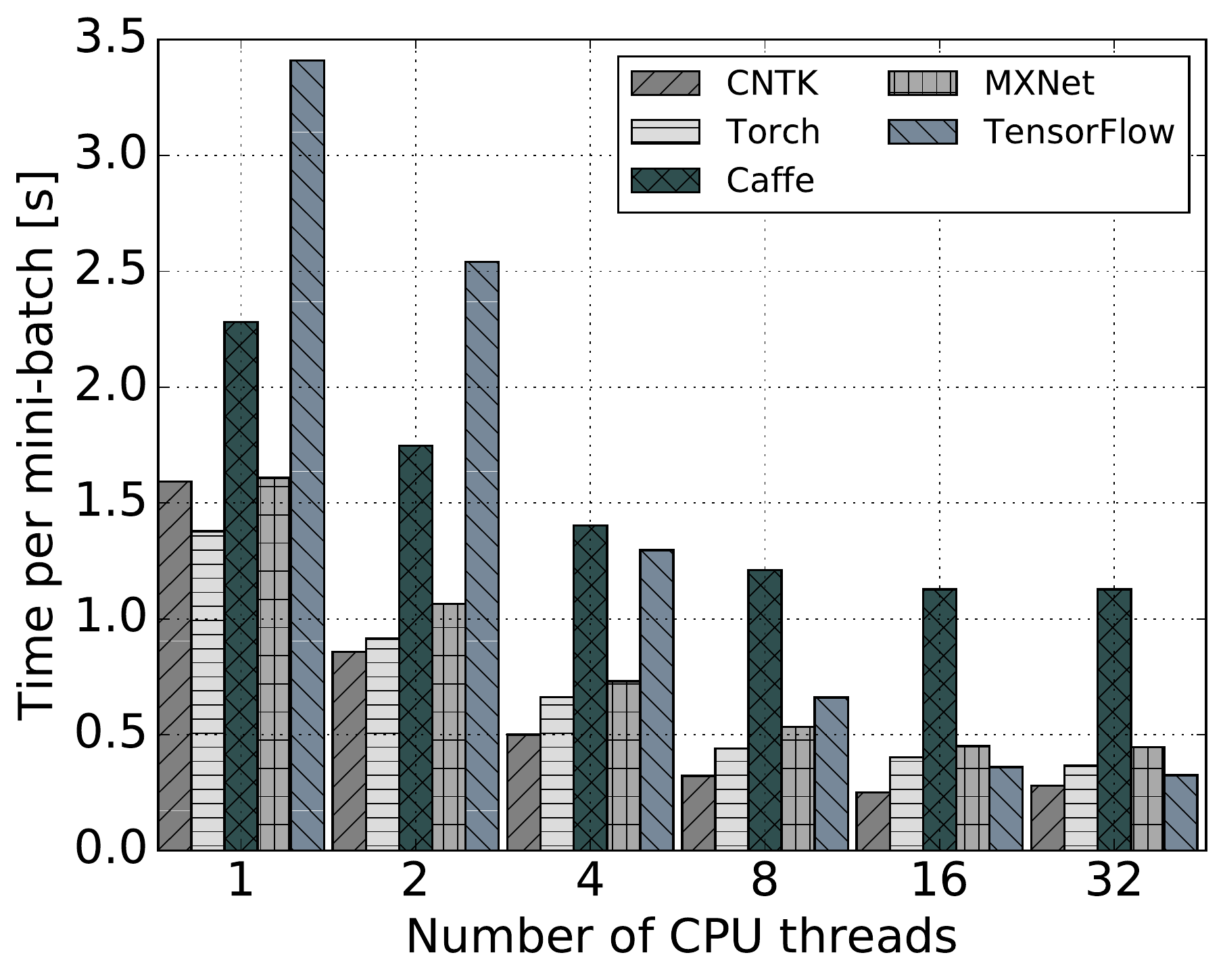}
    \label{fig:realcpulinesfcn5b}
  }
  \caption{FCN-R performance comparison on CPU platform with a mini-batch size of 1024. (The lower the better.)}\label{fig:realcpulinesfcn5}
\end{figure}

\textit{FCN-R}.
The performance comparison is shown in Fig. \ref{fig:realcpulinesfcn5}. On E5-2630v3 with a single thread used, Torch achieves the best performance, followed by CNTK and MXNet, both of which are on the par and they are much better than Caffe and TensorFlow. However, with the parallelization of CPU (i.e., 2-16 threads used), CNTK becomes outstanding, and TensorFlow performs better than Caffe when CPU threads are set to be larger than 2. TensorFlow has a better CPU scalability than others, and even the number of used threads is larger than the number of physical CPU cores, they can achieve some performance improvement. The results on i7-3820 are similar.

\begin{figure}[htbp]
  \centering
  \subfigure[Results on i7-3820.]
  {
    \includegraphics[width=0.46\linewidth]{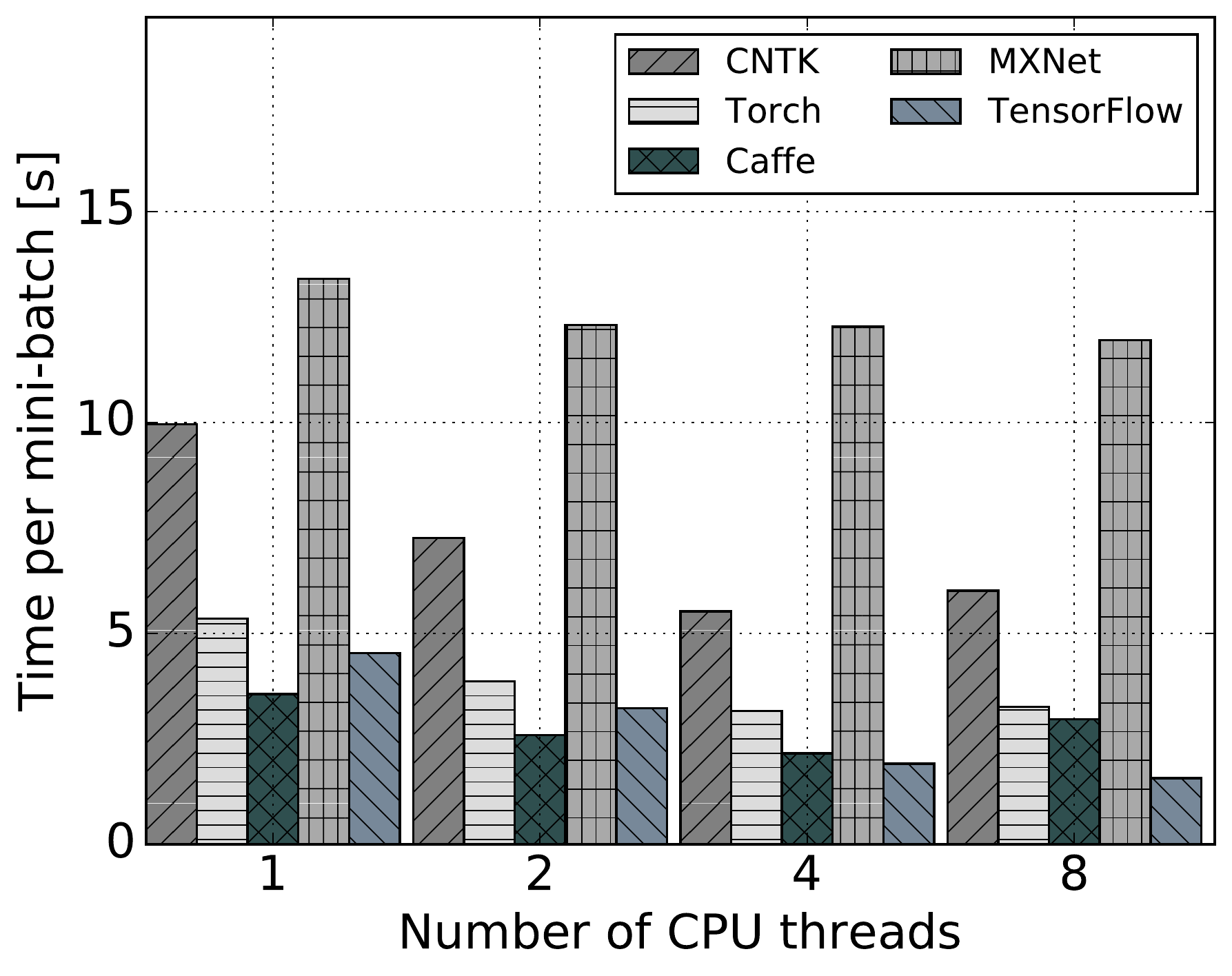}
    \label{fig:realcpubarsalexneta}
  }
  \subfigure[Results on E5-2630.]
  {
    \includegraphics[width=0.46\linewidth]{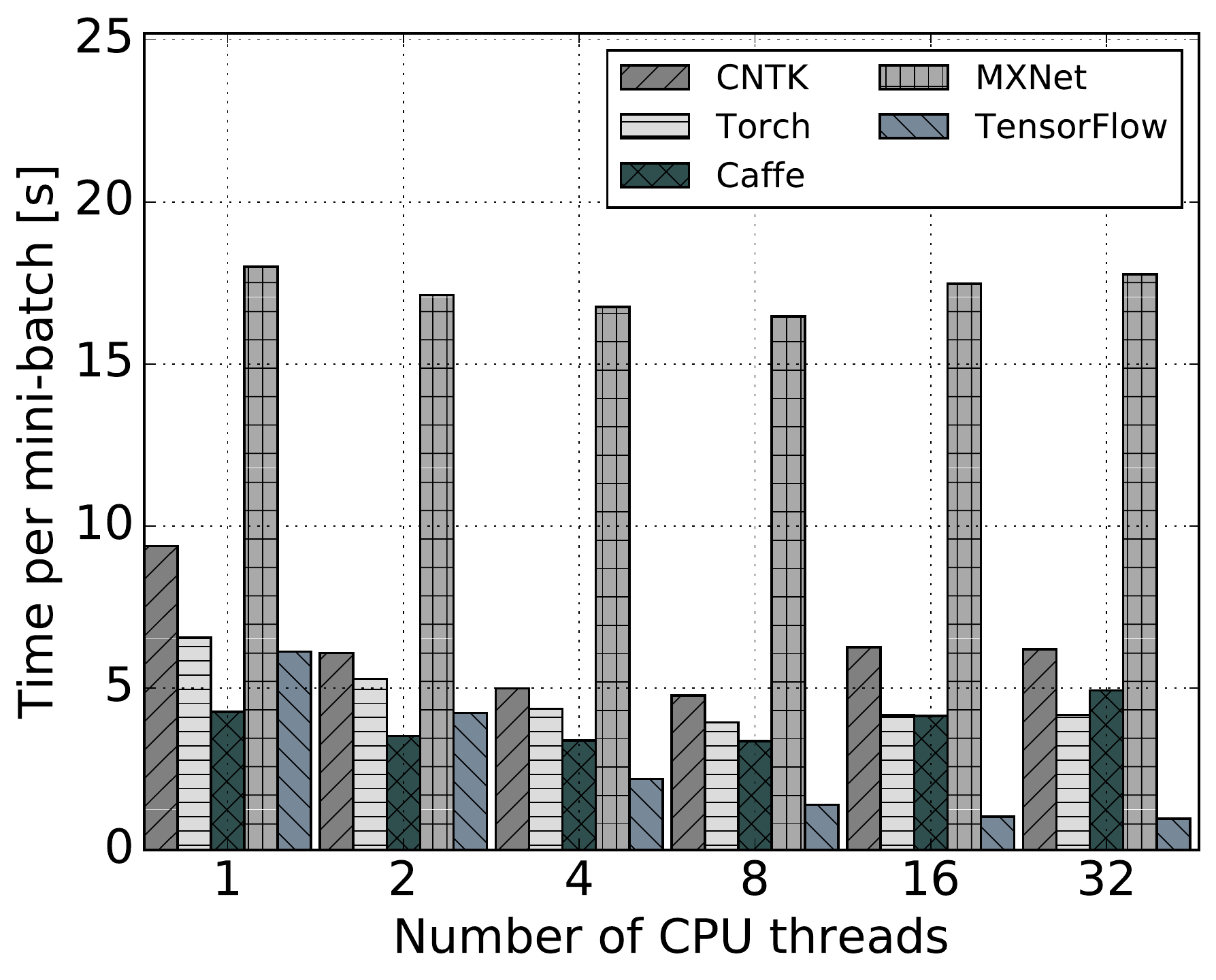}
    \label{fig:realcpubarsalexnetb}
  }
  \caption{AlexNet-R performance comparison on CPU platform with a mini-batch size of 1024. (The lower the better.)}\label{fig:realcpulinesalexnet}
\end{figure}

\textit{AlexNet-R}.
The performance comparison is shown in Fig. \ref{fig:realcpulinesalexnet}. On E5-2630v3, Caffe obtains the best performance with 1 or 2 threads used, but TensorFlow exceeds Caffe when setting the number of threads to be 4 or more. CNTK and Torch have similar results on the CPU parallelization, which are both better than MXNet. On i7-3820, Caffe also achieves the best results on small size of threads settings, and TensorFlow obtains the best with more threads. 

\begin{figure}[htbp]
  \centering
  \subfigure[Results on i7-3820.]
  {
    \includegraphics[width=0.46\linewidth]{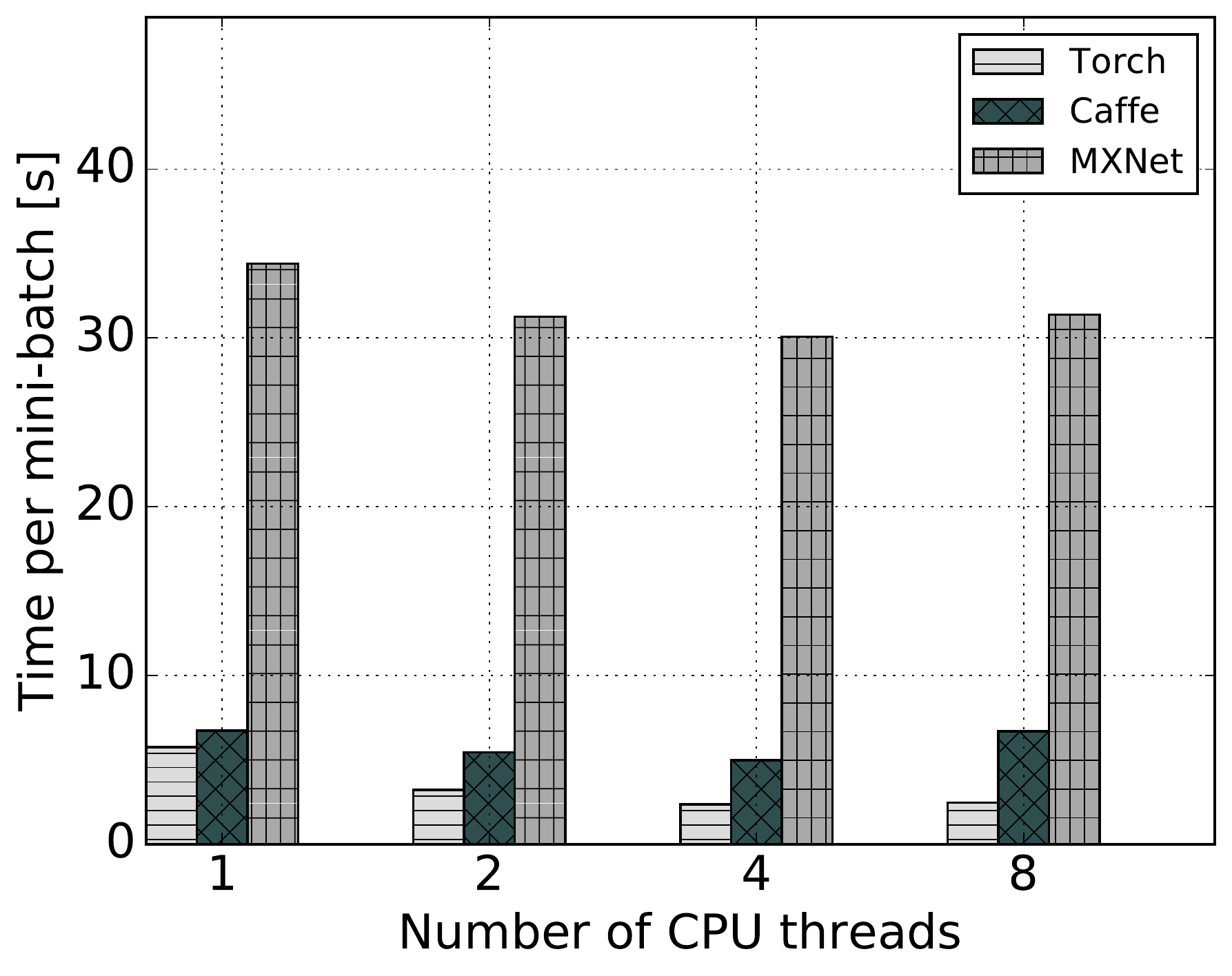}
    \label{fig:realcpubarsresneta}
  }
  \subfigure[Results on E5-2630.]
  {
    \includegraphics[width=0.46\linewidth]{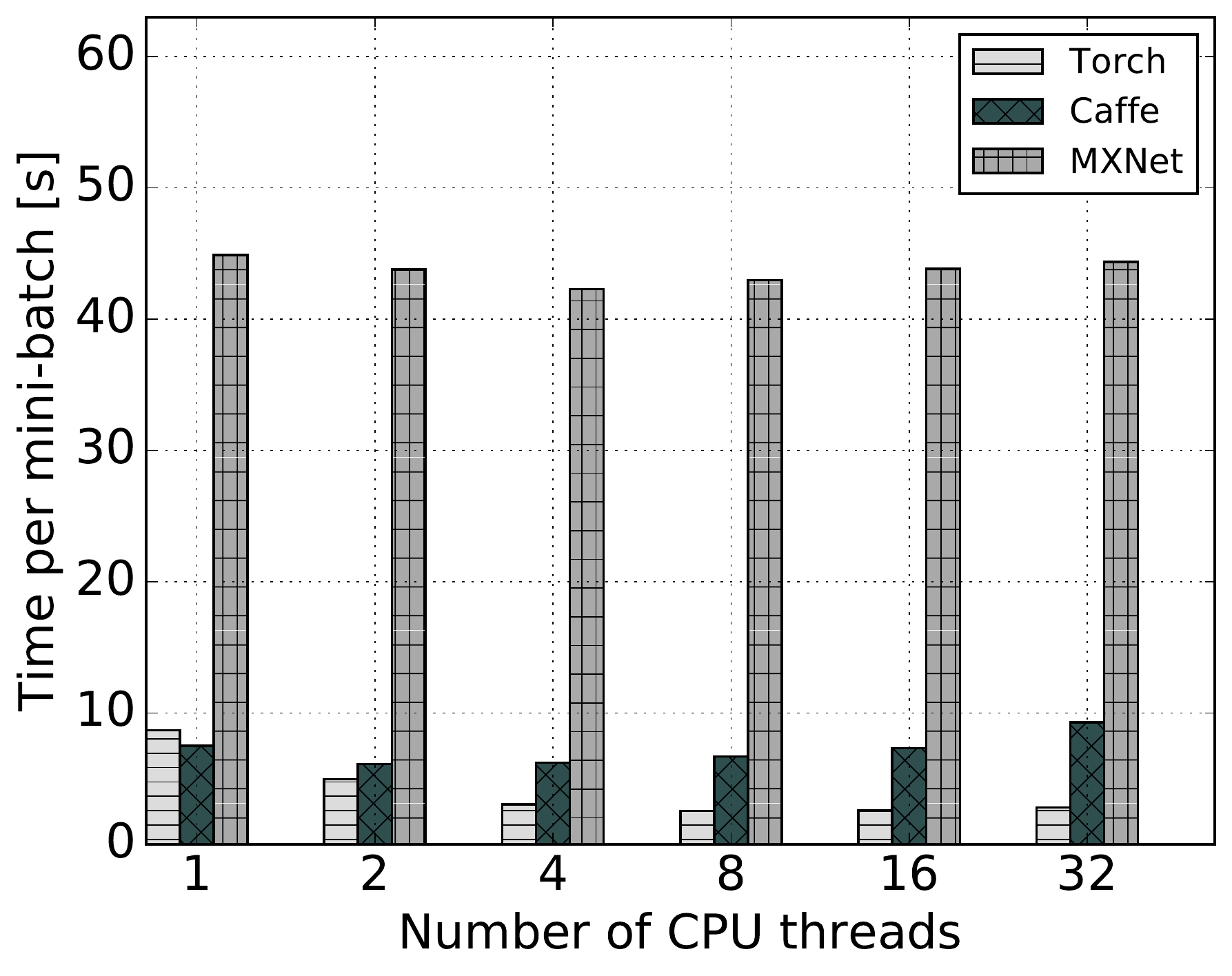}
    \label{fig:realcpubarsresnetb}
  }
  \caption{ResNet-56 performance comparison on CPU platform with a mini-batch size of 128. (The lower the better.)}\label{fig:realcpulinesresnet}
\end{figure}

\textit{ResNet-56}.
TensorFlow does not support the operation of pooling when the stride size is larger than kernel size, so the result of TensorFlow is excluded. The performance comparison is shown in Fig. \ref{fig:realcpulinesresnet}. On both types of CPUs, Torch achieves better time performance than Caffe, while Caffe is better MXNet.

\begin{figure}[htbp]
  \centering
  \subfigure[Results on i7-3820.]
  {
    \includegraphics[width=0.46\linewidth]{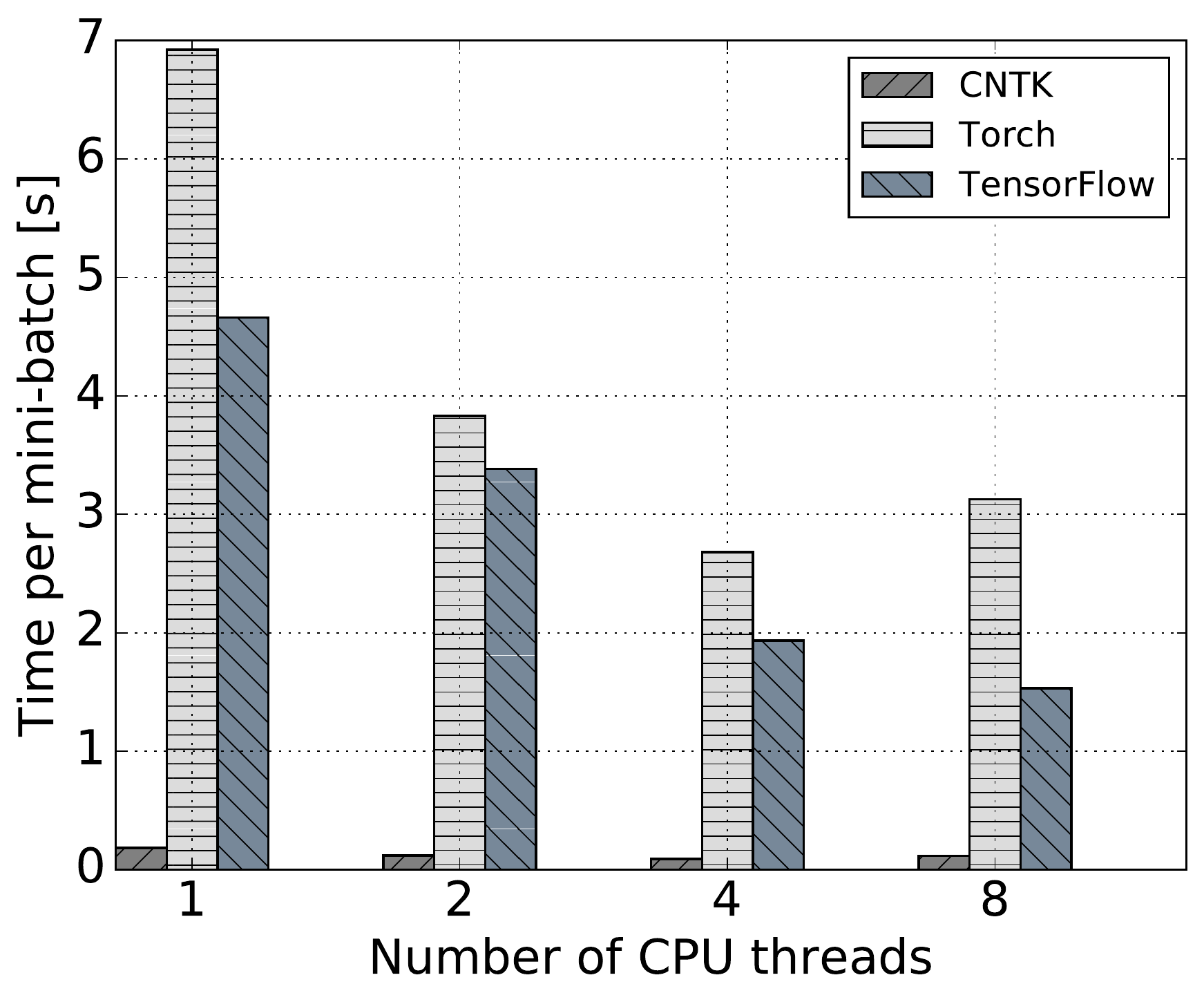}
    \label{fig:realcpubarslstma}
  }
  \subfigure[Results on E5-2630.]
  {
    \includegraphics[width=0.46\linewidth]{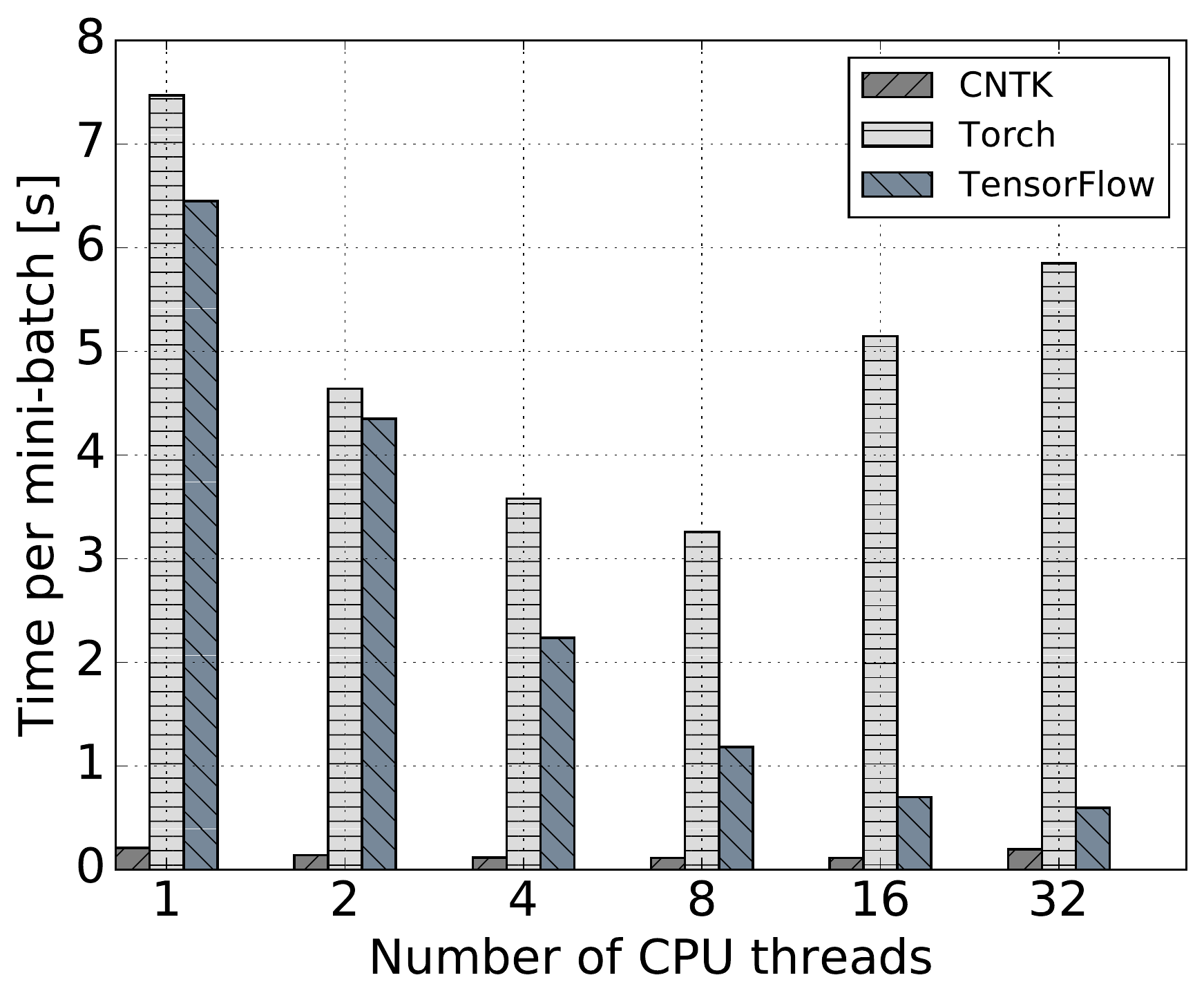}
    \label{fig:realcpubarslstmb}
  }
  \caption{LSTM performance comparison on CPU platform with a mini-batch size of 256. (The lower the better.)}\label{fig:realcpulineslstm}
\end{figure}

\textit{LSTM}.
The performance comparison is shown in Fig. \ref{fig:realcpulineslstm}. MXNet does not support LSTM on CPU version. On both CPUs, CNTK obtains much better performance (up to 5-10 times) than Torch and TensorFlow, while TensorFlow is slightly better than Torch.

\subsection{Single GPU Results}
On single GPU comparison, we also present the results of different mini-batch sizes to show the impact of mini-batch size on performance \cite{bottou2016optimization}.

\subsubsection{Synthetic Data}~\\

\begin{figure*}[htbp]
  \centering     
  \subfigure[Results on Tesla K80.]
  {
    \includegraphics[width=0.3\linewidth]{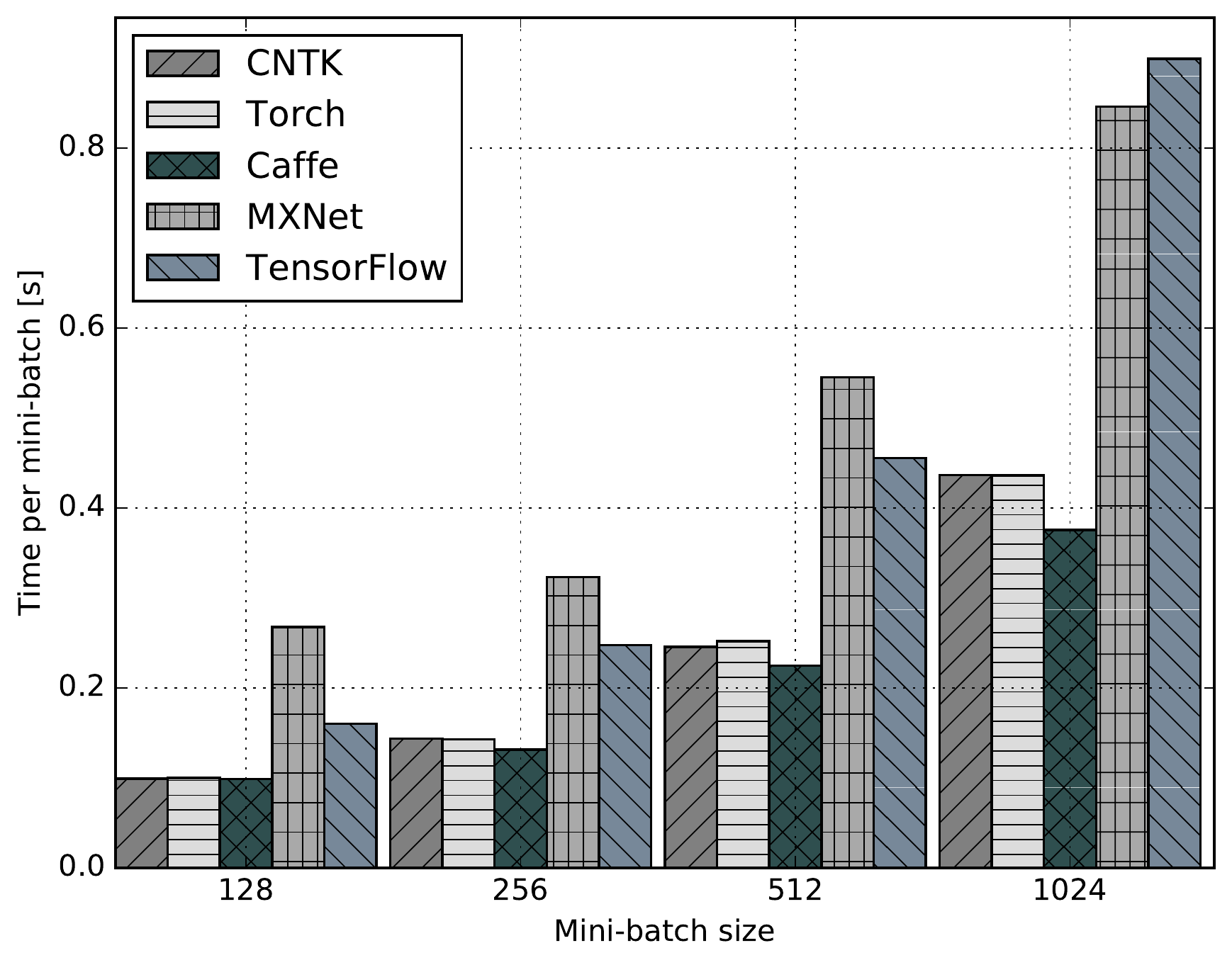}
    \label{fig:a}
  }
  \subfigure[Results on GTX1080.]
  {
    \includegraphics[width=0.3\linewidth]{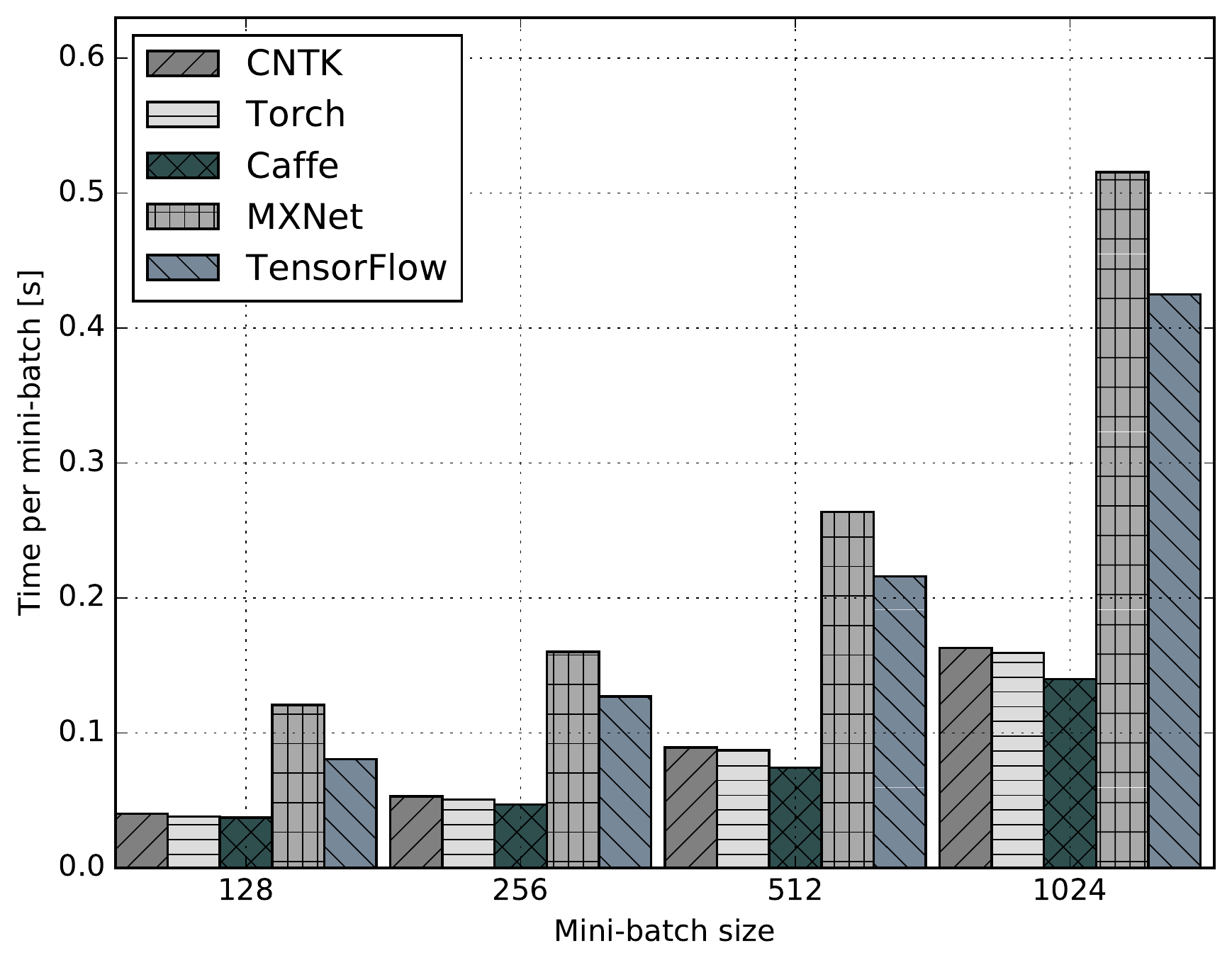}
    \label{fig:b}
  }
  \subfigure[Results on GTX980.]
  {
    \includegraphics[width=0.3\linewidth]{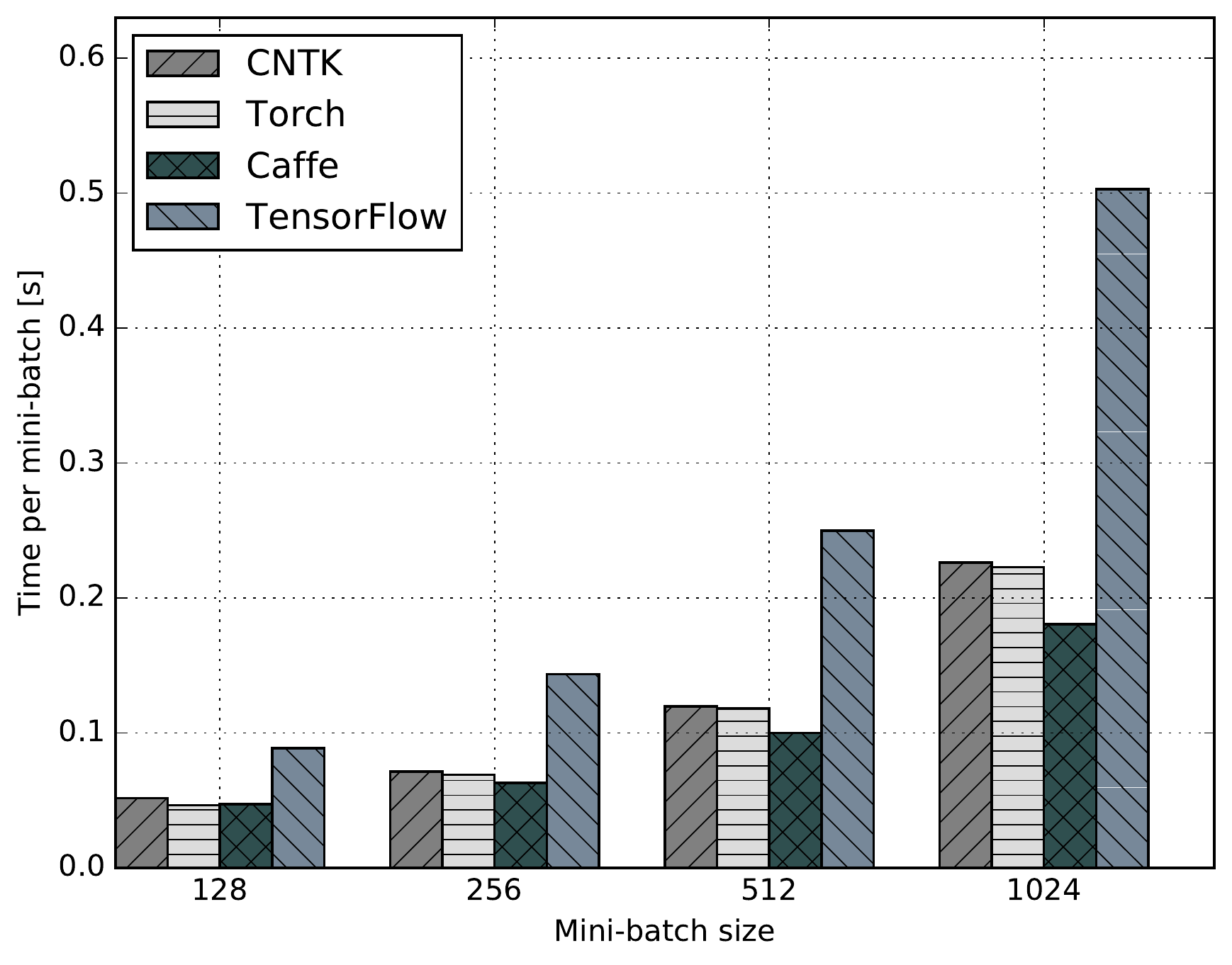}
    \label{fig:c}
  }
  \caption{The performance comparison of FCN-S on GPU platforms.}
  \label{fig:syngpubarsfcn5}
\end{figure*}

\textit{FCN-S}.
The performance comparison among different tools on GPUs is shown in Fig. \ref{fig:syngpubarsfcn5}. Caffe, CNTK and Torch have similar results on all proposed GPUs, which are much better than TensorFlow and MXNet, and Caffe achieves slightly better than CNTK and Torch. MXNet, however, runs out of memory on a GTX980 which has only 4GB memory.

\begin{figure*}[htbp]
  \centering     
  \subfigure[Results on Tesla K80.]
  {
    \includegraphics[width=0.3\linewidth]{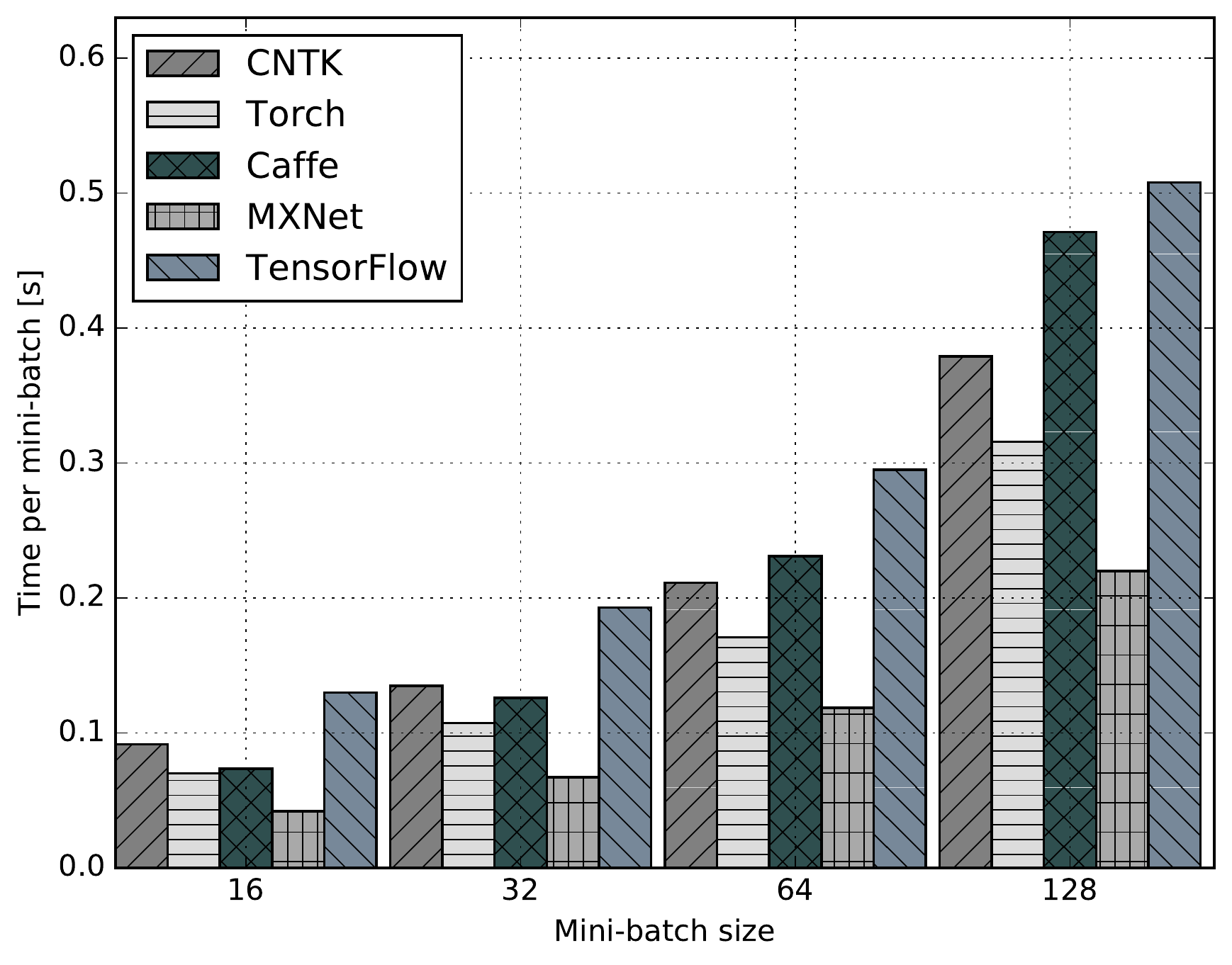}
  }
  \subfigure[Results on GTX1080.]
  {
    \includegraphics[width=0.3\linewidth]{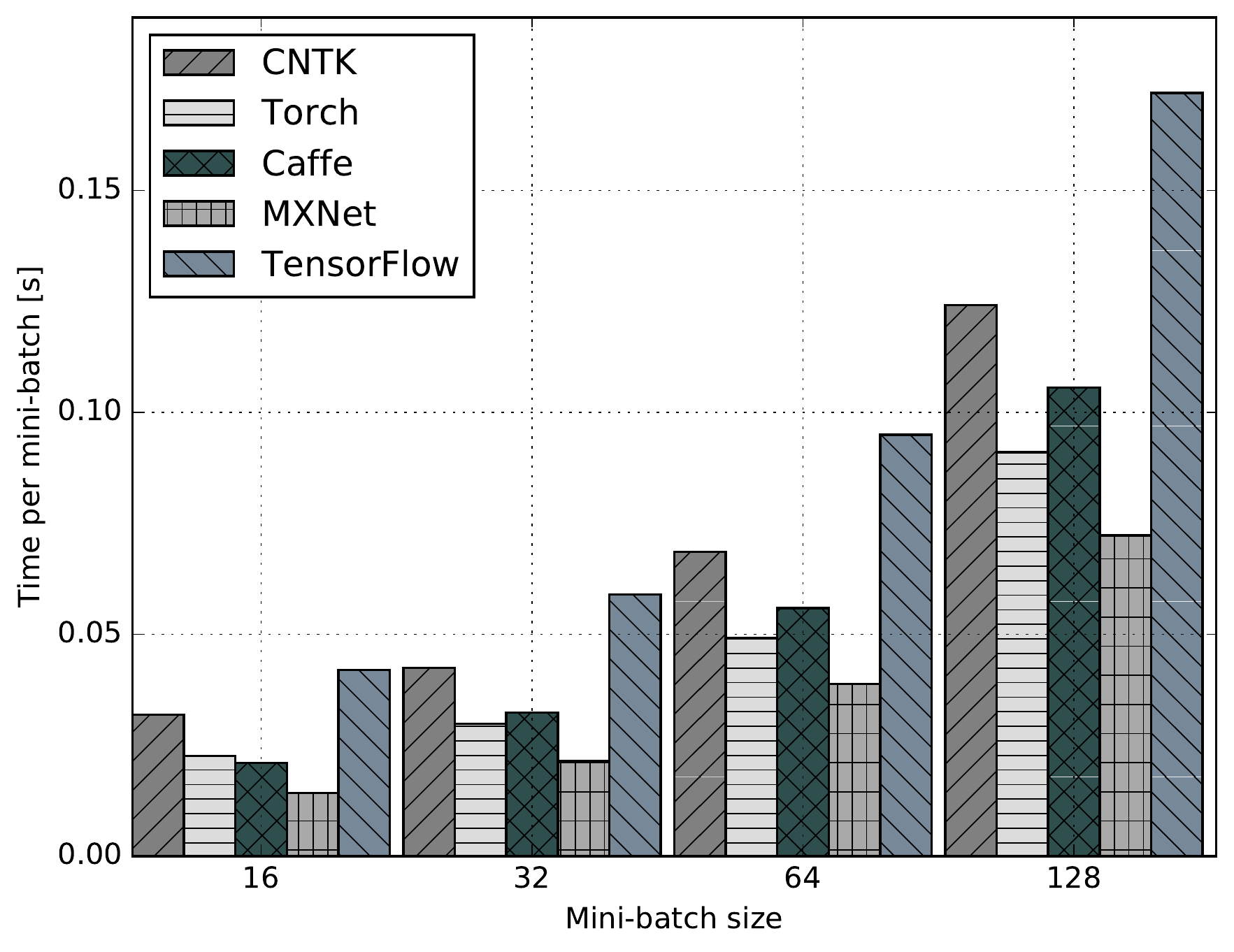}
  }
  \subfigure[Results on GTX980.]
  {
    \includegraphics[width=0.3\linewidth]{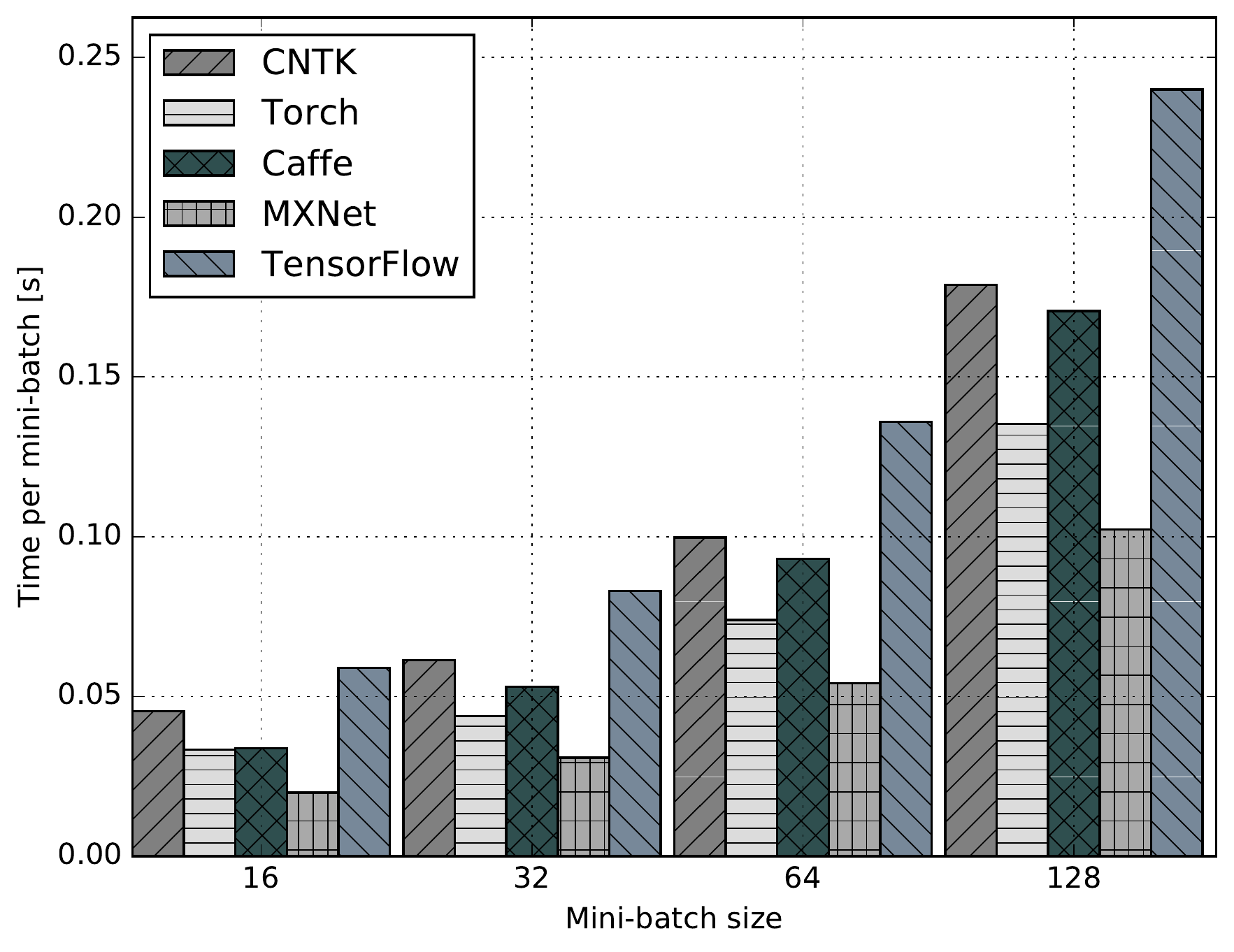}
  }
  \caption{The performance comparison of AlexNet-S on GPU platforms.}
  \label{fig:syngpubarsalexnet}
\end{figure*}

\textit{AlexNet-S}.
The performance comparison among different tools on GPUs is shown in Fig. \ref{fig:syngpubarsalexnet}. In all tested cases, MXNet obtains the best performance, followed by Torch. On Tesla K80, Caffe achieves better performance than CNTK when the size of mini-batch is relatively small while CNTK is slightly better with mini-batch size at both 64 and 128. On GTX1080, Caffe obtains slightly better performance than CNTK in all cases of different mini-batch sizes.

\begin{figure*}[htbp]
  \centering     
  \subfigure[Results on Tesla K80.]
  {
    \includegraphics[width=0.3\linewidth]{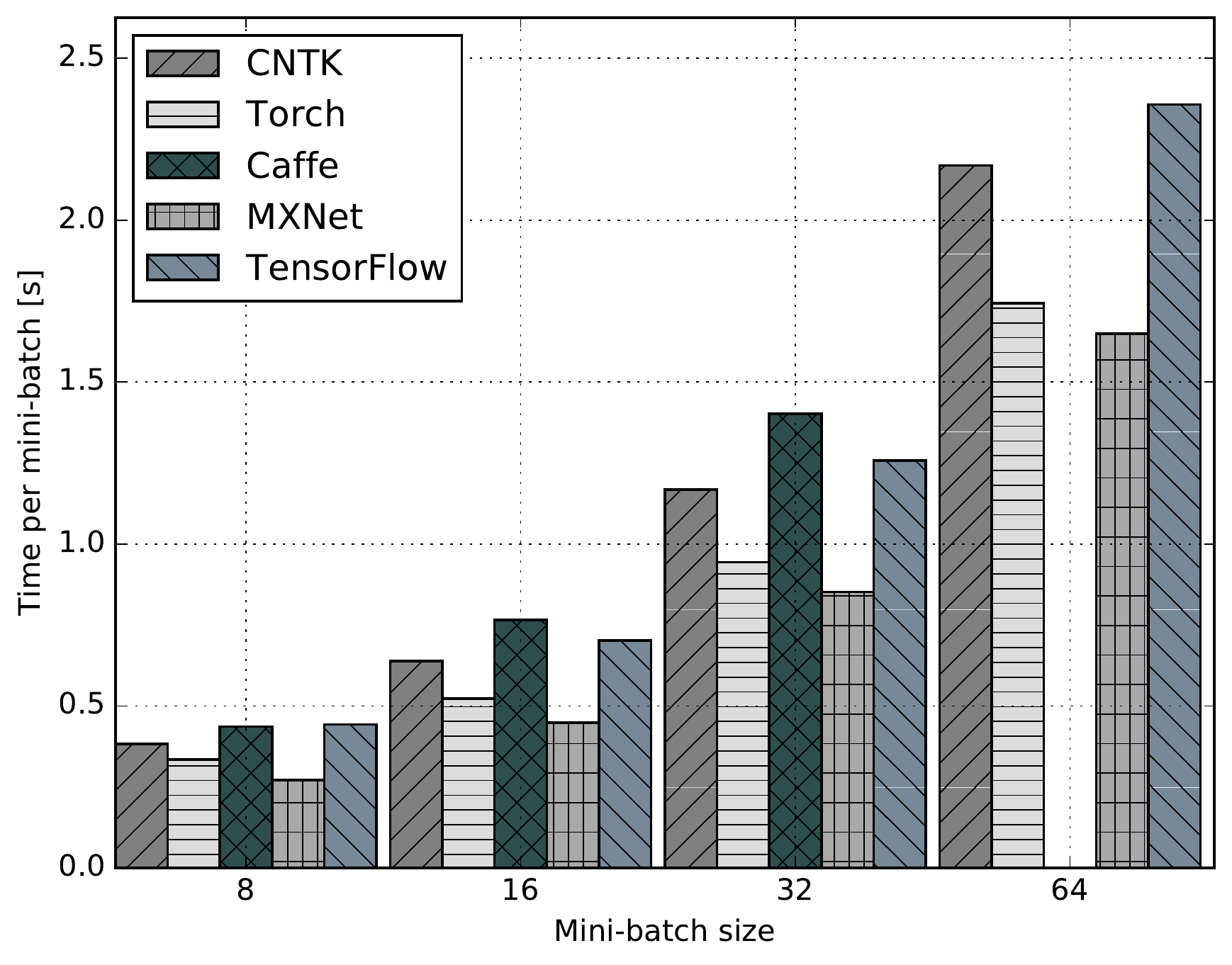}
  }
  \subfigure[Results on GTX1080.]
  {
    \includegraphics[width=0.3\linewidth]{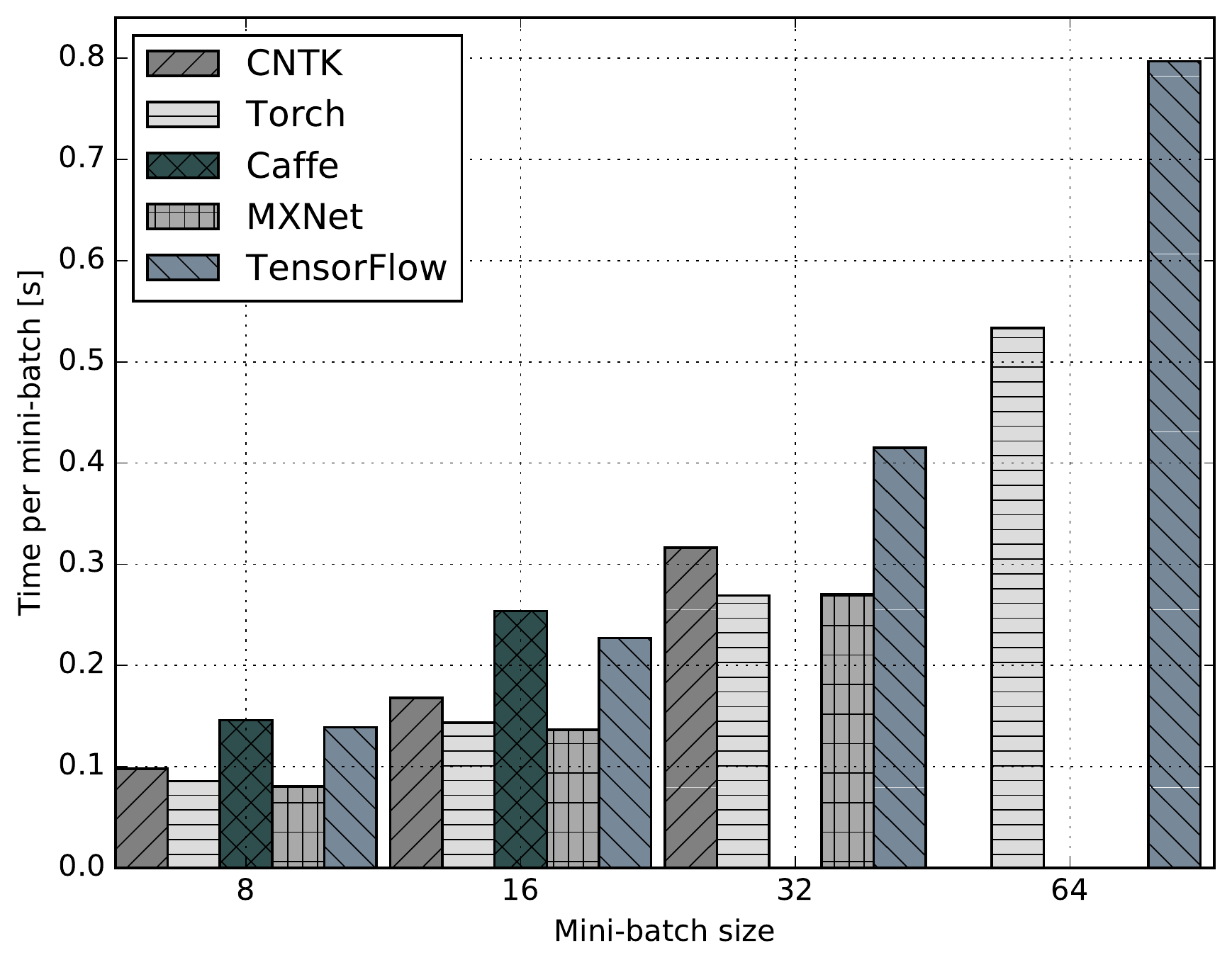}
  }
  \subfigure[Results on GTX980.]
  {
    \includegraphics[width=0.3\linewidth]{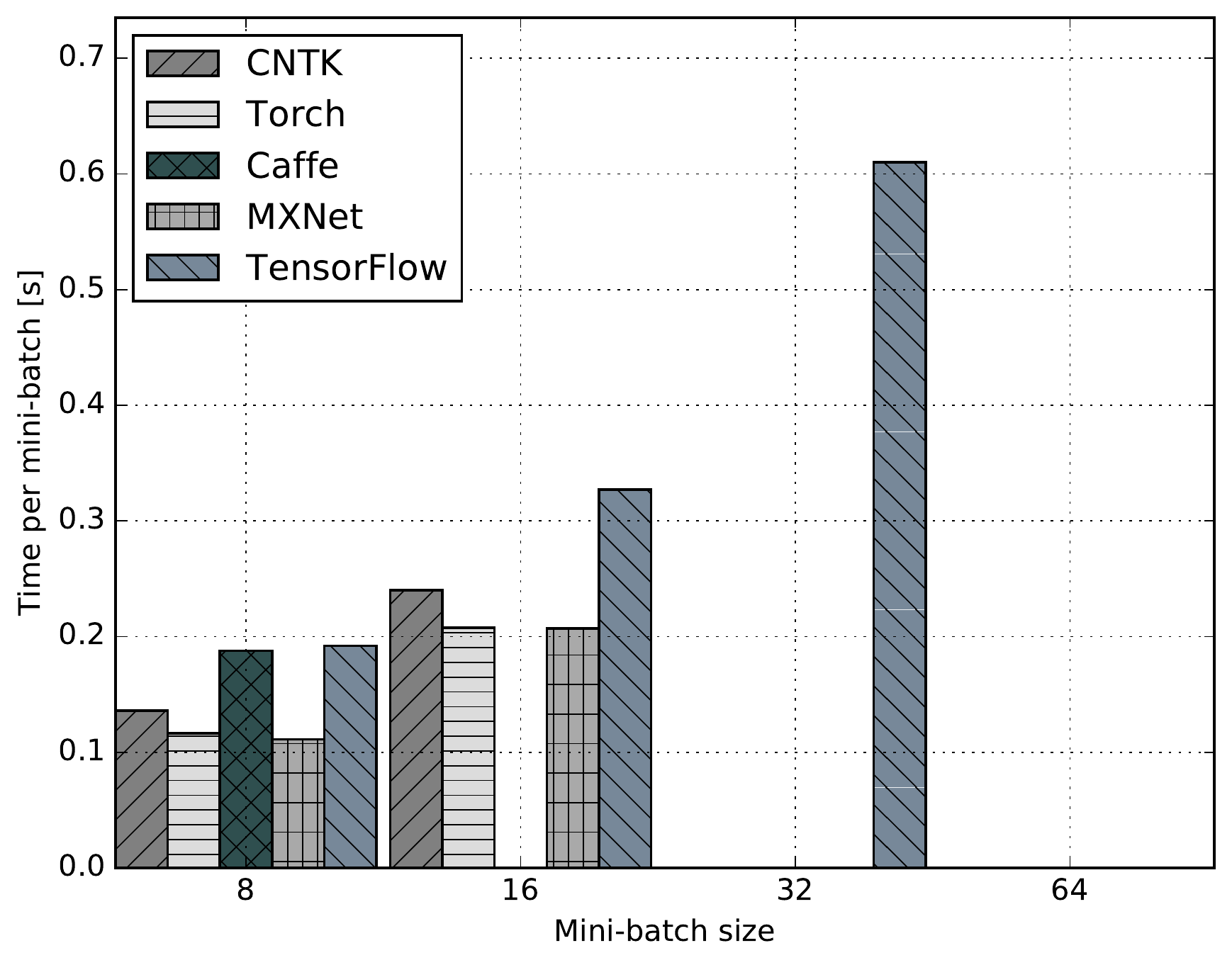}
  }
  \caption{The performance comparison of ResNet-50 on GPU platforms.}
  \label{fig:syngpubarsresnet}
\end{figure*}

\textit{ResNet-50}.
The performance comparison among different tools on GPUs is shown in Fig. \ref{fig:syngpubarsresnet}. The performance of MXNet is better than other tools, but it runs out of memory with larger mini-batch size. Caffe, however, has the relatively worse results. TensorFlow obtains slightly better performance than Caffe, but worse than CNTK. On the GPU cards with smaller memory, there are some cases that cannot be performed. For example, Caffe runs out of memory at the mini-batch size of 32 and 64, and CNTK also runs out of memory with 64 samples per mini-batch on GTX1080.

\subsubsection{Real Data}~\\

\begin{figure*}[htbp]
  \centering     
  \subfigure[Results on Tesla K80.]
  {
    \includegraphics[width=0.3\linewidth]{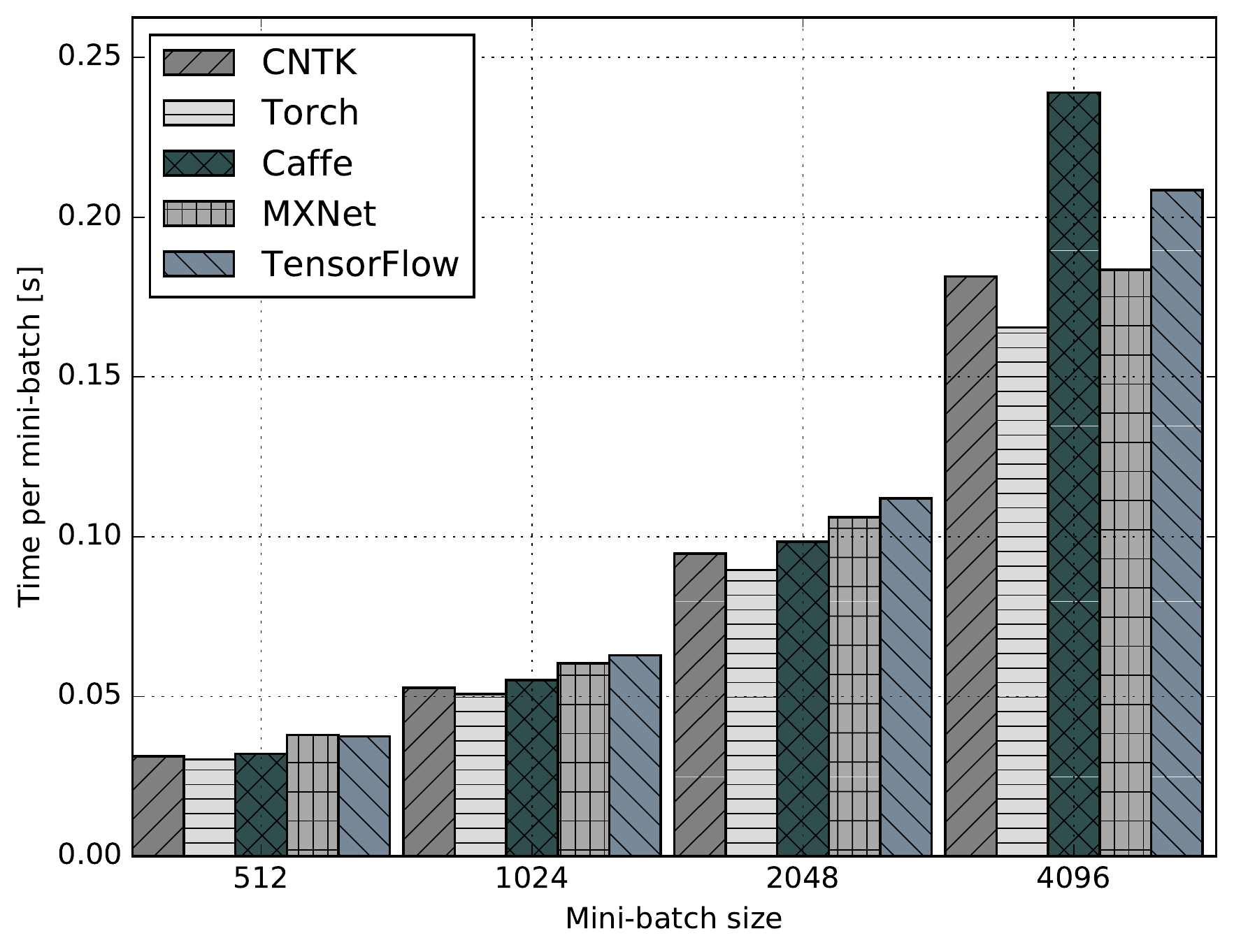}
  }
  \subfigure[Results on GTX1080.]
  {
    \includegraphics[width=0.3\linewidth]{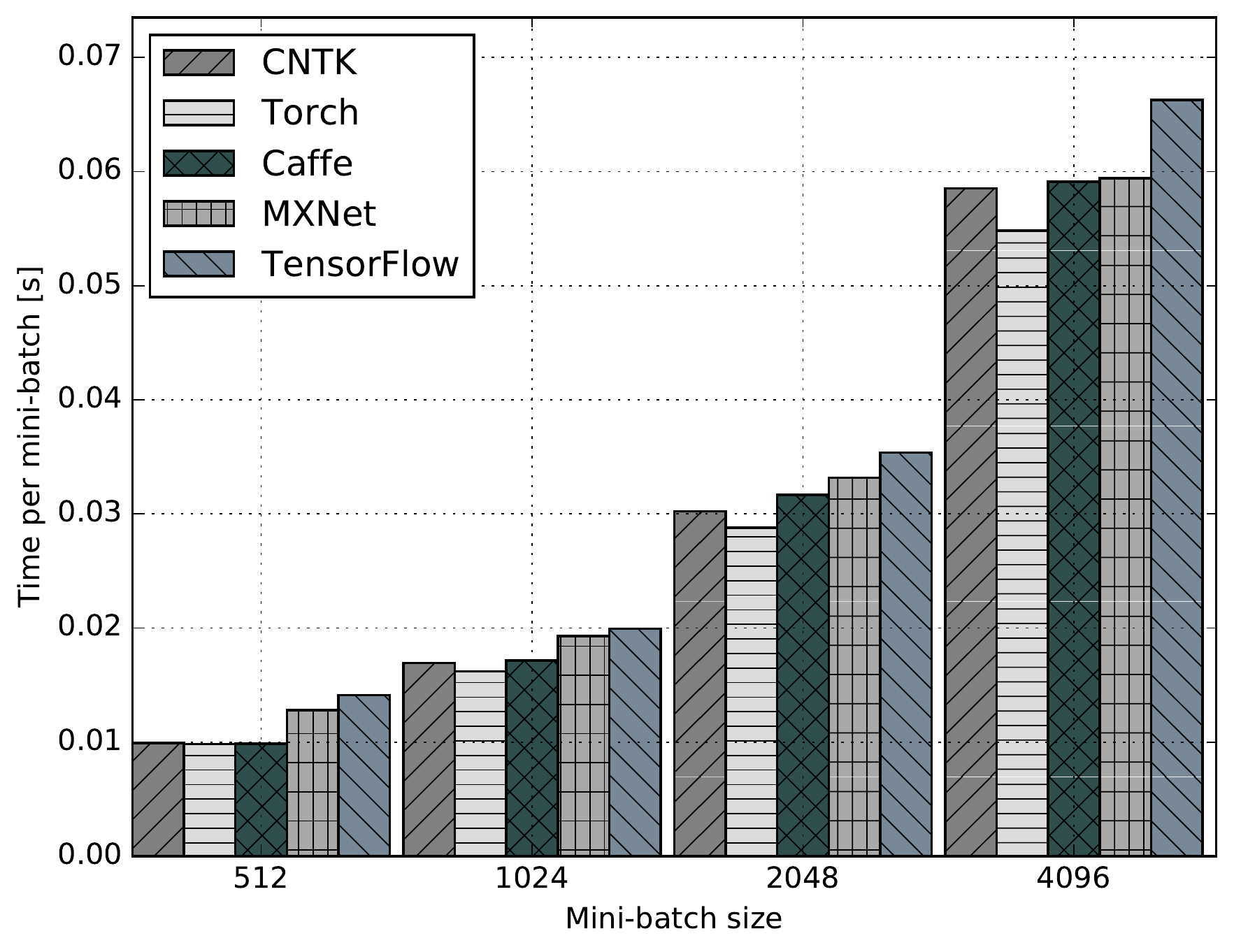}
  }
  \subfigure[Results on GTX980.]
  {
    \includegraphics[width=0.3\linewidth]{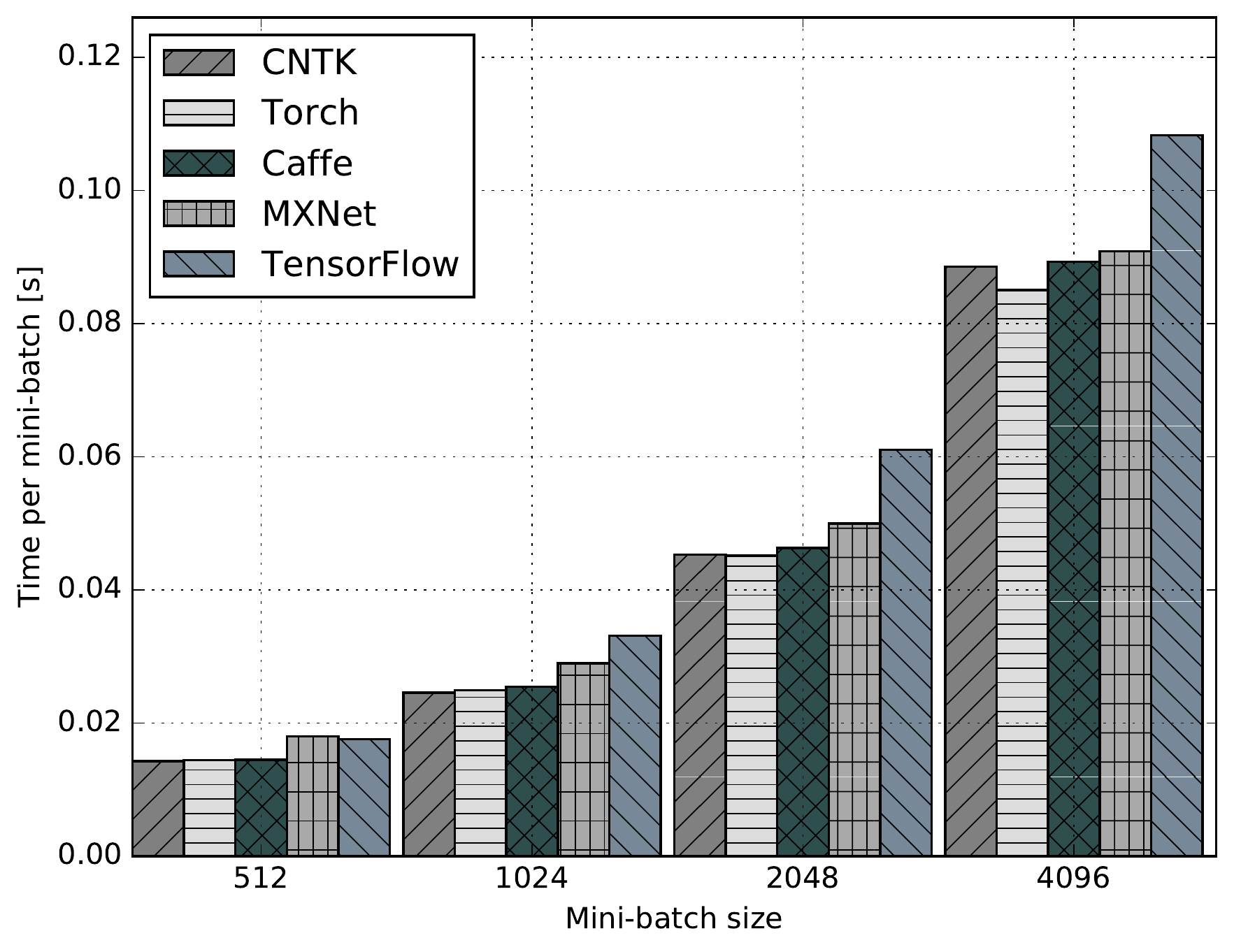}
  }
  \caption{The performance comparison of FCN-R on GPU platforms.}
  \label{fig:realgpubarsfcn5}
\end{figure*}

\textit{FCN-R}.
The performance comparison among different tools on GPUs is shown in Fig. \ref{fig:realgpubarsfcn5}. Torch achieves the best performance in all cases. Caffe, CNTK and MXNet have very similar results with the size of mini-batch becoming larger, which are better than TensorFlow. With the mini-batch size increasing, the speed efficiency also increases except a special case of Caffe with mini-batch size of 4096 which is worse than smaller size.

\begin{figure*}[htbp]
  \centering     
  \subfigure[Results on Tesla K80.]
  {
    \includegraphics[width=0.3\linewidth]{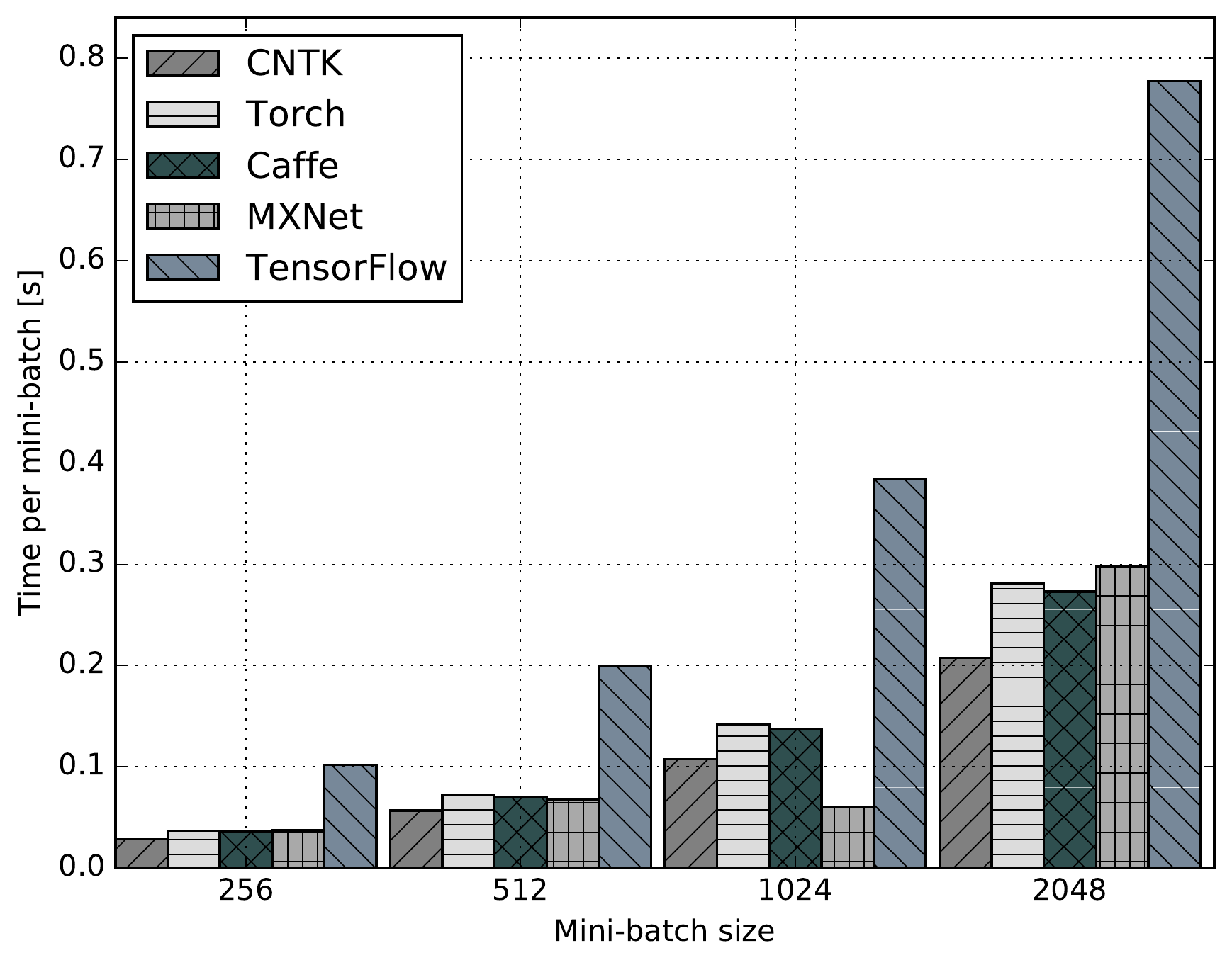}
  }
  \subfigure[Results on GTX1080.]
  {
    \includegraphics[width=0.3\linewidth]{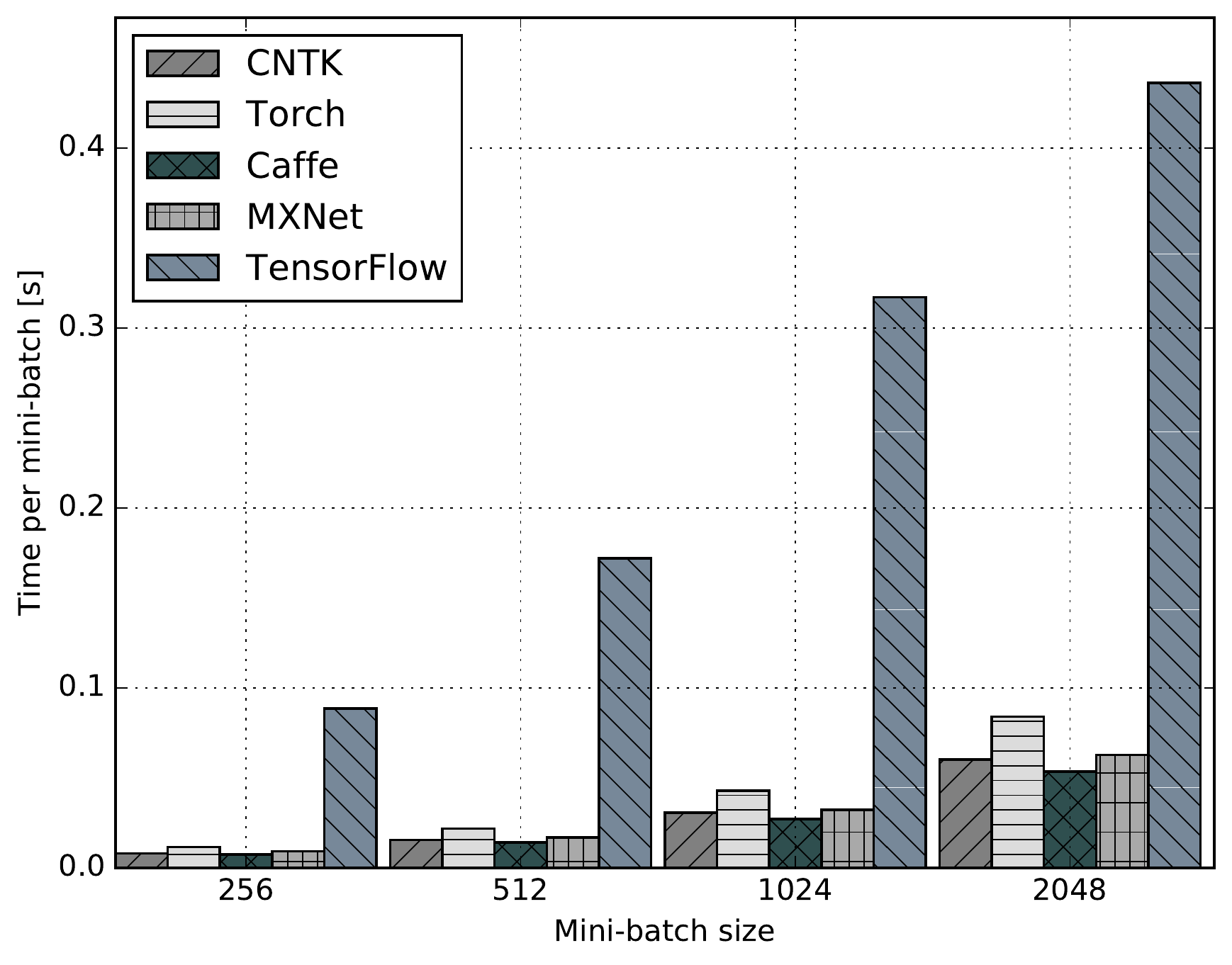}
  }
  \subfigure[Results on GTX980.]
  {
    \includegraphics[width=0.3\linewidth]{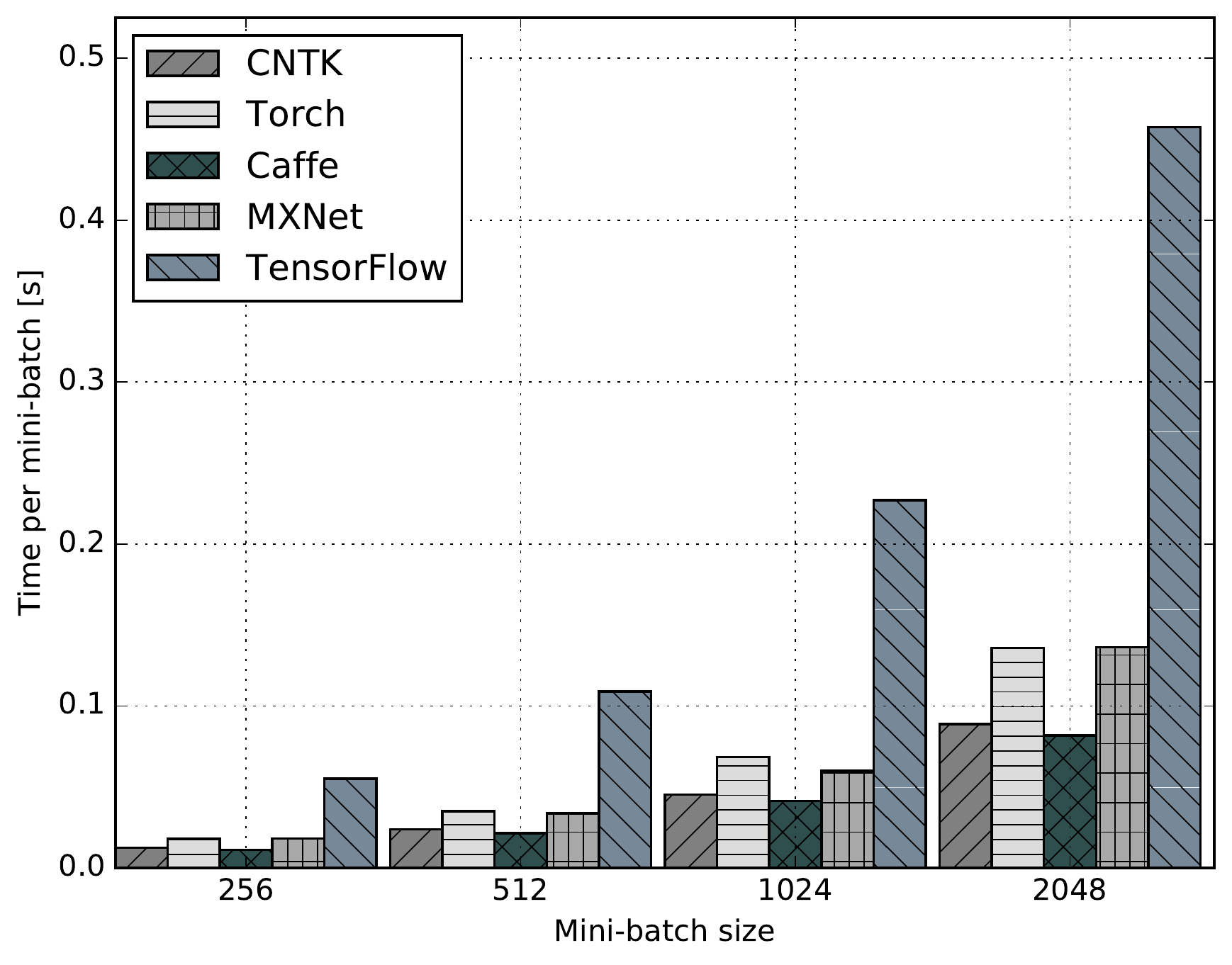}
  }
  \caption{The performance comparison of AlexNet-R on GPU platforms.}
  \label{fig:realgpubarsalexnet}
\end{figure*}

\textit{AlexNet-R}.
The performance comparison among different tools on GPUs is shown in Fig. \ref{fig:realgpubarsalexnet}. On K80 platform, CNTK performs the best, while Caffe and Torch are slightly better than MXNet. TensorFlow, however, results in the worst time efficiency. The performance of MXNet becomes worse when the size of mini-batch is set to be larger than 1024. On the contrary, Caffe results in the best performance in AlexNet, and CNTK and MXNet have very close efficiency which is slightly better than Torch.

\begin{figure*}[htbp]
  \centering     
  \subfigure[Results on Tesla K80.]
  {
    \includegraphics[width=0.3\linewidth]{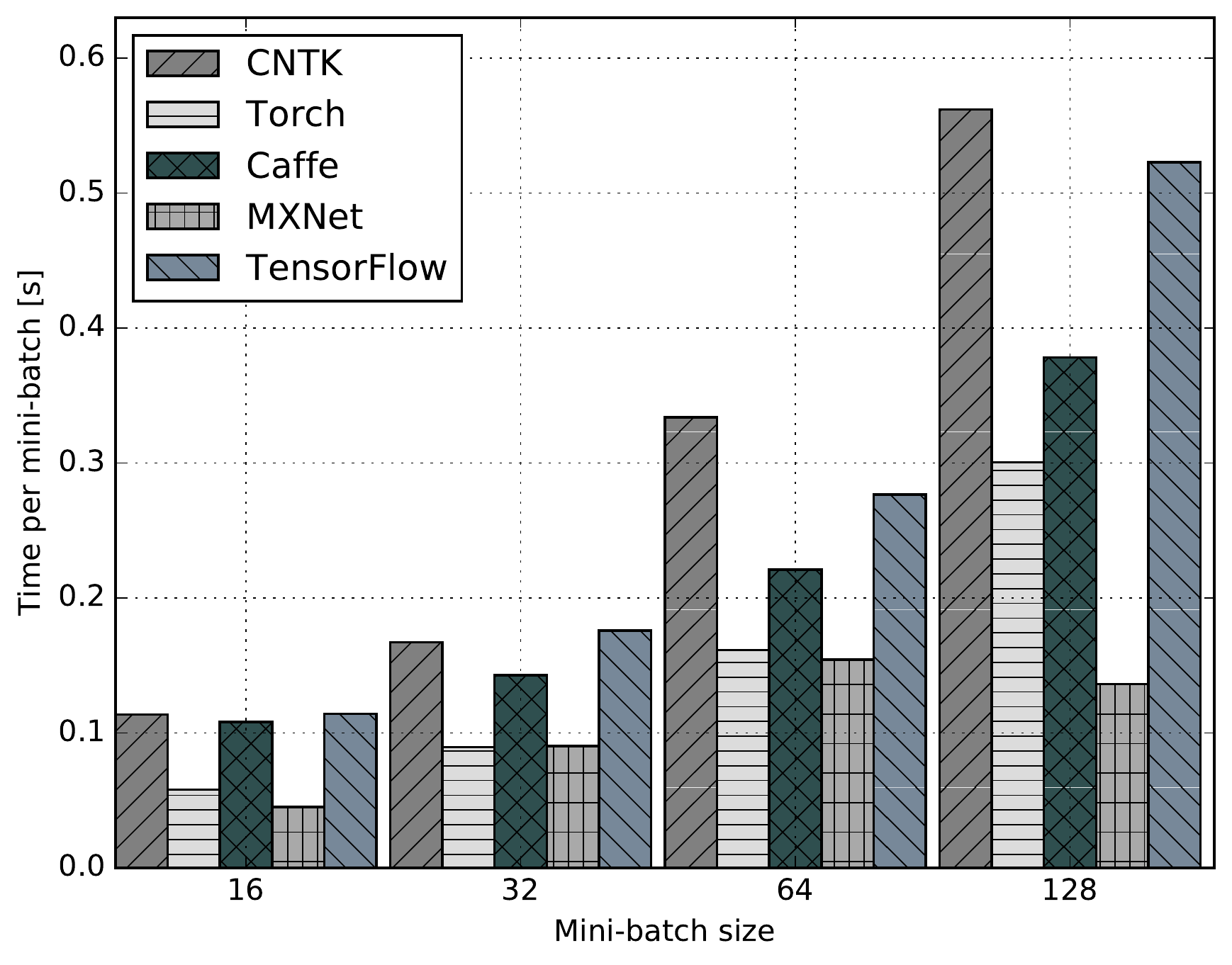}
  }
  \subfigure[Results on GTX1080.]
  {
    \includegraphics[width=0.3\linewidth]{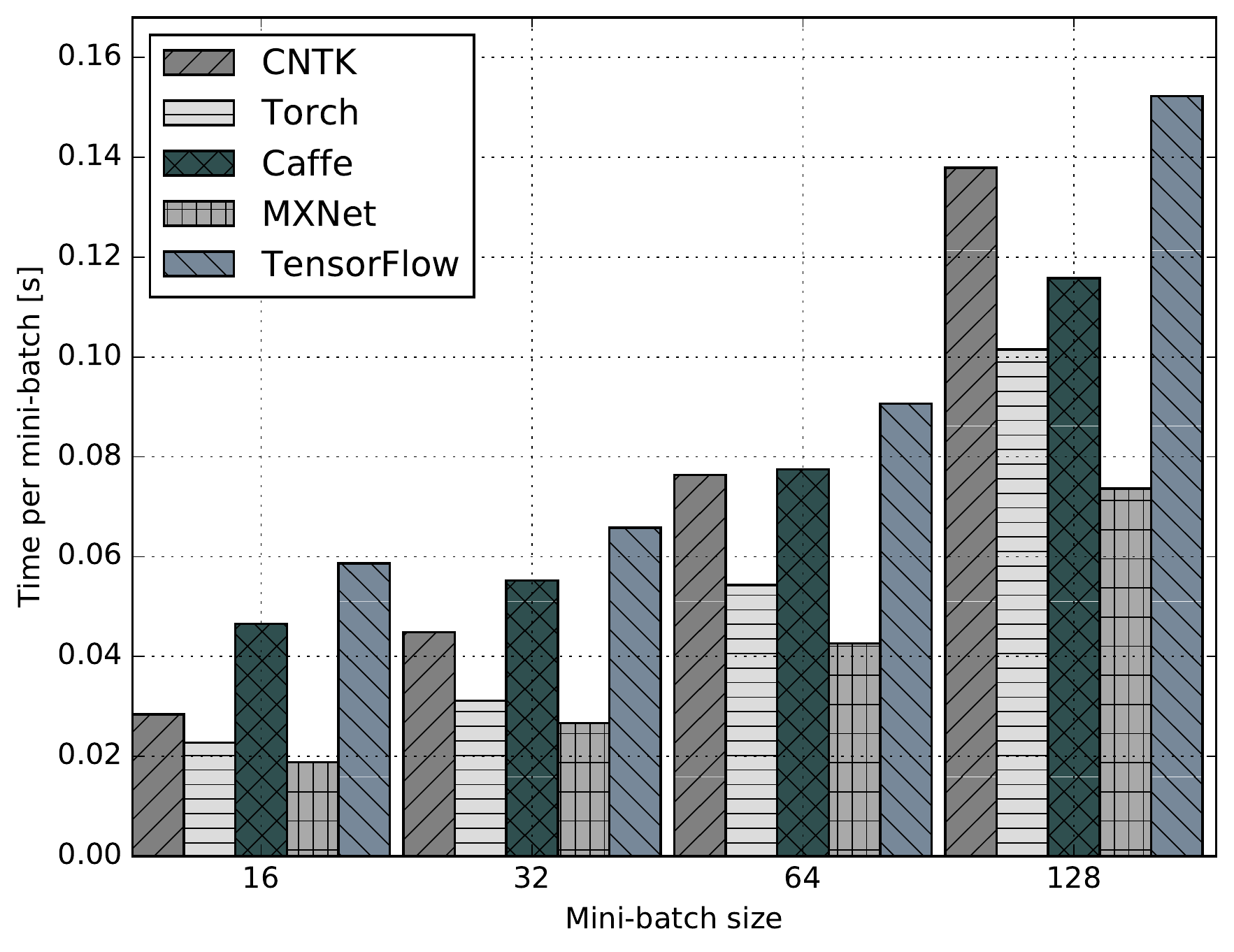}
  }
  \subfigure[Results on GTX980.]
  {
    \includegraphics[width=0.3\linewidth]{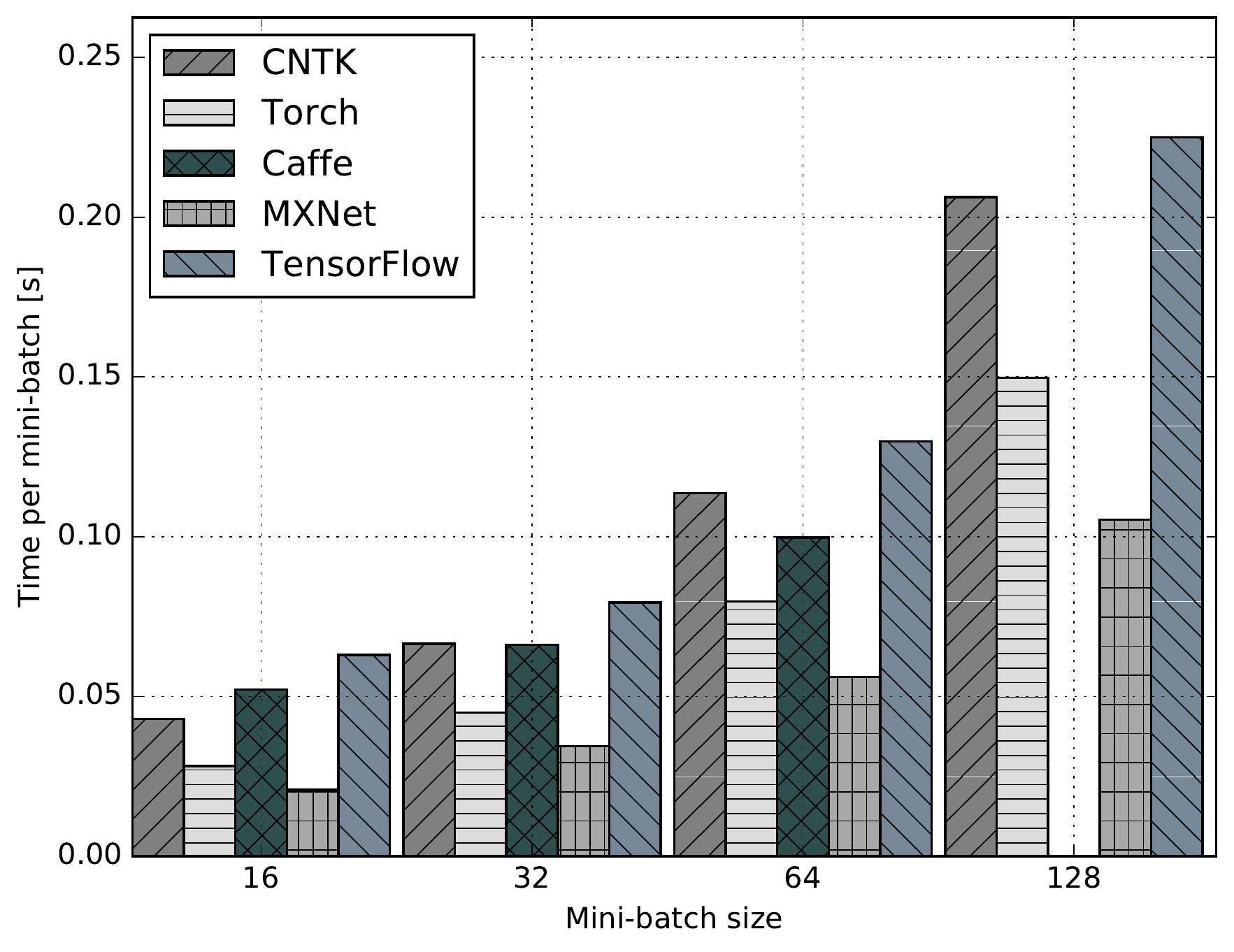}
  }
  \caption{The performance comparison of ResNet-56 on GPU platforms.}
  \label{fig:realgpubarsresnet}
\end{figure*}

\textit{ResNet-56}.
The performance comparison among different tools on GPUs is shown in Fig. \ref{fig:realgpubarsresnet}. Torch results in the best performance, followed by MXNet, while CNTK and TensorFlow have similar results which are slightly worse than Caffe on K80 card. In contrast, MXNet achieves the best performance, while Caffe, CNTK and Torch have similar result which is better than TensorFlow. On the smaller memory size of GPU, Caffe runs out of memory with the mini-batch size of 128, but it achieves better results than CNTK and TensorFlow with smaller mini-batch size of 64 or less on GTX980 card.

\begin{figure*}[htbp]
  \centering     
  \subfigure[Results on Tesla K80.]
  {
    \includegraphics[width=0.3\linewidth]{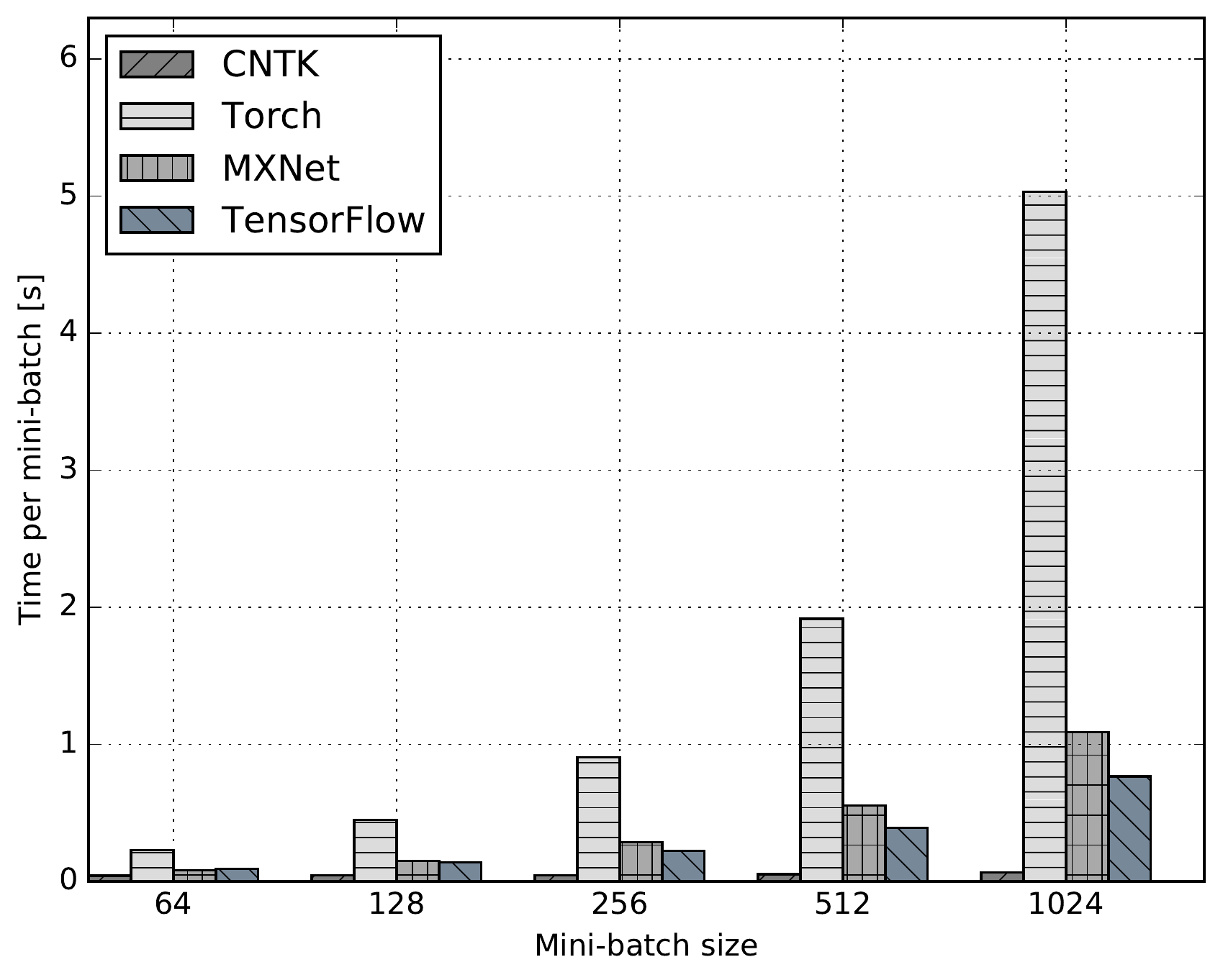}
  }
  \subfigure[Results on GTX1080.]
  {
    \includegraphics[width=0.3\linewidth]{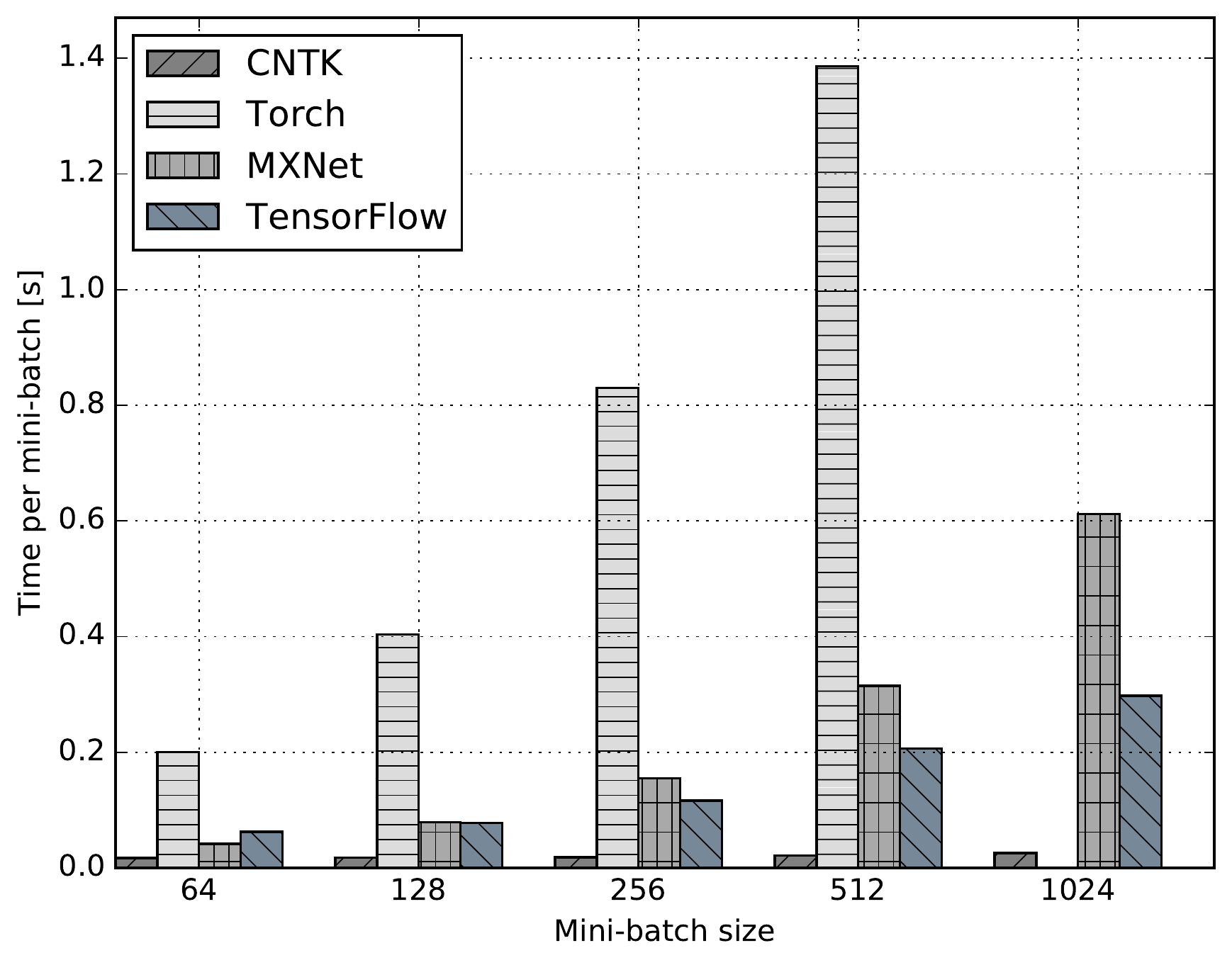}
  }
  \subfigure[Results on GTX980.]
  {
    \includegraphics[width=0.3\linewidth]{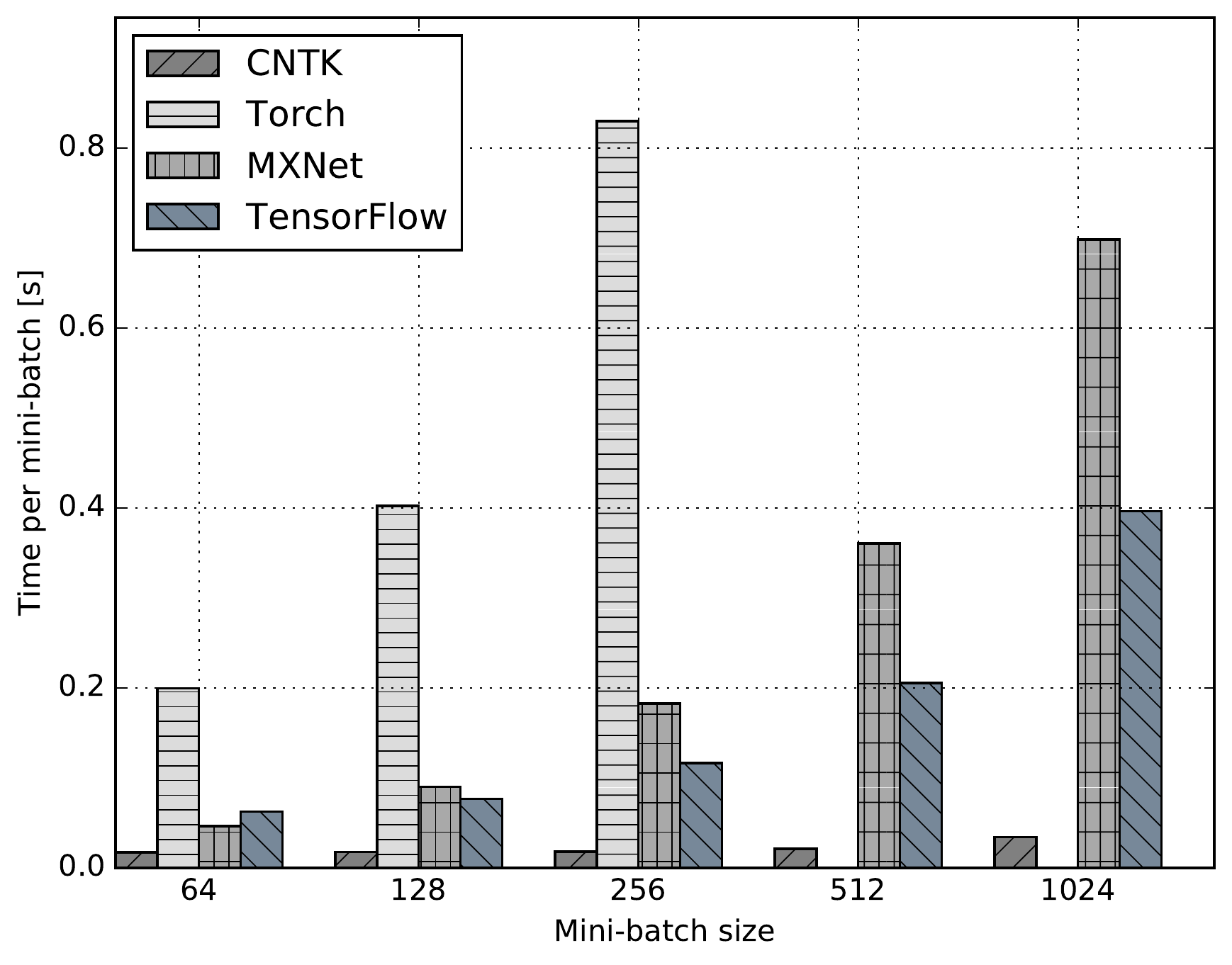}
  }
  \caption{The performance comparison of LSTM on GPU platforms.}
  \label{fig:realgpubarslstm}
\end{figure*}

\textit{LSTM}.
The performance comparison among different tools on GPUs is shown in Fig. \ref{fig:realgpubarslstm}. CNTK surpasses all other tools with many times. Taken the mini-batch size of 1024 on K80 as an example, CNTK only takes about 65 milliseconds, while MXNet, TensorFlow and Torch have up to about 1000, 767 and 5000 milliseconds respectively.

\subsection{Multi-GPU Results}
\begin{figure*}[htbp]
  \centering     
  \subfigure[Performance comparison of FCN-R on mulit-GPU platform.]
  {
    \includegraphics[width=0.45\linewidth]{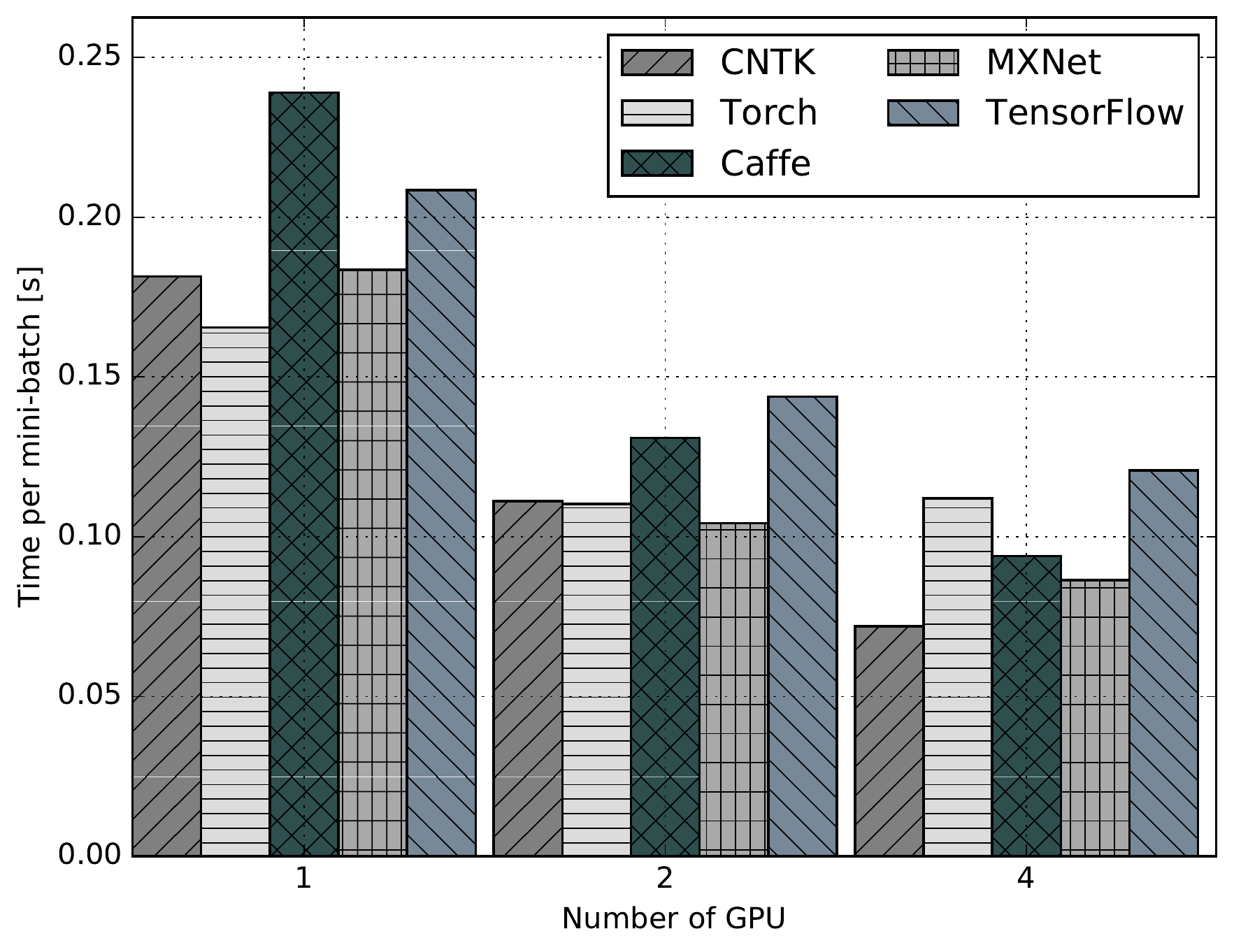}
    \label{fig:multigpufcn5a}
  }
  \subfigure[Convergent speed on multi-GPU platform.]
  {
    \includegraphics[width=0.47\linewidth]{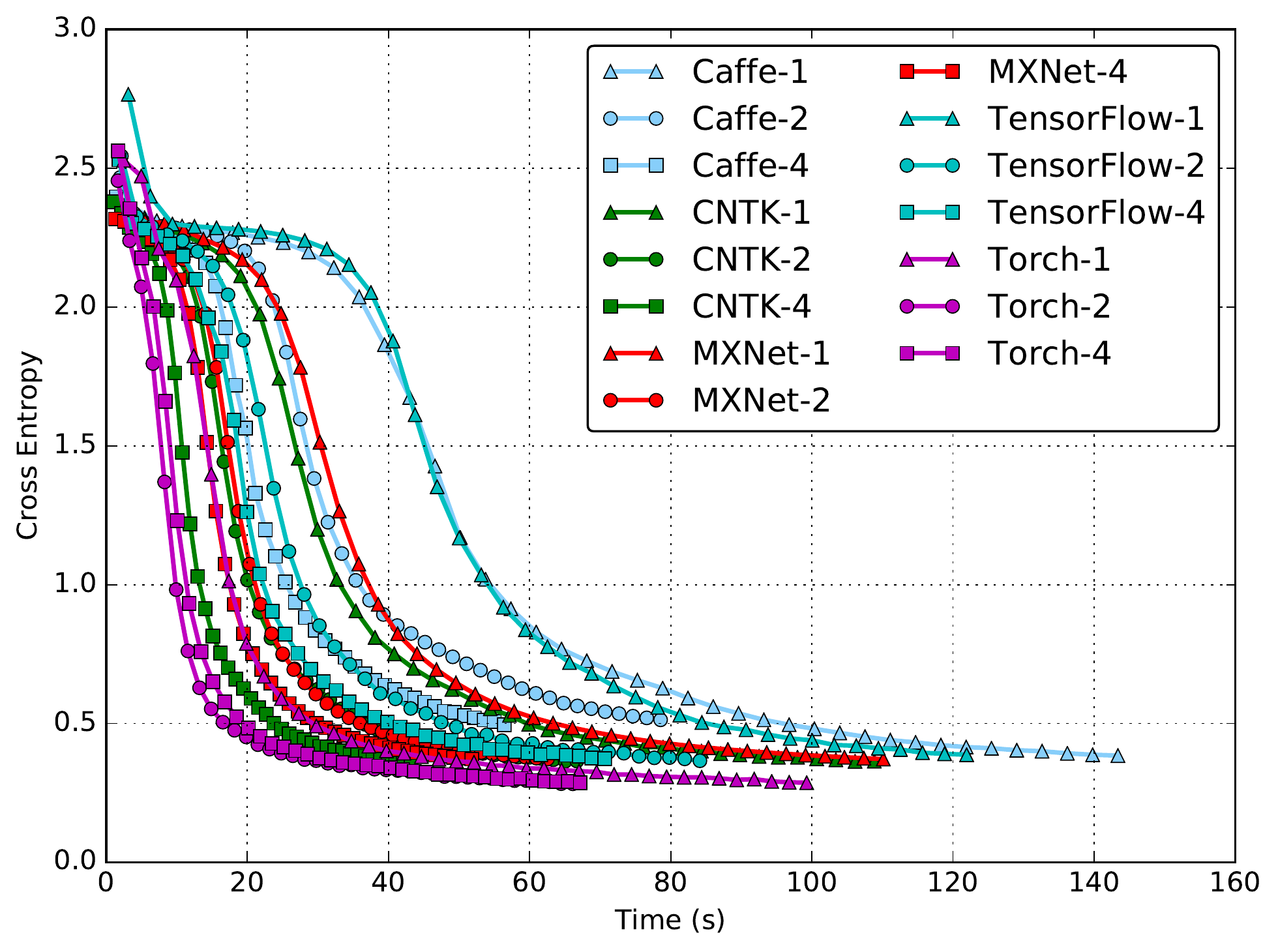}
    \label{fig:multigpufcn5b}
  }
  \caption{Results of FCN-R with a mini-batch size of 4096 on multiple GPUs. (The suffix number of the item in the legend represents the number of GPU used by that tool.)}
  \label{fig:multigpufcn5}
\end{figure*}

\textit{FCN-R}.
The size of mini-batch is set to 4096 in our test, and the results are shown in Fig. \ref{fig:multigpufcn5}. From Fig. \ref{fig:multigpufcn5a}, we can see the speeds of Caffe, CNTK and MXNet are very similar on a single GPU, which are slightly better than TensorFlow and Torch. With the number of GPUs doubled, the scalability of CNTK and MXNet are the best, and they both achieve about 35\% speedup, while caffe achieves about 28\% speedup which is better than that of Torch and TensorFlow with about 10\%. When we scale from 2 GPUs to 4 GPUs, TensorFlow and Torch don't achieve further speedup.

Regarding the convergence performance, we find that the convergence rate is faster with the number of GPU increasing from Fig. \ref{fig:multigpufcn5b}. On a single GPU, the training with Torch is converged fastest, and Caffe, CNTK and MXNet are faster than TensorFlow. When the number of GPU increases to 4, the convergence rates of CNTK and MXNet tend to be close to Torch, while Caffe and TensorFlow are converged relatively slow.

\begin{figure*}[htbp]
  \centering     
  \subfigure[Performance comparison of AlexNet-R on mulit-GPU platform.]
  {
    \includegraphics[width=0.45\linewidth]{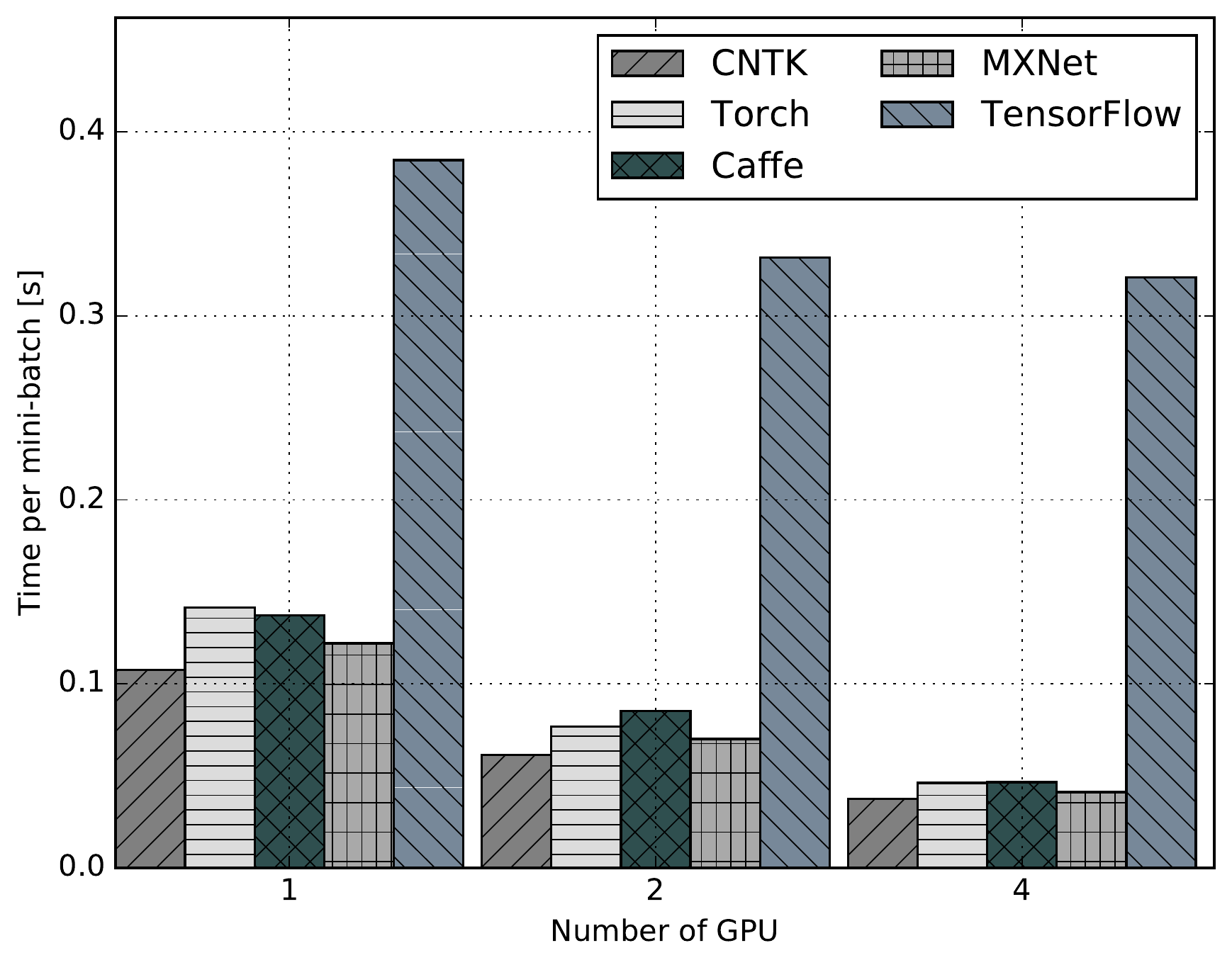}
    \label{fig:multigpualexneta}
  }
  \subfigure[Convergent speed on multi-GPU platform.]
  {
    \includegraphics[width=0.47\linewidth]{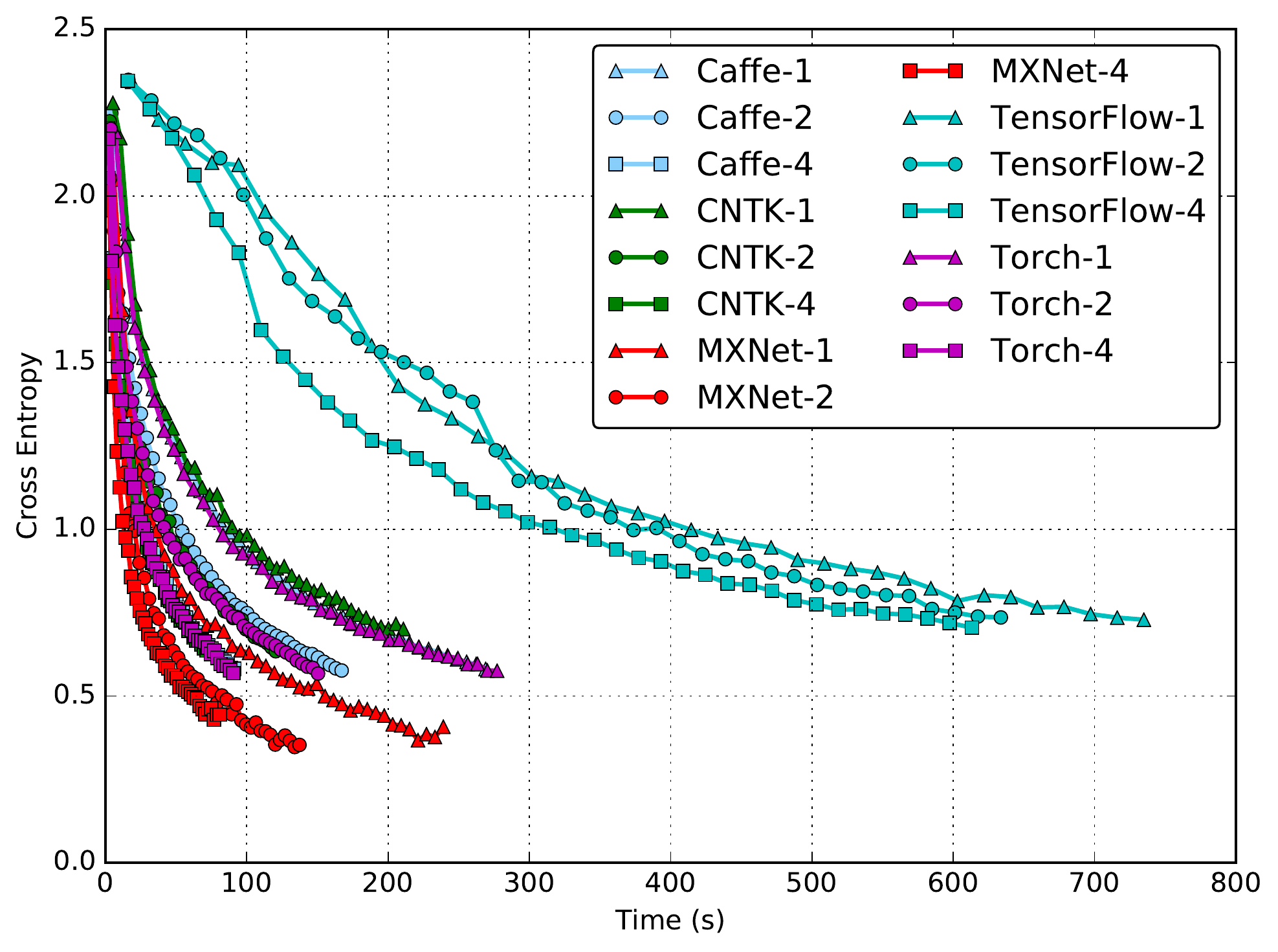}
    \label{fig:multigpualexnetb}
  }
  \caption{Results of AlexNet-R with a mini-batch size of 1024 on multiple GPUs. (The suffix number of the item in the legend represents the number of GPU used by that tool.)}
  \label{fig:multigpualexnet}
\end{figure*}

\textit{AlexNet-R}.
The results are shown in Fig. \ref{fig:multigpualexnet} with a mini-batch size of 1024.  On a single GPU, CNTK, MXNet and Torch display the similar performance, which is much better than that of Caffe and TensorFlow. With the number of GPU doubled, all the tools achieve up to 40\% speedup except TensorFlow which improves the performance by only about 30 percent.

Similar to the convergence performance on FCN-R, the convergence rate is obviously accelerated with larger number of GPUs. On all the cases of different number of GPUs, MXNet and Torch achieve the fastest convergence rate, while CNTK is slightly slower, which is still much faster than that of Caffe and TensorFlow.

\begin{figure*}[htbp]
  \centering     
  \subfigure[Performance comparison of ResNet-56 on mulit-GPU platform.]
  {
    \includegraphics[width=0.45\linewidth]{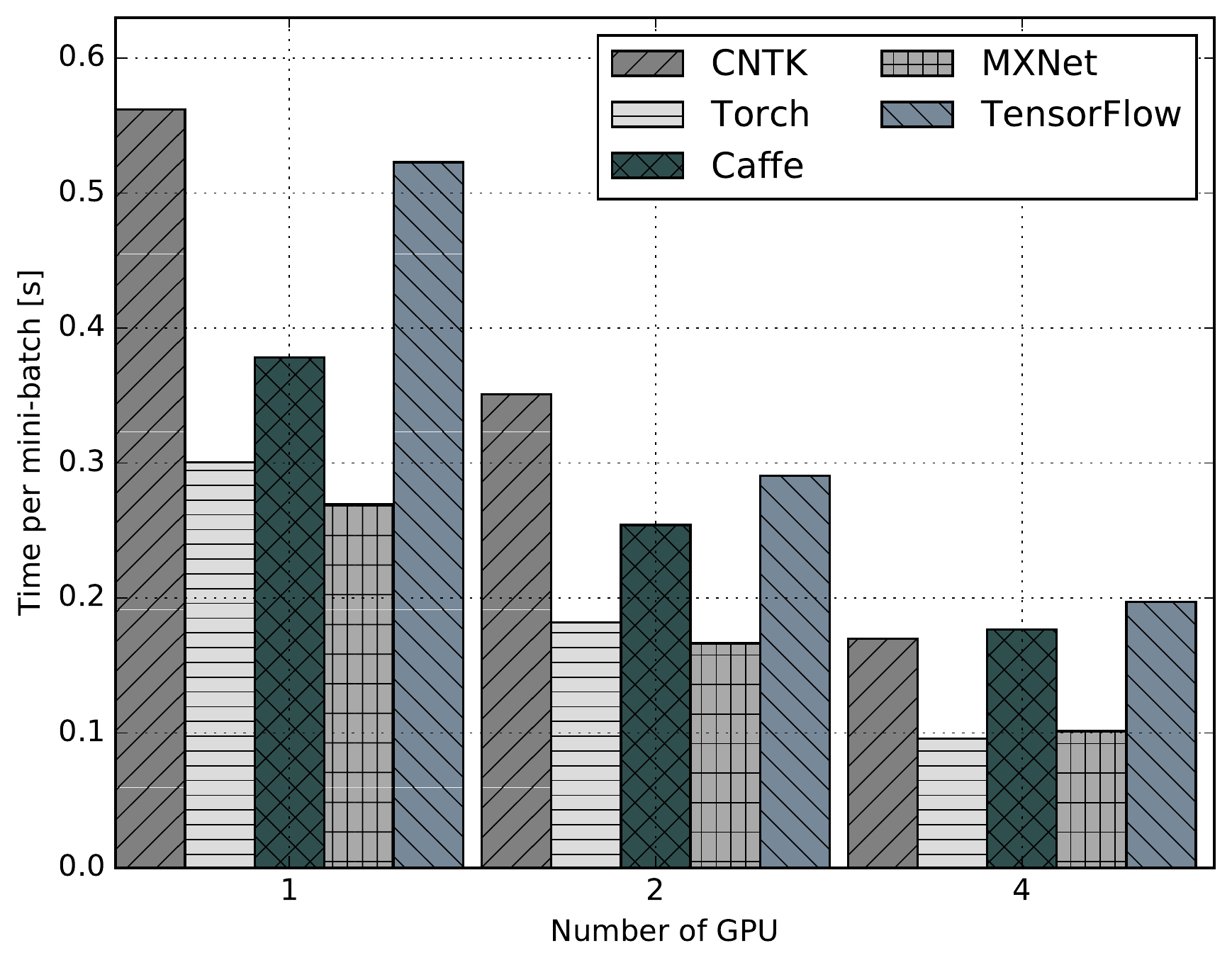}
    \label{fig:multigpuresneta}
  }
  \subfigure[Convergent speed on multi-GPU platform.]
  {
    \includegraphics[width=0.47\linewidth]{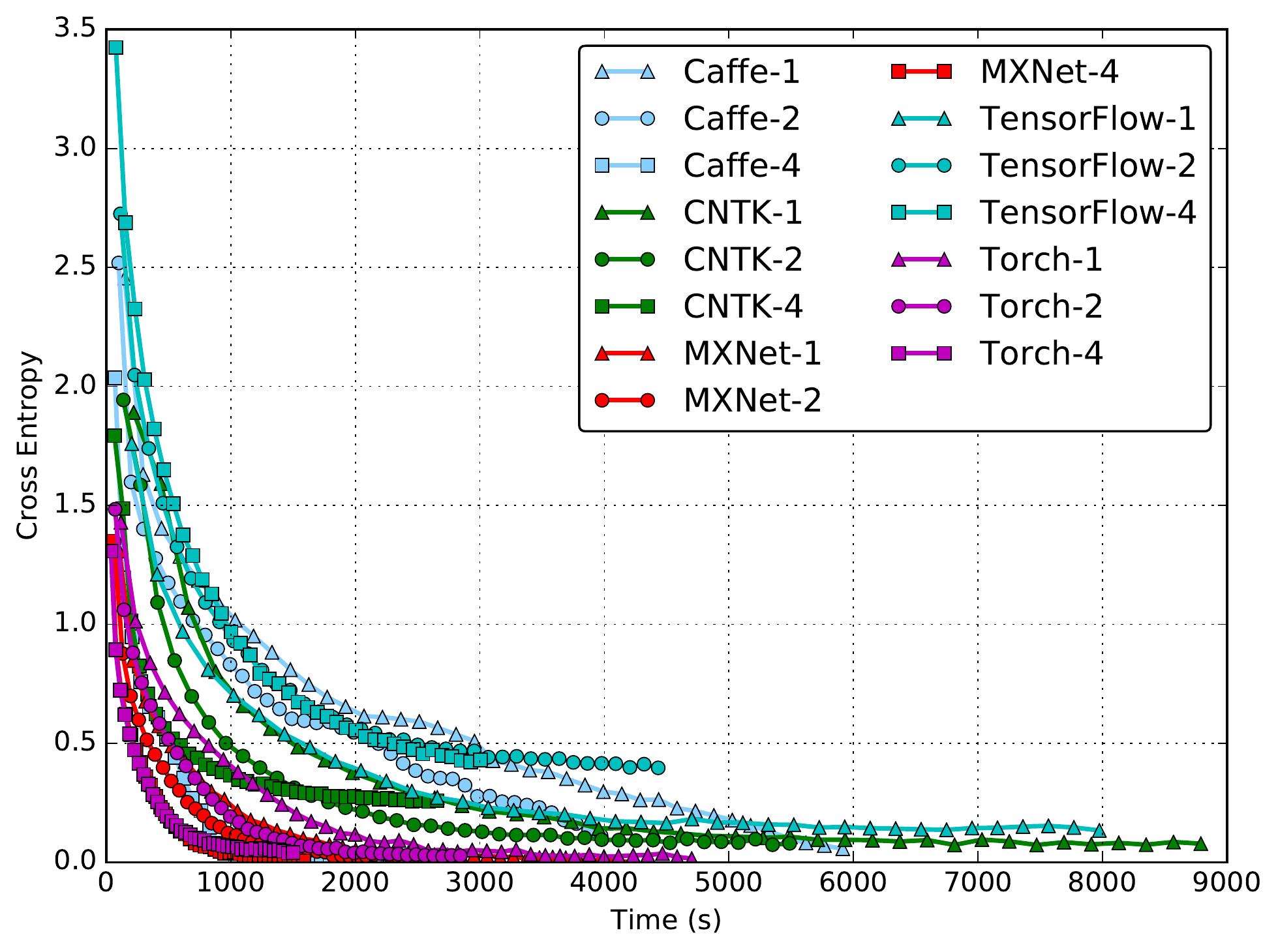}
    \label{fig:multigpuresnetb}
  }
  \caption{Results of ResNet-56 with a mini-batch size of 128 on multiple GPUs. (The suffix number of the item in the legend represents the number of GPU used by that tool.)}
  \label{fig:multigpuresnet}
\end{figure*}

\textit{ResNet-56}.
The results are shown in Fig. \ref{fig:multigpuresnet} with the mini-batch size of 128. And the corresponding convergence rates are shown in Fig. \ref{fig:multigpuresnetb}. In this larger size of network, MXNet achieves the best efficiency if there is only one GPU, while Torch tends to be more efficient on multiple GPUs.

Regarding the convergence speed, MXNet and Torch have better results than the other three tools, and Caffe is the slowest. Caffe, however, achieves better convergence rate with the number of GPUs increases when compared to TensorFlow. It can be noted that the speed of convergence is not changed much when increasing the number of GPUs from 2 to 4 on TensorFlow.

\section{Discussion} \label{discussion}

Considering parallelization on CPUs, the number of computation threads are recommended to be not larger than the number of physical CPU cores. Since it needs additional CPU resource to do the thread scheduling during the computing progress, if the CPU resourses are all exploited for computation, it is difficult to achieve higher performance. However, with the help of BLAS library of Eigen \cite{eigen2016}, which is optimized for SIMD instructions \cite{vanhoucke2011improving}, the performance of TensorFlow is better than others with the number of CPU cores increased.

On FCNs, Caffe, CNTK and Torch perform a little better than MXNet and TensorFlow with one GPU. In general, training a network involves a two-phase computation (i.e., feed-forward and backward propagation). In feed-forward phase, matrix multiplications are the most time-consuming operations, and cuBLAS API: cublasSgemm is adopted by all the four tools. However, if we want to multiply matrix $A$ and the transpose of matrix $B$, we can set the second parameter of cublasSgemm API to CUBLAS\_OP\_T, which will apply in-place matrix transpose and result in up to 3 times slower performance as compared to matrix multiply without transpose (e.g., $C=A\times B^T$, where $A\in R^{1024\times 26752}$ and $B\in R^{2048\times 26752}$). This is because in-place matrix transpose is very time-consuming. CNTK and TensorFlow construct its data structure to call cublasSgemm use CUBLAS\_OP\_N, while Caffe and Torch use CUBLAS\_OP\_T. In the phase of backward propagation, it needs to use matrix multiplication to calculate the gradients and use element-wise matrix operation to update the parameters. When it comes to low efficiency computation of $A$ times transposition of $B$ by calling cuBLAS, it may be better if we transpose $B$ first (out-place if the GPU has enough memory) and then apply matrix multiply. Furthermore, the cublasSgemm API provides the full support to backward propagation because it adds a scaled (parameter beta as scalar) matrix after matrix multiplication. So if we merge gradients calculation and update operation into a single GPU kernel, the calculation efficiency could be improved. To optimize the efficiency of FCNs, it is possible to use cublasSgemm API without transpose and use cublasSgemm to calculate gradients and do update operations at the same time.

On CNNs, all the toolkits use cuDNN library to do convolution operations. Though the same API calls are used, the parameters may result in different GPU kernels to use. We found that FFT is a better solution for convolution operations compared to performing convolution directly in many cases. After FFT of a matrix, convolution calculation can be transformed into inner product operation which is much faster.

On RNNs with LSTM, CNTK performs much better than TensorFlow and Torch, both of which achieve similar performance. To launch training procedure of LSTM, Torch executes lots of pointwise kernels of basic operations such as multiplication, addition, sigmoid, etc. on tensors designed by itself. Regarding the workload of kernels, Torch gives more than 50 blocks of which size is batch size. In this way Torch can somehow fully utilize the computation resources of GPU. We know that TensorFlow organizes all the computations as graph operations \cite{abadi2015tensorflow}, which is also indicated by the kernels launched during the training procedure to some extent. There are a large number of TensorAssignOp kernels with different types of operations also including scalar\_sum, scalar\_product etc. on tensors. As for CNTK, some specific kernels for training LSTM designed by itself are also very different from those of Torch and TensorFlow. Its brain scripts are very flexible to customize the neural networks.

Regarding the data parallelization by using multiple GPUs, the scaling is heavily influenced by process of gradient aggregation which requires data transfer through PCIe. The highest throughput of PCIe3.0 with Telsa K80 is about 8GB/s in our testing platform, which means that it takes 0.0256 seconds to transfer the gradients from GPU to CPU in our FCN-R case. But the computation time of one mini-batch is only about 100ms. Therefore, it is critical to cut down the overhead of transferring data between GPU and CPU. The performance is very different among different tools which is much related to the strategies of parallel design. The gradient averaging and updating are performed on GPU side in Caffe, but it uses a tree reduction strategy. E.g., if there are 4 GPUs for training, two pairs of GPUs will exchange gradients (i.e., GPU 0 exchanges with GPU 1 and GPU 2 exchanges with GPU 3) first, and then GPU 0 exchanges with GPU 2. After that GPU 0 takes the responsibility of calculating updated model and then transfers the model to GPU 1, and then 0 transfers the model to 1 and 2 transfers the model to 3 responsibility in parallel. So the performance of scalability of Caffe heavily depends on PCI-e topology of the system \cite{caffemultigpu}. The authors of CNTK add 1-bit stochastic gradient descent technique \cite{seide20141} into the framework, which means that the PCI-e time of exchanging gradients can be shorten by many times. Therefore, even though by using large network, the scalability of CNTK is also good. MXNet also shows good scalability on this kind of networks because it calculates the gradient aggregation on GPU, which not only reduces the PCI-e time of transferring gradients frequently, but also exploits GPU resource to do the computation in parallel. TensorFlow, however, calculates the gradient aggregation and updated model on the CPU side, which not only needs much time in tranferring gradients through PCI-e, but also updates the model in a serial algorithm by using one CPU. So the scalability of TensorFlow is not as good as other tools. Torch is similar to TensorFlow in scaling to multiple GPUs. Both gradient aggregation and update are performed in the CPU side, but Torch uses the parallel algorithm to utilize all the available CPU resources. Thus, the performance of scalability is slightly better than that of TensorFlow, but still worse than Caffe, CNTK and MXNet.

With GPU computing resources, all the deep learning tools mentioned achieve very high speedups compared to their CPU-only versions. This is not surprising because the performance of matrix multiplication and FFT on the tested GPUs are significantly better than on CPUs.

\section{Conclusion and Future Work} \label{conclusionandfuturework}
This work aims to evaluate the running performance of a set of modern deep learning software tools and see how they perform on different types of neural networks and different hardware platforms. Our experimental results show that all tested tools can make good use of GPUs to achieve significant speedup over their CPU counterparts. However, there is no single software tool that can consistently outperform others, which implies that there exist some opportunities to further optimize the performance.

We have two directions of future work on the agenda. First of all, we plan to include more deep learning software tools (such as Paddle from Baidu) and hardware platforms (such as AMD's GPU, and Intel Xeon Phi) into this benchmarking study. Secondly, we plan to evaluate the scalability of the tools on a high-performance GPU cluster.

\bibliographystyle{IEEEtran}
\Urlmuskip=0mu plus 1mu
\bibliography{arxiv-v7.bbl}

\section{Appendix} \label{appendix}
\subsection{Revision History}
This version (v7):
\begin{itemize}
\item Revise ResNet-50 configuration in MXNet.
\item Add faster implementation of ResNet-56 in TensorFlow with multiple GPUs.
\end{itemize}

Version 6 (v6, 25 Jan 2017):
\begin{itemize}
\item Includes results of multiple GPUs.
\item Include MXNet into our evaluation.
\item Update all tools to the latest major versions.
\item Include results of real data sets: MNIST and Cifar10.
\end{itemize}

Version 5 (v5, 19 Sep 2016):
\begin{itemize}
\item Revise a bug of AlexNet configuration in TensorFlow.
\item Add an update operation in AlexNet with Torch.
\end{itemize}

Version 4 (v4, 11 Sep 2016):
\begin{itemize}
\item Remedy the bug of ResNet-50 configuration in TensorFlow.
\item Change time measurement method of Caffe from ``caffe time'' to ``caffe train''.
\item Add an option of ``prefetch=true'' to configuration file of CNNs in CNTK.
\end{itemize}

Version 3 (v3, 3 Sep 2016):
\begin{itemize}
\item Correct the CUDA version on Table \ref{table:hardwaresetup} and re-test the experiments.
\item Revise minor difference of network configuration and delete some extra operations like dropout on AlexNet.
\item On RNN of CNTK, we remove an extra LSTM classification task which is not included in other tools and change the configuration file with ``SimpleNetworkBuilder'' to customized brain scripts.
\end{itemize}

\end{document}

%% file: totalresultsSingleGPU.tex
\begin{table*}[htbp]
\begin{threeparttable}
\centering
\caption{Comparative experiment results (time per mini-batch in second)}
\label{table:totalresultsSingleGPU}
\begin{tabular}{|l|l||l|l|l|l||l|l|l|l|l|l||l|l|l||}
\cline{1-15}
 & & \multicolumn{4}{c||}{Desktop CPU (Threads used)}& \multicolumn{6}{c||}{Server CPU (Threads used)}& \multicolumn{3}{c||}{Single GPU} \\\cline{1-15}
 & & 1 & 2 & 4 & 8& 1 & 2 & 4 & 8 & 16 & 32& G980 & G1080 & K80 \\\cline{1-15}
 & Caffe& 1.324 & 0.790 & \textbf{0.578} & 15.444& 1.355 & 0.997 & 0.745 & \textbf{0.573} & 0.608 & 1.130& 0.041 & \textbf{0.030} & 0.071 \\\cline{2-15}
 & CNTK& 1.227 & 0.660 & \textbf{0.435} & -& 1.340 & 0.909 & 0.634 & 0.488 & \textbf{0.441} & 1.000& 0.045 & \textbf{0.033} & 0.074 \\\cline{2-15}
 FCN-S& TF& 7.062 & 4.789 & 2.648 & \textbf{1.938}& 9.571 & 6.569 & 3.399 & 1.710 & 0.946 & \textbf{0.630}& 0.060 & \textbf{0.048} & 0.109 \\\cline{2-15}
 & MXNet& 4.621 & 2.607 & 2.162 & \textbf{1.831}& 5.824 & 3.356 & 2.395 & 2.040 & \textbf{1.945} & 2.670& - & \textbf{0.106} & 0.216 \\\cline{2-15}
 & Torch& 1.329 & 0.710 & \textbf{0.423} & -& 1.279 & 1.131 & 0.595 & 0.433 & \textbf{0.382} & 1.034& 0.040 & \textbf{0.031} & 0.070 \\\cline{1-15}
\hline \hline
 & Caffe& 1.606 & 0.999 & \textbf{0.719} & -& 1.533 & 1.045 & \textbf{0.797} & 0.850 & 0.903 & 1.124& 0.034 & \textbf{0.021} & 0.073 \\\cline{2-15}
 & CNTK& 3.761 & 1.974 & \textbf{1.276} & -& 3.852 & 2.600 & 1.567 & 1.347 & \textbf{1.168} & 1.579& 0.045 & \textbf{0.032} & 0.091 \\\cline{2-15}
 AlexNet-S& TF& 6.525 & 2.936 & 1.749 & \textbf{1.535}& 5.741 & 4.216 & 2.202 & 1.160 & \textbf{0.701} & 0.962& 0.059 & \textbf{0.042} & 0.130 \\\cline{2-15}
 & MXNet& 2.977 & 2.340 & 2.250 & \textbf{2.163}& 3.518 & 3.203 & 2.926 & 2.828 & \textbf{2.827} & 2.887& 0.020 & \textbf{0.014} & 0.042 \\\cline{2-15}
 & Torch& 4.645 & 2.429 & \textbf{1.424} & -& 4.336 & 2.468 & 1.543 & 1.248 & \textbf{1.090} & 1.214& 0.033 & \textbf{0.023} & 0.070 \\\cline{1-15}
\hline \hline
 & Caffe& 11.554 & 7.671 & \textbf{5.652} & -& 10.643 & 8.600 & 6.723 & \textbf{6.019} & 6.654 & 8.220& - & \textbf{0.254} & 0.766 \\\cline{2-15}
 & CNTK& - & - & - & -& - & - & - & - & - & -& 0.240 & \textbf{0.168} & 0.638 \\\cline{2-15}
 RenNet-50& TF& 23.905 & 16.435 & 10.206 & \textbf{7.816}& 29.960 & 21.846 & 11.512 & 6.294 & \textbf{4.130} & 4.351& 0.327 & \textbf{0.227} & 0.702 \\\cline{2-15}
 & MXNet& 48.000 & 46.154 & 44.444 & \textbf{43.243}& 57.831 & 57.143 & 54.545 & 54.545 & \textbf{53.333} & 55.172& 0.207 & \textbf{0.136} & 0.449 \\\cline{2-15}
 & Torch& 13.178 & 7.500 & \textbf{4.736} & 4.948& 12.807 & 8.391 & 5.471 & 4.164 & \textbf{3.683} & 4.422& 0.208 & \textbf{0.144} & 0.523 \\\cline{1-15}
\hline \hline
 & Caffe& 2.476 & 1.499 & \textbf{1.149} & -& 2.282 & 1.748 & 1.403 & 1.211 & 1.127 & \textbf{1.127}& 0.025 & \textbf{0.017} & 0.055 \\\cline{2-15}
 & CNTK& 1.845 & 0.970 & 0.661 & \textbf{0.571}& 1.592 & 0.857 & 0.501 & 0.323 & \textbf{0.252} & 0.280& 0.025 & \textbf{0.017} & 0.053 \\\cline{2-15}
 FCN-R& TF& 2.647 & 1.913 & 1.157 & \textbf{0.919}& 3.410 & 2.541 & 1.297 & 0.661 & 0.361 & \textbf{0.325}& 0.033 & \textbf{0.020} & 0.063 \\\cline{2-15}
 & MXNet& 1.914 & 1.072 & 0.719 & \textbf{0.702}& 1.609 & 1.065 & 0.731 & 0.534 & 0.451 & \textbf{0.447}& 0.029 & \textbf{0.019} & 0.060 \\\cline{2-15}
 & Torch& 1.670 & 0.926 & \textbf{0.565} & 0.611& 1.379 & 0.915 & 0.662 & 0.440 & 0.402 & \textbf{0.366}& 0.025 & \textbf{0.016} & 0.051 \\\cline{1-15}
\hline \hline
 & Caffe& 3.558 & 2.587 & \textbf{2.157} & 2.963& 4.270 & 3.514 & 3.381 & \textbf{3.364} & 4.139 & 4.930& 0.041 & \textbf{0.027} & 0.137 \\\cline{2-15}
 & CNTK& 9.956 & 7.263 & \textbf{5.519} & 6.015& 9.381 & 6.078 & 4.984 & \textbf{4.765} & 6.256 & 6.199& 0.045 & \textbf{0.031} & 0.108 \\\cline{2-15}
 AlexNet-R& TF& 4.535 & 3.225 & 1.911 & \textbf{1.565}& 6.124 & 4.229 & 2.200 & 1.396 & 1.036 & \textbf{0.971}& \textbf{0.227} & 0.317 & 0.385 \\\cline{2-15}
 & MXNet& 13.401 & 12.305 & 12.278 & \textbf{11.950}& 17.994 & 17.128 & 16.764 & \textbf{16.471} & 17.471 & 17.770& 0.060 & \textbf{0.032} & 0.122 \\\cline{2-15}
 & Torch& 5.352 & 3.866 & \textbf{3.162} & 3.259& 6.554 & 5.288 & 4.365 & \textbf{3.940} & 4.157 & 4.165& 0.069 & \textbf{0.043} & 0.141 \\\cline{1-15}
\hline \hline
 & Caffe& 6.741 & 5.451 & \textbf{4.989} & 6.691& 7.513 & \textbf{6.119} & 6.232 & 6.689 & 7.313 & 9.302& - & \textbf{0.116} & 0.378 \\\cline{2-15}
 & CNTK& - & - & - & -& - & - & - & - & - & -& 0.206 & \textbf{0.138} & 0.562 \\\cline{2-15}
 RenNet-56& TF& - & - & - & -& - & - & - & - & - & -& 0.225 & \textbf{0.152} & 0.523 \\\cline{2-15}
 & MXNet& 34.409 & 31.255 & \textbf{30.069} & 31.388& 44.878 & 43.775 & \textbf{42.299} & 42.965 & 43.854 & 44.367& 0.105 & \textbf{0.074} & 0.270 \\\cline{2-15}
 & Torch& 5.758 & 3.222 & \textbf{2.368} & 2.475& 8.691 & 4.965 & 3.040 & \textbf{2.560} & 2.575 & 2.811& 0.150 & \textbf{0.101} & 0.301 \\\cline{1-15}
\hline \hline
 & Caffe& - & - & - & -& - & - & - & - & - & -& - & - & - \\\cline{2-15}
 & CNTK& 0.186 & 0.120 & \textbf{0.090} & 0.118& 0.211 & 0.139 & 0.117 & 0.114 & \textbf{0.114} & 0.198& 0.018 & \textbf{0.017} & 0.043 \\\cline{2-15}
 LSTM& TF& 4.662 & 3.385 & 1.935 & \textbf{1.532}& 6.449 & 4.351 & 2.238 & 1.183 & 0.702 & \textbf{0.598}& 0.133 & \textbf{0.065} & 0.140 \\\cline{2-15}
 & MXNet& - & - & - & -& - & - & - & - & - & -& 0.089 & \textbf{0.079} & 0.149 \\\cline{2-15}
 & Torch& 6.921 & 3.831 & \textbf{2.682} & 3.127& 7.471 & 4.641 & 3.580 & \textbf{3.260} & 5.148 & 5.851& 0.399 & \textbf{0.324} & 0.560 \\\cline{1-15} 
\end{tabular}

\begin{tablenotes}
      \item[]Note: The mini-batch sizes for FCN-S, AlexNet-S, ResNet-50, FCN-R, AlexNet-R, ResNet-56 and LSTM  are 64, 16, 16, 1024, 1024, 128 and 128 respectively.
    \end{tablenotes}
  \end{threeparttable}
\end{table*}

%% file: totalresultsMultiGPU.tex
\begin{table}[htbp]
\centering
\begin{threeparttable}
\caption{Comparative experiment results between single GPU and multiple GPUs (time per mini-batch in second)}
\label{table:totalresultsMultiGPU}
\begin{tabular}{|l|l|l|l|l|}
\cline{1-5}
 & & \multicolumn{3}{c|}{\# of GK210} \\\cline{1-5}
 & & 1 & 2 & 4 \\\cline{1-5}
 & Caffe& 0.239 & 0.131 & \textbf{0.094} \\\cline{2-5}
 & CNTK& 0.181 & 0.111 & \textbf{0.072} \\\cline{2-5}
 FCN-R& TF& 0.208 & 0.144 & \textbf{0.121} \\\cline{2-5}
 & MXNet& 0.184 & 0.104 & \textbf{0.086} \\\cline{2-5}
 & Torch& 0.165 & \textbf{0.110} & 0.112 \\\cline{1-5}
\hline \hline
 & Caffe& 0.137 & 0.085 & \textbf{0.047} \\\cline{2-5}
 & CNTK& 0.108 & 0.062 & \textbf{0.037} \\\cline{2-5}
 AlexNet-R& TF& 0.385 & 0.332 & \textbf{0.321} \\\cline{2-5}
 & MXNet& 0.122 & 0.070 & \textbf{0.041} \\\cline{2-5}
 & Torch& 0.141 & 0.077 & \textbf{0.046} \\\cline{1-5}
\hline \hline
 & Caffe& 0.378 & 0.254 & \textbf{0.177} \\\cline{2-5}
 & CNTK& 0.562 & 0.351 & \textbf{0.170} \\\cline{2-5}
 RenNet-56& TF& 0.523 & 0.291 & \textbf{0.197} \\\cline{2-5}
 & MXNet& 0.270 & 0.167 & \textbf{0.101} \\\cline{2-5}
 & Torch& 0.301 & 0.182 & \textbf{0.096} \\\cline{1-5}
\end{tabular}

\begin{tablenotes}
      \item[]Note: The mini-batch sizes for FCN-R, AlexNet-R and ResNet-56 are 4096, 1024 and 128 respectively.
    \end{tablenotes}
  \end{threeparttable}
\end{table}